\documentclass[twocolumn,astrosymb,twocolappendix]{aastex7}
\usepackage{xspace}
\usepackage{CJKutf8}
\usepackage{bm}
\usepackage{appendix}
\usepackage{amsmath,amssymb}

\newcommand{\sbunit}{\mathrm{mag\ arcsec}^{-2}}

\newcommand{\elvesdwarf}{\textsc{ELVES-Dwarf}\xspace}

\newcommand{\code}[1]{\textbf{\texttt{#1}}}
\newcommand{\sersic}{S\'ersic\xspace}

\newcommand{\kms}{\mathrm{km\ s^{-1}}}
\newcommand{\rvir}{$R_{\rm vir}$\xspace}
\newcommand{\nsat}{$N_{\rm sat}$\xspace}

\newcommand{\numhosts}{39\xspace}
\newcommand{\numcands}{207\xspace}
\newcommand{\numsats}{39\xspace}

\newcommand{\catalogurl}{\url{https://elves-surveys.github.io/elves-dwarf}\xspace}

\usepackage{xcolor}
\usepackage{enumitem}

\defcitealias{Anand2021}{An21}
\defcitealias{Carlin2016}{C16}
\defcitealias{Carlin2019}{C19}
\defcitealias{Carlin2021}{C21}
\defcitealias{Carlin2024}{C24}
\defcitealias{Chazov2026}{C26}
\defcitealias{Carlsten2021}{Ca21}
\defcitealias{Collins2024}{Co24}
\defcitealias{Dalcanton2009}{D09}
\defcitealias{Danieli2023}{D23}
\defcitealias{Doliva-Dolinsky2025}{Do25}
\defcitealias{Fingerhut2010}{F10}
\defcitealias{Garling2020}{G20}
\defcitealias{Greco2018}{G18}
\defcitealias{Karachentsev1999DDO044}{K99}
\defcitealias{Karachentsev2013}{K13}
\defcitealias{Karachentsev2015LeoSpur}{K15}
\defcitealias{Karachentsev2017_DDO161}{K17}
\defcitealias{Karachentsev2020}{K20}
\defcitealias{Karachentsev2022}{K22}
\defcitealias{Koribalski2018}{K18}
\defcitealias{Martin2016}{M16}
\defcitealias{Li2025}{Paper I}
\defcitealias{Manwadkar2022}{M22}
\defcitealias{McConnachie2012}{M12}
\defcitealias{McNanna2024}{Mc24}
\defcitealias{McQuinn2014}{M14}
\defcitealias{McQuinn2015}{M15}
\defcitealias{McQuinn2017}{M17}
\defcitealias{Medoff2025}{Me25}
\defcitealias{Moustakas2023}{M23}
\defcitealias{Ogami2024}{O24}
\defcitealias{Pace2024}{P24}
\defcitealias{Sabbi2018}{S18}
\defcitealias{Sand2015}{S15}
\defcitealias{Sand2024}{S24}
\defcitealias{Tully2013}{T13}
\defcitealias{Tully2016}{T16}
\defcitealias{Tully2023}{T23}

\shorttitle{ELVES-Dwarf Survey II}
\shortauthors{Li et al.}
\graphicspath{{./}{figures/}}

\begin{document}
\begin{CJK*}{UTF8}{gbsn}

\title{\elvesdwarf. II. A Systematic Search for Satellite Systems of Dwarf Galaxies in the Local Volume}
\correspondingauthor{Jiaxuan Li}

\author[0000-0001-9592-4190]{Jiaxuan Li (李嘉轩)}
\email[show]{jiaxuanl@stanford.edu}
\affiliation{Department of Astrophysical Sciences, 4 Ivy Lane, Princeton University, Princeton, NJ 08540, USA}
\affiliation{Kavli Institute for Particle Astrophysics and Cosmology, Stanford University, Stanford, CA 94305, USA}

\author[0000-0002-5612-3427]{Jenny E. Greene}
\email{jgreene@astro.princeton.edu}
\affiliation{Department of Astrophysical Sciences, 4 Ivy Lane, Princeton University, Princeton, NJ 08540, USA}

\author[0000-0002-1841-2252]{Shany Danieli}
\email{sdanieli@tauex.tau.ac.il}
\affiliation{School of Physics and Astronomy, Tel Aviv University, Tel Aviv 69978, Israel}

\author[0000-0002-5382-2898]{Scott G. Carlsten}
\email{scarlsten@gmail.com}
\affiliation{Department of Astrophysical Sciences, 4 Ivy Lane, Princeton University, Princeton, NJ 08540, USA}

\author[0000-0002-2733-4559]{John Moustakas}
\email{jmoustakas@siena.edu}
\affiliation{Department of Physics and Astronomy, Siena College, 515 Loudon Rd., Loudonville, NY 12110, USA}

\author[0000-0002-7007-9725]{Marla Geha}
\email{marla.geha@yale.edu}
\affiliation{Department of Astronomy, Yale University, New Haven, CT 06520, USA}

\author[0000-0002-5011-5178]{Masayuki Tanaka}
\email{masayuki.tanaka@nao.ac.jp}
\affiliation{National Astronomical Observatory of Japan, Osawa 2-21-1, Mitaka, Tokyo 181-8588, Japan}

\author[0000-0001-6115-0633]{Fangzhou Jiang}
\email{fangzhou.jiang@pku.edu.cn}
\affiliation{Kavli Institute for Astronomy and Astrophysics, Peking University, Beijing 100871, China}

\author[0000-0003-0853-6427]{Ping Chen}
\email{ping.chen@zju.edu.cn}
\affiliation{Institute for Advanced Study in Physics, Zhejiang University, Hangzhou 310027, People’s Republic of China}

\author[0009-0003-7044-9751]{Sufia Birmingham}
\email{sb8685@alumni.princeton.edu}
\affiliation{Department of Astrophysical Sciences, 4 Ivy Lane, Princeton University, Princeton, NJ 08540, USA}




\begin{abstract}
We present the Exploration of Local VolumE Satellites of Dwarf Galaxies (\elvesdwarf) survey, a systematic census of satellite systems around dwarf hosts in the Local Volume. Our final sample comprises \numhosts predominantly isolated hosts with stellar masses $10^{7}<M_\star<10^{10}\,M_\odot$, including 32 hosts searched uniformly in this work and 7 drawn from the literature. We search for satellite candidates within the projected virial radius of each host using $155~\mathrm{deg}^2$ of Legacy Surveys imaging data. We determine satellite membership using surface brightness fluctuation distances from Subaru/HSC, Magellan/IMACS, and Gemini/GMOS imaging, supplemented by literature TRGB distances and radial velocities. From \numcands candidates, we confirm \numsats satellites with $M_\star>10^5\,M_\odot$ around \numhosts hosts. Above our fiducial completeness threshold of $M_\star\gtrsim10^{5.7}\,M_\odot$ and within the projected virial radius, 21 hosts have no confirmed satellites, 10 have one, six have two, and two have four, revealing substantial host-to-host scatter in satellite abundance. Overall, the observed satellite abundances and stellar mass functions are broadly consistent with predictions from the cosmological simulation TNG50 and galaxy formation models calibrated using Milky Way satellites. The projected radial distribution of the satellites is also consistent with theoretical expectations and with satellite populations around Milky Way-mass hosts. In contrast, the quenched fraction of satellites around dwarf hosts is substantially lower than around Milky Way-mass hosts, suggesting that environmental quenching is less efficient in dwarf halos. \elvesdwarf provides the first large, homogeneous, distance-confirmed sample of satellites around dwarf hosts and establishes a foundation for understanding galaxy formation and evolution in less-dense environments. 
\end{abstract}


\section{Introduction}\label{sec:intro}

Dwarf galaxies, the smallest and the most dark-matter-dominated systems in the Universe, are sensitive probes of the $\Lambda$CDM cosmological paradigm and provide critical constraints on the physics that shapes galaxy formation and evolution \citep{Bullock2017,Sales2022}. The abundance and internal properties, such as dark matter profile, stellar population, and gas content, encode critical information to answer a wide range of questions, from dark matter models \citep{Klypin1999,Moore1999,Nadler2021,Drlica-Wagner2022} to the influence of reionization, baryonic feedback, and environmental effects on dwarf galaxies \citep{Bullock2000,Hopkins2011,Wetzel2015}. Much of this insight has come from the rich satellite populations of the Milky Way (MW) and M31, which reveal a remarkable diversity in their distribution, star formation, gas content, and chemical enrichment histories \citep[e.g.,][]{Mateo1998,Weisz2011,McConnachie2012,Simon2019,Sacchi2021,Savino2025,Tan2025,Geha2026}. More recently, systematic searches for satellites around MW analogs beyond the Local Group ($<1-2$~Mpc) have further extended our view. Targeted studies \citep[e.g.,][]{Danieli2017,Tanaka2018,Smercina2018,Crnojevic2019,Bennet2019,Mutlu-Pakdil2024} and large surveys such as the Exploration of Local VolumE Satellites survey \citep[ELVES;][]{CarlstenELVES2022} and the Satellites Around Galactic Analogs survey \citep[SAGA;][]{SAGA-I,SAGA-III} have enabled detailed population-level comparisons across a wide range of environments, placing the MW in a broader cosmological context and revealing how satellite systems vary across host properties \citep[e.g.,][]{Muller2018,ELVES-I,Greene2023,Danieli2023,SAGA-IV,SAGA-V,Zhu2025}.

A key prediction of the $\Lambda$CDM model is that dwarf galaxies, despite residing in much lower-mass dark matter halos than the MW, should host their own satellites \citep[e.g.,][]{Sales2013,Wheeler2015,Dooley2017a,Dooley2017b,Munshi2019,Santos-Santos2022,Nadler2023}. The Magellanic Clouds \citep{Sales2017} provide the nearest example of such a hierarchy: the Large Magellanic Cloud (LMC) hosts the Small Magellanic Cloud (SMC) as a bright satellite and is likely associated with several ultra-faint dwarfs \citep{Kallivayalil2018,Erkal2020,Santos-Santos2021,Battaglia2022}. However, the LMC group represents only a single example and is embedded within the gravitational potential of the Local Group and has been strongly influenced by both the MW and the broader Local Group environment, making comparisons to simple $\Lambda$CDM expectations much less straightforward. Systematically characterizing the satellite populations of dwarf hosts, ideally in isolation, provides a powerful probe of $\Lambda$CDM on small scales and offers insight into how satellite formation and evolution proceed in less-dense environments.

Satellites of dwarf galaxies have been identified through deep imaging studies of individual systems \citep{Martinez-Delgado2012,Sand2015,Zhang2021_NGC6822,Kim2022,Muller2023,Sand2024,McNanna2024,Stierwalt2026}, including the Magellanic Analog Dwarf Companions And Stellar Halos (MADCASH) survey \citep{Carlin2016,Carlin2019,Hargis2020,Carlin2024}, the Large Binocular Telescope Satellites Of Nearby Galaxies Survey (LBT-SONG) survey \citep{Garling2021,Davis2021,Davis2024}, and the DELVE-DEEP survey \citep{Doliva-Dolinsky2025,Medoff2025}. However, many of these satellite detections lack direct distance measurements, leaving their physical association with the host uncertain. The heterogeneous depth, completeness, and radial coverage of existing searches make it difficult to assemble a uniform and statistically meaningful sample. A complementary approach is to estimate satellite abundance statistically by subtracting background galaxy counts from photometric samples \citep[e.g.,][]{Tanaka2018,Wang2021,Li2022,Bhattacharyya2024,Wang2025,Pan2026}, though this sacrifices information on individual systems. \citet{Paudel2026} identified rich dwarf-galaxy groups using SDSS and DESI spectroscopy, but the flux limits of these surveys restrict the search to relatively bright companions. A systematic search extending to fainter satellites, with direct distance measurements and well-characterized completeness, is therefore needed to build a robust sample of satellites of dwarf galaxies.

Recently, two surveys have begun to systematically build statistical samples of satellite populations of dwarf galaxies. The Exploration of Local VolumE Satellites of Dwarf Galaxies (\elvesdwarf) survey \citep{Li2025,Li2026_DDO161} is designed to characterize the satellite populations of primarily \textit{isolated} dwarf galaxies with stellar masses $10^8 < M_\star < 10^{10}\,M_\odot$ in the Local Volume ($D < 10$~Mpc). Isolation is required for the hosts to avoid ambiguity when assigning satellites to the hosts. \elvesdwarf first identifies satellite candidates from wide-field imaging surveys, then obtains surface brightness fluctuation (SBF, \citealt{Tonry1988}) distances using deeper ground-based imaging data to confirm candidates. Such a combination of integrated light searches and SBF distances has proven to be very efficient for searching and confirming satellites as well as isolated dwarf galaxies in the Local Volume \citep{Cohen2018,Carlsten2019,Carlsten2021,CarlstenELVES2022,Li2024,ELVES_Field_I,ELVES_Field_II}. In \citet{Li2025}, we present the first results from the \elvesdwarf survey, including eight isolated hosts and six newly confirmed satellites. These results show no evidence for a ``missing satellite'' problem for dwarf hosts and reveal that the majority of satellites around dwarf hosts are star-forming. 

In parallel, the Identifying Dwarfs of MC Analog GalaxiEs (ID-MAGE) survey \citep{Hunter2025,Hunter2026} covers a similar range of host mass and distance as \elvesdwarf but includes both isolated and non-isolated hosts. In ID-MAGE, satellite candidates are confirmed through a combination of spectroscopic observations, \ion{H}{1} observations, and SBF measurements.

In this paper, we present the full sample of \elvesdwarf, covering \numhosts hosts (including 7 hosts from the literature) and \numsats confirmed satellites. This paper is structured as follows. The dwarf host sample is described in Section \ref{sec:hosts} and satellite candidate detection is described in Section \ref{sec:detection}. We describe our follow-up observations for satellite candidates in Section \ref{sec:followup}, and describe SBF distance measurements in Section \ref{sec:sbf}. Section \ref{sec:results} shows the main results, including satellite abundance, stellar mass function, radial distribution, and star formation properties. We discuss the implications of these results in Section \ref{sec:discussion}. Appendices present the detailed survey footprint, completeness, satellite properties, as well as SBF distances to two host galaxies in our sample.

We adopt a flat $\Lambda$CDM cosmology with present-day matter density $\Omega_{\rm m}=0.3$, dark-energy density $\Omega_{\Lambda}=0.7$, and Hubble constant $H_0=70\ {\rm km\ s^{-1}\ Mpc^{-1}}$. All photometry is reported in the AB magnitude system \citep{Oke1983}. Stellar masses are calculated assuming a \citet{Kroupa2001} initial mass function. We correct for Milky Way foreground extinction using the dust maps of \citet{SFD1998}, recalibrated by \citet{Schlafly2011}. Solar absolute magnitudes are taken from \citet{Willmer2018}. When showing the LMC and SMC as comparison systems, we adopt stellar masses of $M_\star=2\times10^9\,M_\odot$ and $M_\star=3.2\times10^8\,M_\odot$, respectively \citep{Skibba2012LMC}. Virial masses and radii are defined following \citet{Bryan1998}, corresponding to an overdensity of $\Delta_{\rm vir}\simeq101$ relative to the critical density of the Universe, $\rho_{\rm c}=3H_0^2/8\pi G$. The virial mass and radius are computed using the Python package \code{colossus}\footnote{\url{https://bdiemer.bitbucket.io/colossus/index.html}} \citep{Colossus}.

\section{Dwarf Host Sample}\label{sec:hosts}
The \elvesdwarf survey is designed to systematically characterize satellites of dwarf galaxies in the Local Volume ($<12$~Mpc). We therefore start by constructing the parent host galaxy sample, then follow up on a fraction of the systems in this parent sample. In this section, we first describe the sample selection criteria, then present the properties of the selected host galaxies. 

\subsection{Host Sample Selection}
We build the parent host galaxy sample from the Updated Nearby Galaxy Catalog \citep[UNGC;][]{Karachentsev2013}\footnote{We use the version of UNGC available as of May 2026; \url{https://www.sao.ru/lv/lvgdb/}}, which compiles distances, radial velocities, photometry, and other basic properties for galaxies in the Local Volume and is regularly updated. We also cross-check host properties using the Extragalactic Distance Database \citep[EDD;][]{Tully2009}. We then select the host galaxies based on their distance, stellar mass, environment, and coverage in existing wide-field imaging surveys.

\textit{(a) Distance.} Host galaxies must lie beyond the Local Group so that a statistical sample of dwarf hosts can be assembled and their satellites can be identified in integrated light, while remaining close enough to enable SBF distance measurements with deep ground-based imaging. We therefore focus on hosts at $\sim$4--12 Mpc, complementing resolved-star surveys of satellites around nearer dwarf galaxies \citep[e.g.,][]{Mutlu-Pakdil2024,Carlin2024,McNanna2024}. Distances in the UNGC are compiled from a variety of methods, including the tip of the red giant branch (TRGB), the Tully--Fisher relation (TF), and SBF. Among these, TRGB distances provide the most precise and robust distance scale for nearby galaxies \citep{Lee1993,Tully2013,Jang2017,Anand2021-EDD}; we therefore prioritize TRGB measurements whenever multiple estimates are available. For the primary host sample, we require the hosts to have TRGB distances, which enable reliable satellite association and consistent derivation of host and satellite properties.

\textit{(b) Stellar mass and foreground contamination:} We restrict the host sample to dwarf galaxies with approximate stellar masses in the range $8 \lesssim \log (M_\star/M_\odot) \lesssim 10$. At lower masses, the expected number of satellites above our detection threshold becomes very small (\citealt{Dooley2017a}), whereas more massive hosts have already been systematically surveyed by programs such as ELVES and SAGA \citep[e.g.,][]{CarlstenELVES2022,SAGA-III}. For the initial selection, we use the $K_s$-band luminosity as a practical stellar-mass proxy because near-infrared light is relatively insensitive to recent star formation and dust extinction \citep{Bell2003}. We require $8 < \log (L_{K_s}/L_{\odot}) < 10$, corresponding approximately to $8 < \log (M_\star/M_\odot) < 10$ for an assumed $M_\star/L_{K_s}=1$. The final stellar masses used throughout this work are derived independently from optical photometry from the Siena Galaxy Atlas (\citealt{Moustakas2023}; see Section \ref{sec:host_properties}). To reduce contamination from Milky Way stars, Galactic dust, and cirrus, we further require Galactic latitude $|b|>15^\circ$ and $E(B-V)<0.15$.

\textit{(c) Isolation:} A defining feature of the \elvesdwarf host sample is that the hosts are required to be predominantly \textit{isolated}, following the selection adopted in \citet{Li2025} and \citet{Li2026_DDO161}. This requirement is motivated by both observational and theoretical considerations. Observationally, isolation minimizes ambiguity in assigning candidate satellites to a unique host and avoids systems in which the dwarf host is itself a satellite of a more massive galaxy. Theoretically, it enables cleaner comparisons with models of satellite populations around low-mass central galaxies, without the added complexity of group preprocessing or ``satellite-of-satellite'' configurations. In this respect, our selection differs from surveys such as MADCASH, LBT-SONG, and ID-MAGE, which include dwarf hosts spanning a broader range of environments \citep{Hunter2025,Hunter2026}.

We quantify isolation using the tidal indices reported in \citet{Karachentsev2013}. These indices estimate the external tidal field from neighboring galaxies using a proxy of the form
$\Theta \propto \log \left(\sum_i M_i/D_i^3\right) + C,$
where $M_i$ and $D_i$ are the stellar mass and three-dimensional separation of the $i$-th perturber, and $C=-10.96$ is a normalization constant. Negative values of $\Theta$ indicate that the host is outside the nominal zero-velocity surface of its dominant neighboring system and therefore resides in a weak tidal environment. The index $\Theta_1$ measures the influence of the most dominant perturber, and $\Theta_5$ includes the sum of the contributions from the five strongest perturbers. We require both $\Theta_1<0$ and $\Theta_5<0$ when selecting the hosts. We further visually inspect each host to ensure that it is not obviously embedded in an uncataloged nearby association. This conservative isolation cut selects hosts with relatively simple environmental histories and minimizes ambiguity when assigning satellites to the hosts. 

\textit{(d) Imaging coverage:} Finally, we require each host to have Legacy Surveys\footnote{\url{https://www.legacysurvey.org/dr10/description/}} Data Release 10 (DR10) imaging coverage over its projected virial radius, with available data in $g$ and at least one redder band, either $r$ or $i$. 

\vspace{1em}
Applying the criteria described above yields a parent sample of 65 dwarf hosts. Removing the isolation requirement increases this number to 117, while additionally removing the requirement for having Legacy Surveys imaging coverage increases it to 123. We emphasize that this parent sample of 65 hosts, which is conditioned on isolation, having reliable TRGB distances, and Legacy Surveys imaging coverage, is not volume-complete. 

\begin{figure*}
    \centering
    \includegraphics[width=1\linewidth]{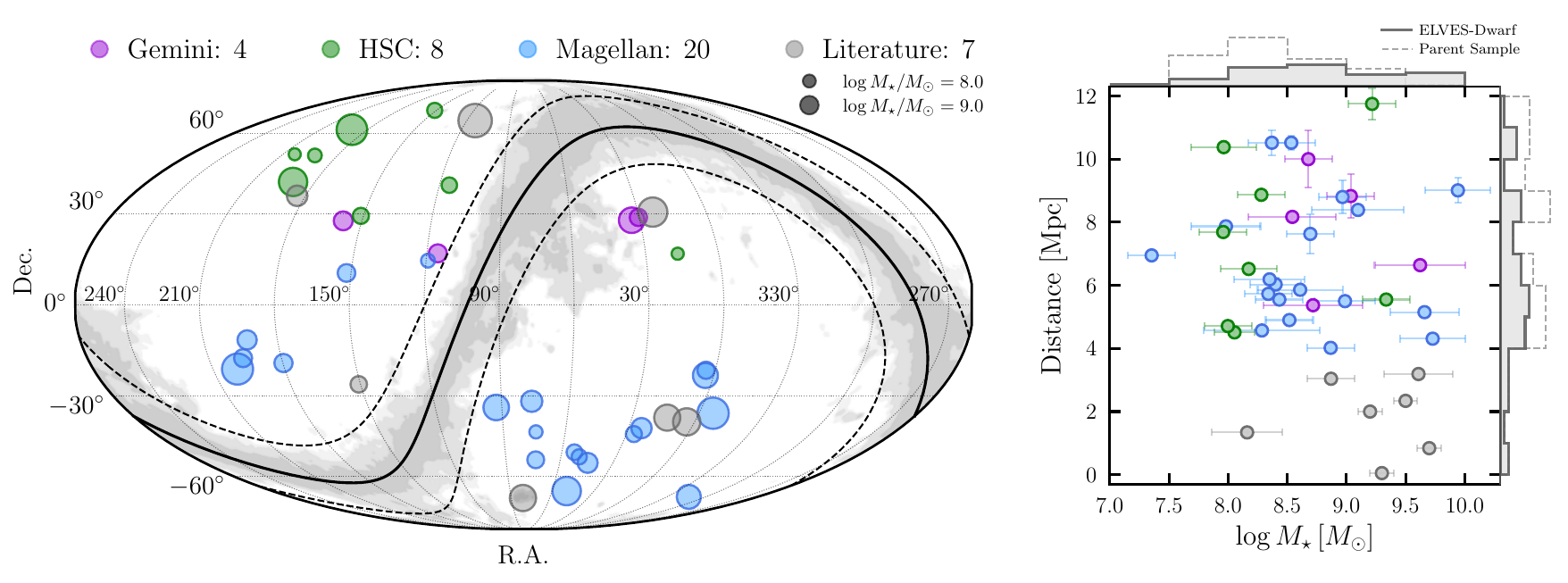}
    \caption{Overview of the \elvesdwarf host galaxy sample presented in this work. \textit{Left:} Sky distribution of the host galaxies, color-coded by the follow-up imaging data used for satellite distance confirmation. Symbol size scales with host stellar mass. The gray background shows the Milky Way foreground extinction from \citet{SFD1998}. \textit{Right:} Stellar mass versus distance for the host-galaxy sample. The data points share the same color coding as in the left panel. Detailed host galaxy properties are listed in Table \ref{tab:hosts}. The distribution of stellar mass and distance of the \elvesdwarf host sample and the parent sample are shown in the top and right histograms, respectively. We note that the parent sample, selected to be isolated and have reliable TRGB distances, is not volume-complete.}
    \label{fig:hosts_sky}
\end{figure*}

We then selected a subset of hosts for satellite searches and follow-up observations with ground-based facilities, primarily Magellan and Gemini (see \S \ref{sec:followup}). This observed \elvesdwarf sample does not uniformly sample the parent population in host stellar mass (see the right panel of Figure \ref{fig:hosts_sky}): we targeted relatively fewer low-mass hosts because their expected yield of satellites above our detection threshold is very low. In practice, this observed subset was shaped by the constraints of the observing program, including allocated observing time, weather losses, target visibility, and scheduling, as well as the availability of archival Subaru/HSC imaging, including the data presented in \citet{Li2025}. For the same practical reasons, the early phase of the \elvesdwarf survey included several systems that do not strictly satisfy all of the host-selection criteria. These include hosts without TRGB distances: NGC~855, for which we adopt an SBF distance, and NGC~1800 and NGC~4700, for which only Tully--Fisher distances are available. To better determine the distances to NGC~1800 and NGC~4700, we measure their SBF distances in Appendix \ref{ap:ngc4700}, and recalculate their tidal indices using the new distances. We also include NGC~5068, which is slightly non-isolated ($\Theta_1=0.05$). These systems are therefore added to the parent sample, retained in the present analysis, and flagged accordingly in Table~\ref{tab:hosts}. We note that some more distant systems have TRGB distances with substantial uncertainties. For example, NGC~3274 has a TRGB distance of $D=10.0\pm0.9$ Mpc from \citet{Sabbi2018}, reflecting the limited depth of the available color--magnitude diagram from HST.

To place the \elvesdwarf sample in a broader context, extend the dynamic range in host properties, and maintain continuity with previous work, we augment our observed sample with seven dwarf hosts from the literature that have been searched to comparable depth and radial coverage as \elvesdwarf; they are LMC, M33, NGC~3109, NGC~300, NGC~55, NGC~2403, and NGC~4214. The references adopted for each host are listed in Table~\ref{tab:hosts}. Among these systems, NGC~3109 is classified as isolated, while the remaining hosts are non-isolated. For these literature host systems, we include only confirmed satellites with $M_\star>10^5\,M_\odot$ and remove lower-mass ultra-faint satellites from the compiled sample. This cut has the largest impact on M33, around which several lower-mass satellites are known but none have stellar masses above $M_\star = 10^5\,M_\odot$. A compilation of the relevant literature searches is provided in Appendix~A of \citet{Li2025} and in Table~\ref{tab:sats} of this paper. Together, these systems define the \elvesdwarf host sample analyzed in this paper, comprising \numhosts hosts in total.

\subsection{Host Galaxy Properties}\label{sec:host_properties}

Table~\ref{tab:hosts} summarizes the full \elvesdwarf host sample, including the basic properties of each host galaxy, the imaging data used for satellite searches and follow-up, and the relevant references for the host properties. The sky distribution of the \elvesdwarf hosts is shown in Figure~\ref{fig:hosts_sky}, with points color-coded by the follow-up facility. The right panel of Figure~\ref{fig:hosts_sky} shows the distribution of host stellar mass and distance.

Host stellar mass is a key quantity in this work because it sets both the projected virial radius over which satellites are searched (see \S\ref{sec:detection}) and the basis for comparison with theoretical predictions. The $K_s$-band luminosities listed in Table~\ref{tab:hosts}, which are used for the initial host selection, are primarily based on the relatively shallow data from the Two Micron All Sky Survey \citep[2MASS;][]{Jarrett2000,Karachentsev2013}. Many of the dwarf hosts in our sample are low surface brightness, and they are either undetected or only marginally detected in 2MASS, making the corresponding stellar mass estimates uncertain. Stellar masses based on Wide-field Infrared Survey Explorer (WISE) $3.4\,\mu{\rm m}$ photometry are available for many nearby galaxies from \citet{Leroy2019}, but this catalog does not cover all of the hosts in our sample. We therefore attempt to derive a homogeneous set of stellar-mass estimates from deep optical imaging.

For this purpose, we use photometry from the Siena Galaxy Atlas \citep[SGA;][]{Moustakas2023,Moustakas2025_inprep}, which provides integrated photometry for nearby galaxies based on Legacy Surveys imaging using elliptical isophotal modeling with careful masking of foreground stars and background sources. In this work, we use the updated SGA-2025 catalog\footnote{\url{https://sga.readthedocs.io/en/latest/}}, which is based on Legacy Surveys DR11. We estimate the integrated photometry of each host galaxy with an elliptical curve-of-growth analysis. Specifically, we fit the cumulative magnitude profile, $m(<r)$, measured in concentric elliptical apertures with an empirical function
\begin{equation}
m(<r)=m_{\rm tot}+\Delta m\left[1-\exp\left(-\alpha_1\left(r/r_0\right)^{-\alpha_2}\right)\right],
\end{equation}
where $m_{\rm tot}$ is the asymptotic total magnitude, $r_0=10\arcsec$ is a fixed scale radius, and $\alpha_1$ and $\alpha_2$ control the shape of the curve of growth. We restrict the fit to $r>5\arcsec$ to reduce the influence of the innermost apertures, where seeing, centering uncertainties, and nuclear star clusters can bias the fitting result. The best-fit value of $m_{\rm tot}$ in each band is adopted as the total integrated magnitude. 

After obtaining the integrated magnitudes in the $griz$ bands, we first correct for Milky Way foreground extinction and then estimate stellar masses using optical color--mass-to-light ratio ($M_\star/L$) relations. Specifically, we compute $M_\star/L_g$ using the calibrations of \citet{Into2013}, \citet{Bell2003}, and \citet{delosReyes2025}. For the \citet{Bell2003} and \citet{Into2013} relations, we use $g-i$ color when $i$-band imaging is available from the Legacy Surveys, and otherwise adopt $g-r$. For the \citet{delosReyes2025} calibration, we use $g-r$ uniformly for the full sample. Our fiducial host stellar masses are based on the \citet{delosReyes2025} relation and are listed in Table~\ref{tab:hosts}, together with the $g-r$ colors of the hosts. In Appendix~\ref{ap:mstar}, we compare these stellar mass estimates with stellar masses inferred from $K_s$-band luminosities and, where available, with the WISE-based measurements of \citet{Leroy2019}. The different estimators are broadly consistent, with a typical scatter of $\sim0.2$ dex, although the WISE-based masses are systematically lower than our fiducial values. The stellar mass uncertainties reported in Table~\ref{tab:hosts} additionally include the propagated uncertainties in the distance and in the color used to derive $M_\star/L_g$. To account for systematic uncertainties in the photometry, stellar-population assumptions, and color--$M_\star/L$ calibrations, we set a minimal systematic uncertainty of 0.2 dex. We note that for very nearby hosts, including the LMC, M33, NGC~55, and NGC~300, the Legacy Surveys imaging suffers from significant over-subtraction of the sky background; we therefore adopt their stellar-mass estimates directly from the literature rather than from the SGA.



Finally, we estimate the virial radius of each host from its stellar mass. We first estimate the mean halo mass, $M_{\rm vir}$, using the inverse stellar-to-halo mass relation (SHMR) of \citet{RP2017}, which properly accounts for intrinsic scatter in the SHMR. We do not include additional scatter in $M_{\rm vir}$ at fixed stellar mass, as our goal is to assign a characteristic virial radius to each host rather than to model the full conditional halo-mass distribution. We then compute the corresponding virial radius, \rvir, defined as the radius enclosing a mean density $\Delta_{\rm vir}\simeq101$ times the critical density of the Universe at $z=0$, following \citet{Bryan1998}. For simplicity, we do not distinguish between the three-dimensional virial radius and the projected virial radius. The resulting virial radii range from $\sim70$ to $\sim200$ kpc, corresponding to angular radii of $0.5^\circ$--$4.5^\circ$ for hosts beyond the Local Group, and are listed in Table~\ref{tab:hosts}. Propagating the measurement uncertainties in host stellar mass yields typical uncertainties of $\sim0.1$ dex in $M_{\rm vir}$ and $\sim8\%$ in \rvir.


\setlength{\tabcolsep}{2pt}
\begin{deluxetable*}{lrrcccccccccccc}
\tabletypesize{\footnotesize}
\tablewidth{0pt}
\tablecaption{Properties of Dwarf Hosts in This Work \label{tab:hosts}}
\tablehead{
\colhead{Name} & \colhead{$D$} & \colhead{Method} & \colhead{$v_{\rm h}$} & \colhead{$\Theta_1$} & \colhead{$\Theta_5$} & \colhead{$\log L_{\rm K_s}$} & \colhead{$M_V$} & \colhead{$(g - r)_0$} & \colhead{$\log M_\star$} & \colhead{$R_{\rm vir}$} & \colhead{Area} & \colhead{$f_{\rm unmask}$} & \colhead{Data} & \colhead{Ref.} \\
\colhead{} & \colhead{(Mpc)} & \colhead{} & \colhead{(km~s$^{-1}$)} & \colhead{} & \colhead{} & \colhead{($L_\odot$)} & \colhead{(mag)} & \colhead{(mag)} & \colhead{($M_\odot$)} & \colhead{(kpc)} & \colhead{(deg$^{2}$)} & \colhead{} & \colhead{} & \colhead{}
}
\startdata
\multicolumn{15}{c}{Isolated Hosts} \\ \hline
DDO~046 & $10.38\pm0.10$ & TRGB & $364$ & $-1.05$ & $-0.73$ & $8.47$ & $-15.7$ & $0.22$ & $7.96\pm0.28$ & $93$ & 1.31 & 80\% & B--H--B & \citetalias{Tully2013} \\
DDO~047 & $8.17\pm0.10$ & TRGB & $270$ & $-0.69$ & $-0.63$ & $8.83$ & $-17.0$ & $0.34$ & $8.54\pm0.37$ & $116$ & 2.37 & 70\% & D--G--D & \citetalias{Tully2013} \\
DDO~161 & $6.03\pm0.20$ & TRGB & $744$ & $-0.86$ & $-0.48$ & $8.74$ & $-16.9$ & $0.22$ & $8.40\pm0.21$ & 111 & 5.18 & 87\% & D--M--D & \citetalias{Karachentsev2017_DDO161} \\
ESO154-023 & $5.86\pm0.05$ & TRGB & $574$ & $-1.69$ & $-1.13$ & $8.93$ & $-17.2$ & $0.30$ & $8.61\pm0.36$ & $118$ & 4.54 & 96\% & D--M--D & \citetalias{Tully2013} \\
ESO245-005 & $4.57\pm0.10$ & TRGB & $394$ & $-0.73$ & $-0.59$ & $8.53$ & $-16.4$ & $0.27$ & $8.29\pm0.49$ & $105$ & 5.42 & 96\% & D--M--D & \citetalias{Tully2013} \\
ESO252-001 & $6.95\pm0.12$ & TRGB & $660$ & $-1.56$ & $-1.29$ & $8.20$ & $-14.1$ & $0.20$ & $7.35\pm0.20$ & $75$ & 2.18 & 93\% & D--M--D & \citetalias{Tully2013} \\
ESO572-034 & $10.52\pm0.40$ & TRGB & $1114$ & $-0.77$ & $-0.65$ & $8.76$ & $-16.5$ & $0.34$ & $8.37\pm0.20$ & $108$ & 1.62 & 87\% & D--M--D & \citetalias{Tully2013} \\
IC~1959 & $6.19\pm0.10$ & TRGB & $640$ & $-1.49$ & $-1.11$ & $8.49$ & $-16.4$ & $0.33$ & $8.35\pm0.30$ & $108$ & 3.11 & 95\% & D--M--D & \citetalias{Tully2013} \\
IC~5052 & $5.50\pm0.10$ & TRGB & $584$ & $-1.84$ & $-1.29$ & $9.27$ & $-17.6$ & $0.50$ & $8.99\pm0.26$ & $136$ & 6.39 & 68\% & D--M--D & \citetalias{Tully2013} \\
IC~5332 & $9.01\pm0.40$ & TRGB & $701$ & $-1.50$ & $-1.23$ & $9.74$ & $-19.6$ & $0.45$ & $9.94\pm0.28$ & $196$ & 3.68 & 89\% & D--M--D & \citetalias{Anand2021} \\
NGC~45 & $6.64\pm0.08$ & TRGB & $465$ & $-0.87$ & $-0.57$ & $9.33$ & $-19.0$ & $0.40$ & $9.62\pm0.38$ & $173$ & 7.05 & 96\% & D--M,G--D & \citetalias{Tully2013} \\
NGC~59 & $4.90\pm0.10$ & TRGB & $361$ & $-0.37$ & $-0.33$ & $8.66$ & $-16.2$ & $0.56$ & $8.52\pm0.20$ & $114$ & 10.42 & 92\% & D--M--D & \citetalias{Tully2013} \\
NGC~625 & $4.02\pm0.10$ & TRGB & $395$ & $-0.33$ & $-0.28$ & $8.96$ & $-17.3$ & $0.52$ & $8.87\pm0.20$ & $130$ & 13.46 & 90\% & D--M--D & \citetalias{Tully2013} \\
NGC~784 & $5.37\pm0.10$ & TRGB & $193$ & $-1.20$ & $-0.79$ & $8.67$ & $-17.3$ & $0.36$ & $8.72\pm0.42$ & $123$ & 7.30 & 84\% & D--G--D & \citetalias{Tully2013} \\
NGC~855 & $8.83\pm0.70$ & SBF & $591$ & $-0.93$ & $-0.59$ & $9.29$ & $-17.6$ & $0.56$ & $9.04\pm0.20$ & $139$ & 3.99 & 83\% & D--G--D & \citetalias{Tully2013} \\
NGC~1311 & $5.55\pm0.10$ & TRGB & $571$ & $-1.19$ & $-0.91$ & $8.43$ & $-16.3$ & $0.43$ & $8.43\pm0.20$ & $111$ & 3.42 & 95\% & D--M--D & \citetalias{Tully2013} \\
NGC~1313 & $4.31\pm0.10$ & TRGB & $470$ & $-0.68$ & $-0.47$ & $9.57$ & $-19.3$ & $0.35$ & $9.73\pm0.28$ & $180$ & 18.13 & 82\% & D--M--D & \citetalias{Tully2013} \\
NGC~1705 & $5.73\pm0.10$ & TRGB & $628$ & $-1.87$ & $-1.44$ & $8.62$ & $-16.1$ & $0.41$ & $8.34\pm0.20$ & $107$ & 3.67 & 94\% & D--M--D & \citetalias{Tully2013} \\
NGC~1800 & $7.63\pm0.63$ & SBF & $814$ & $-1.70$ & $-1.50$ & $9.10$ & $-16.9$ & $0.48$ & $8.70\pm0.20$ & $122$ & 2.68 & 80\% & D--M--D & TW \\
NGC~2188 & $8.39\pm0.10$ & TRGB & $745$ & $-1.33$ & $-0.88$ & $9.37$ & $-18.3$ & $0.36$ & $9.10\pm0.39$ & $142$ & 3.18 & 65\% & D--M--D & \citetalias{Tully2013} \\
NGC~3109 & $1.34\pm0.10$ & TRGB & $403$ & $-0.33$ & $-0.11$ & $8.58$ & $-15.9$ & $0.34$ & $8.16\pm0.30$ & 100 & \nodata & \nodata & \citetalias{Doliva-Dolinsky2025} & \citetalias{McConnachie2012} \\
NGC~3274 & $10.00\pm0.90$ & TRGB & $541$ & $-1.03$ & $-0.68$ & $8.83$ & $-17.4$ & $0.31$ & $8.68\pm0.20$ & $121$ & 2.43 & 92\% & D--G--D & \citetalias{Sabbi2018} \\
NGC~4605 & $5.55\pm0.10$ & TRGB & $151$ & $-1.07$ & $-0.55$ & $9.70$ & $-18.6$ & $0.50$ & $9.34\pm0.20$ & $155$ & 8.14 & 94\% & B--H,C--B & \citetalias{Tully2013} \\
NGC~4625 & $11.75\pm0.50$ & TRGB & $606$ & $-1.00$ & $-0.41$ & $9.56$ & $-18.1$ & $0.54$ & $9.22\pm0.20$ & $148$ & 2.18 & 92\% & B--H--B & \citetalias{McQuinn2017} \\
NGC~4707 & $6.52\pm0.10$ & TRGB & $468$ & $-0.45$ & $-0.07$ & $8.25$ & $-15.9$ & $0.35$ & $8.17\pm0.24$ & $101$ & 2.55 & 95\% & B--H--B & \citetalias{Tully2016} \\
NGC~5238 & $4.51\pm0.10$ & TRGB & $229$ & $-0.41$ & $-0.22$ & $8.02$ & $-15.2$ & $0.45$ & $8.05\pm0.17$ & $96$ & 3.92 & 94\% & B--H--B & \citetalias{Tully2013} \\
UGC~00685 & $4.71\pm0.08$ & TRGB & $156$ & $-1.51$ & $-1.11$ & $8.03$ & $-15.0$ & $0.47$ & $8.00\pm0.20$ & $95$ & 3.43 & 95\% & D--H--D & \citetalias{Tully2013} \\
UGC~04115 & $7.87\pm0.10$ & TRGB & $343$ & $-1.72$ & $-1.25$ & $8.26$ & $-15.5$ & $0.30$ & $7.98\pm0.29$ & $94$ & 1.94 & 72\% & D--M--D & \citetalias{Tully2013} \\
UGC~05423 & $8.87\pm0.10$ & TRGB & $348$ & $-0.52$ & $-0.40$ & $8.42$ & $-15.6$ & $0.53$ & $8.28\pm0.20$ & $105$ & 1.80 & 94\% & B--H--B & \citetalias{Tully2013} \\
UGC~05427 & $7.69\pm0.10$ & TRGB & $498$ & $-1.49$ & $-0.95$ & $8.20$ & $-15.2$ & $0.37$ & $7.96\pm0.20$ & $93$ & 1.87 & 94\% & D--H--D & \citetalias{Tully2013} \\
UGC~05456 & $10.52\pm0.22$ & TRGB & $527$ & $-0.89$ & $-0.39$ & $8.71$ & $-17.0$ & $0.31$ & $8.53\pm0.20$ & $115$ & 2.00 & 96\% & D--M--D & \citetalias{Tully2013} \\
\hline \multicolumn{15}{c}{Non-Isolated Hosts} \\ \hline
LMC & $0.05\pm0.00$ & Ceph & $278$ & $3.51$ & $3.58$ & $9.42$ & $-18.1$ & \nodata & $9.30\pm0.10$ & 152 & \nodata & \nodata & \citetalias{Pace2024} & \citetalias{Pace2024} \\
M33 & $0.84\pm0.01$ & Ceph & $-182$ & $1.47$ & $1.50$ & $9.62$ & $-18.8$ & \nodata & $9.70\pm0.10$ & 177 & \nodata & \nodata & \citetalias{Pace2024} & \citetalias{Pace2024} \\
NGC~55 & $2.34\pm0.10$ & TRGB & $129$ & $0.04$ & $0.15$ & $9.49$ & $-18.5$ & \nodata & $9.50\pm0.20$ & 170 & \nodata & \nodata & \citetalias{Medoff2025} & \citetalias{Karachentsev2013} \\
NGC~300 & $2.00\pm0.10$ & TRGB & $146$ & $0.13$ & $0.23$ & $9.41$ & $-18.4$ & \nodata & $9.20\pm0.20$ & 147 & \nodata & \nodata & \citetalias{Sand2024} & \citetalias{Dalcanton2009} \\
NGC~2403 & $3.19\pm0.10$ & TRGB & $125$ & $0.20$ & $0.59$ & $9.86$ & $-19.0$ & $0.39$ & $9.61\pm0.29$ & $172$ & \nodata & \nodata & \citetalias{Carlin2024} & \citetalias{Karachentsev2013} \\
NGC~4214 & $3.04\pm0.10$ & TRGB & $291$ & $0.65$ & $0.76$ & $8.98$ & $-17.7$ & $0.38$ & $8.87\pm0.20$ & 130 & \nodata & \nodata & \citetalias{Carlin2021} & \citetalias{Dalcanton2009} \\
NGC~4700 & $8.80\pm0.53$ & SBF & $1411$ & $0.60$ & $0.60$ & $8.90$ & $-18.0$ & $0.37$ & $8.97\pm0.20$ & 135 & 2.68 & 80\% & D--M--D & TW \\
NGC~5068 & $5.15\pm0.10$ & TRGB & $668$ & $0.05$ & $0.15$ & $9.73$ & $-19.1$ & $0.40$ & $9.66\pm0.29$ & $175$ & 13.18 & 86\% & D--M--D & \citetalias{Tully2013} \\
\enddata
\tablecomments{The table lists the properties of the host galaxies, including the distance and its measurement method, heliocentric radial velocity ($v_h$), tidal indices ($\Theta_1$ and $\Theta_5$), $K_s$-band luminosity, $V$-band absolute magnitude, $g-r$ color, estimated stellar mass ($M_\star$), estimated virial radius ($R_{\rm vir}$), unmasked fraction of the searched area, and data sources for each host. Distance methods include Ceph (Cepheids), TRGB (tip of the red giant branch), and SBF (surface brightness fluctuation). The reference column includes the sources of distance and the $K_s$-band luminosity, respectively. The radial velocities are taken from \citetalias{Karachentsev2013}. The photometry of these dwarf hosts in the $g$ and $r$ bands is from the Siena Galaxy Atlas \citep{Moustakas2023} using the Legacy Surveys data \citep{Dey2019}, and we convert them to $V$-band using the filter conversion from \citet{DES-DR2}. The reference column includes the sources of distance and the $K_s$-band luminosity, respectively. The letters stand for the following: B--BASS; D--DECaLS; H$_{\rm LA}$--Subaru/HSC-LA; H--Subaru/HSC; C--CFHT/MegaCam; M--Magellan/IMACS; G--Gemini/GMOS. The references stand for: 
\citetalias{Anand2021}, \citet{Anand2021};
\citetalias{Carlin2021}, \citet{Carlin2021};
\citetalias{Carlin2024}, \citet{Carlin2024};
\citetalias{Dalcanton2009}, \citet{Dalcanton2009};
\citetalias{Doliva-Dolinsky2025}, \citet{Doliva-Dolinsky2025};
\citetalias{Karachentsev2013}, \citet{Karachentsev2013};
\citetalias{Karachentsev2017_DDO161}, \citet{Karachentsev2017_DDO161};
\citetalias{McConnachie2012}, \citet{McConnachie2012};
\citetalias{McQuinn2017}, \citet{McQuinn2017};
\citetalias{Medoff2025}, \citet{Medoff2025};
\citetalias{Pace2024}, \citet{Pace2024};
\citetalias{Sabbi2018}, \citet{Sabbi2018};
\citetalias{Sand2015}, \citet{Sand2015};
\citetalias{Sand2024}, \citet{Sand2024};
\citetalias{Tully2013}, \citet{Tully2013};
\citetalias{Tully2016}, \citet{Tully2016}; TW, this work.}
\end{deluxetable*}


\section{Satellite Candidate Detection and Photometry}\label{sec:detection}

The detection and photometric characterization of satellite candidates follow the same methodology as in \citet{Li2025}. Here, we summarize the main steps of the procedure and highlight several important updates in this work.

\subsection{Satellite Candidate Detection}\label{sec:detection_method}

We search for satellite candidates around each dwarf host in Table~\ref{tab:hosts} using imaging from Legacy Surveys Data Release 10\footnote{\url{https://www.legacysurvey.org/dr10/description/}} \citep{Dey2019}. The Legacy Surveys combine the Beijing-Arizona Sky Survey \citep[BASS;][]{BASS} in the northern sky ($\mathrm{Dec}>+32^\circ$), the Dark Energy Camera Legacy Survey (DECaLS) in the south, and additional public imaging data taken with the Dark Energy Camera \citep{DECam}. Because the imaging depth and seeing vary across these datasets, with BASS generally shallower and having poorer seeing than the DECaLS data, we allow modest host-by-host adjustments to the parameters used in the detection pipeline.

For each host, we retrieve the coadded $g$- and $r$-band images for all Legacy Surveys bricks overlapping the projected virial radius. Across the 32 hosts searched uniformly by \elvesdwarf (excluding 7 hosts taken from literature), the total area used for satellite search is $155~\mathrm{deg}^2$. Then we run a low-surface-brightness galaxy detection pipeline based on the methods developed by \citet{Greco2018}, \citet{Carlsten2021}, and \citet{CarlstenELVES2022}. The pipeline is designed to identify diffuse, extended dwarf galaxy candidates while retaining sensitivity to higher-surface-brightness satellites. 

We first remove bright and compact sources in the image to enhance low-surface-brightness features. To do this, we identify Milky Way stars brighter than 19 mag in both $g$ and $r$ using Gaia Data Release 2 \citep{GAIA2018} and mask out the surrounding area around stars, including the wing and ghost of the bright stars. The fraction of the remaining area after star masking varies from 65\% to 96\%, depending primarily on Galactic latitude (see $f_{\rm unmask}$ in Table~\ref{tab:hosts}). We then identify bright sources and their diffuse wings through surface-brightness thresholding and replace those regions with noise matched to the local sky background. Compact sources and noise peaks are further identified using \code{sep} \citep{Barbary2016} and replaced with noise, producing a cleaned image in which diffuse low-surface-brightness structures remain.

To further increase sensitivity to extended features, we smooth the cleaned images with a Gaussian kernel with FWHM $\sim$2.5--3.5 times the FWHM of the point spread function (PSF). We then run \code{SExtractor} \citep{SExtractor} on both the cleaned and smoothed images in both bands, using a Gaussian filter with FWHM $=5$ pixels and a low detection threshold of $\sim$2--2.5\,$\sigma$. We require each candidate to be detected in both $g$ and $r$, which effectively removes image artifacts such as bright-star ghosts and chip edges. The exact configuration is adjusted slightly from host to host to account for variations in image quality. Relative to the ELVES survey, we adopt a deeper effective detection threshold in order to avoid missing the small number of satellites expected around dwarf hosts, at the expense of a higher false-positive rate in the initial detection stage.

The raw detection catalogs are then filtered through several steps to reduce false positives. We first visually inspect all detections and remove obvious artifacts, including Galactic cirrus, shredded stellar ghosts, galaxy outskirts, and co-addition defects, as well as clear background systems such as galaxy mergers, tidal features, edge-on disks, and galaxies with spiral structure or strong color gradients \citep[see also the appendix of][]{Carlsten2020}. We then cross-match the remaining objects against NED \citep{NED}, SIMBAD \citep{SIMBAD}, UNGC \citep{Karachentsev2013}, and DESI DR1 \citep{DESI-DR1}, rejecting any source with a heliocentric velocity differing from that of the host by more than $\Delta v_h = 300\ \mathrm{km\ s^{-1}}$ (see Appendix \ref{ap:vel_thresh}). We have not performed a dedicated cross-match with the ALFALFA or FASHI \ion{H}{1} catalogs, so some candidates may have published \ion{H}{1} velocities that are not included in the current compilation. Fewer than 5\% of the initial automated detections survive this stage. We also use UNGC to identify any extended and bright satellites missed by the automated detection and add them back into the candidate list when appropriate.

Even after this cleaning, many of the remaining objects are background galaxies whose appearance in Legacy Surveys imaging is broadly consistent with that of a diffuse dwarf, and the distances to these objects are all unknown. To further reduce this contamination, we apply a physically motivated cut based on the dwarf galaxy mass--size relation: if a candidate lies at the distance of the host, a genuine satellite should fall near the mass--size locus occupied by normal dwarf galaxies, whereas background interlopers are typically too compact for their inferred stellar mass. Using the \sersic-fitting photometry described in Section~\ref{sec:photometry}, we place each candidate on the stellar mass--size plane assuming the host distance and reject objects whose sizes lie more than $2\sigma$ below the mean satellite mass--size relation measured for the ELVES sample \citep{ELVES-I}. This cut removes systems that would be implausibly compact if they were true satellites at the host distance, eliminating roughly 50\%--70\% of the remaining candidates. 

This size-based cut assumes that satellites of dwarf hosts follow a mass--size relation similar to that of satellites around MW-mass hosts. Observationally, dwarf-galaxy sizes show only weak environmental dependence from the field to group environments \citep{ELVES-I,Asali2025}, supporting the use of the ELVES relation for this purpose. Moreover, the scatter about the average mass--size relation is approximately Gaussian \citep{ELVES-I}, so a one-sided $2\sigma$ cut would roughly retain 97.7\% of genuine satellites. The resulting $\sim2.3\%$ incompleteness due to the size cut is sufficiently small for the purposes of this work, and we therefore do not correct for it.

\begin{figure*}
    \centering
    \includegraphics[width=0.9\linewidth]{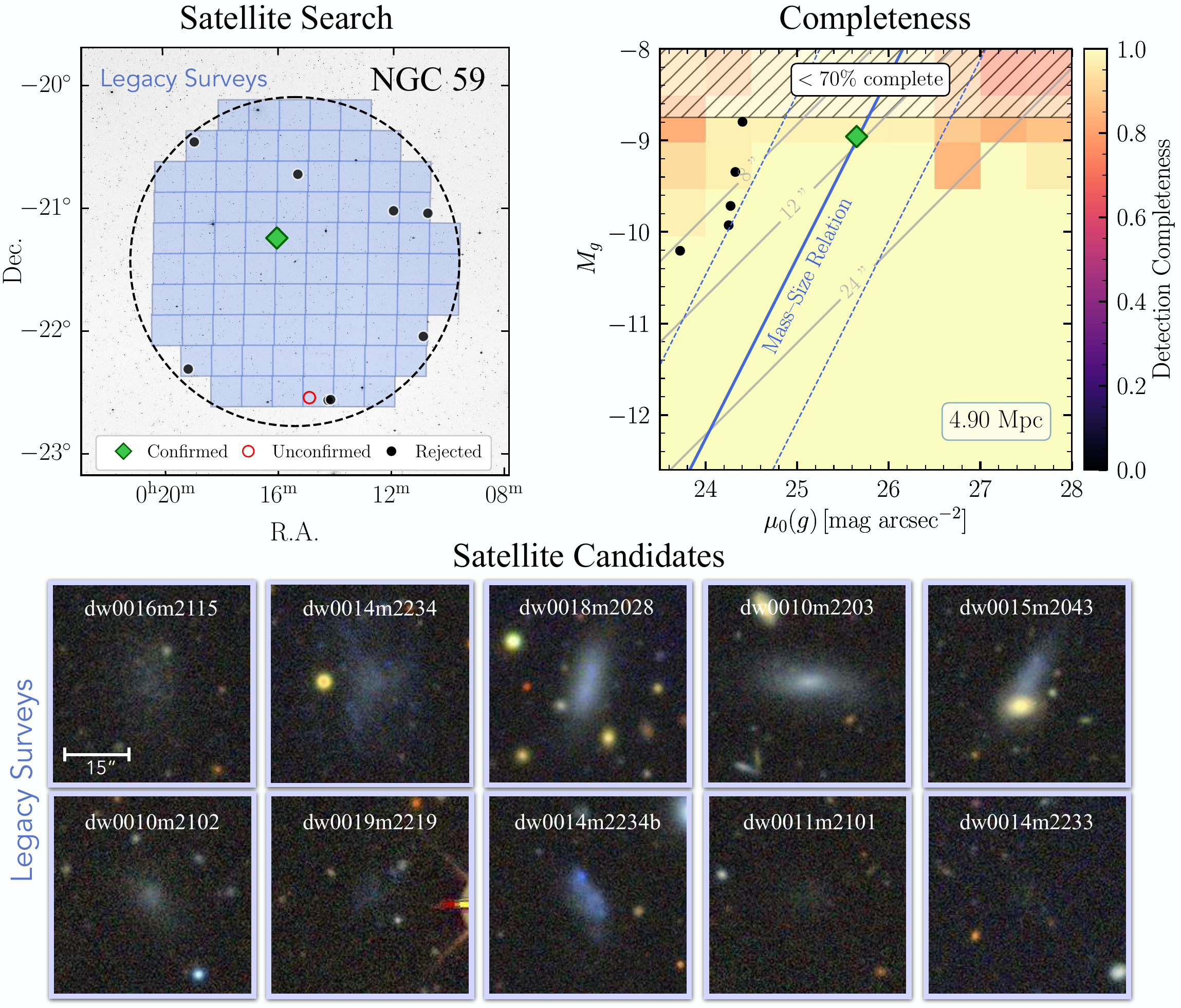}
    \caption{Example satellite candidate search around NGC~59, a host galaxy at $D=4.90$~Mpc. \textit{Top left}: Spatial distribution of ten satellite candidates within the Legacy Surveys footprint used for the search, shown in blue. The black dashed circle marks the projected virial radius, $R_{\rm vir}\approx114\,\rm kpc$. Green diamonds, black circles, and red open circles denote confirmed satellites, rejected candidates, and unconfirmed candidates, respectively. \textit{Top right}: Detection completeness from mock galaxy injection tests as a function of $g$-band central surface brightness, $\mu_0(g)$, and absolute magnitude, $M_g$. Gray diagonal lines show constant apparent size. The blue solid line shows the average mass--size relation for dwarf satellites from \citet{ELVES-I}, and the blue dashed lines show $1\sigma$ scatter in the mass--size relation. The search is highly complete for satellites with $M_g<-8.75$~mag at the distance of NGC~59. Only six candidates fall within the displayed $M_g$--$\mu_0(g)$ range and are plotted; the other four lie outside this range. \textit{Bottom}: Legacy Surveys color-composite cutouts of the satellite candidates of NGC~59. Each cutout is 45\arcsec{} on a side. }
    \label{fig:det_demo}
\end{figure*}

As an example, the top-left panel of Figure~\ref{fig:det_demo} shows the spatial distribution of satellite candidates detected around one representative host, NGC~59. The blue tiles indicate the Legacy Surveys imaging footprint used for the search, with the identified candidates overplotted. The black dashed circle marks the projected virial radius of the host. In \S\ref{sec:sbf}, we measure SBF distances to these candidates using deep, high-resolution imaging obtained as part of our follow-up observing campaign (\S\ref{sec:followup}).

\subsection{Previously Reported Candidates and Literature Information}

Several hosts in our sample have been searched previously, and we review these studies both to acknowledge overlapping detections and to incorporate relevant literature information. For the seven literature host systems described in Section~\ref{sec:hosts}, we include all confirmed satellites within the projected virial radius and above $M_\star>10^5\,M_\odot$. For hosts searched directly by \elvesdwarf, our detection pipeline defines the candidate list: previously reported but unconfirmed objects are included only if they independently satisfy our candidate-selection criteria, while previously confirmed satellites are retained even if they were missed by the automated pipeline. The catalog also incorporates the candidate lists for eight hosts analyzed in \citet{Li2025}.

A few published searches overlap with ours. \citet{Karachentsev2025b} reported dw1321m2201 and dw1324m2017 around NGC~5068 and dw0312m6643 around NGC~1313; all three candidates were also identified independently by our detection pipeline. \citet{Muller2020} previously searched around NGC~45 and reported no satellite candidates. The ID-MAGE survey \citep{Hunter2025,Hunter2026} has 14 hosts overlapping with our sample, but its candidate catalogs were not publicly available when this work was carried out; therefore, we did not incorporate ID-MAGE candidates into our analysis.

A distinct literature addition is SMDG0740+4032, which \citet{Chazov2026} confirmed as a satellite of DDO~046 using its radial velocity. This object was not included in the previous \elvesdwarf candidate catalog of \citet{Li2025}, and its location was not covered by the HSC imaging used for SBF follow-up. We therefore add it to the present catalog based on its velocity confirmation.

For all candidates, we also collect existing distance measurements from the literature and from databases including EDD, SIMBAD, and UNGC. For example, both satellite candidates around NGC~784 have published TRGB distances, from \citet{Tully2023} and \citet{McQuinn2014}, respectively. These external measurements are adopted when available and are listed in the final satellite catalog (Table~\ref{tab:sats}).


\subsection{Completeness}\label{sec:completeness}

Quantifying the detection completeness is essential for determining the satellite mass limits of our survey and for comparing the observed satellite population with theoretical models. We estimate the completeness by injecting mock dwarf galaxies into the Legacy Surveys imaging and attempting to recover them with the same detection pipeline used for the real data, following \citet{CarlstenELVES2022,Li2022,Li2025}. The mock galaxies are built as single-component \sersic models with $n=1$ and span a broad range of properties chosen to approximate the observed satellite population: $-12.5<M_g<-7.5$, $23.5<\mu_0(g)<28.0\ \sbunit$, $0.3<g-r<0.9$, and $0<\varepsilon<0.5$. For each brick in the Legacy Surveys, we inject 12 mock galaxies at a time into the coadded images, run the full detection pipeline, and match the recovered sources to the input catalog. Repeating this procedure many times yields the completeness, defined as the fraction of injected objects that are recovered. We find that the completeness depends primarily on total magnitude and central surface brightness, with much weaker dependence on color and ellipticity. Objects falling within the bright-star masks are excluded from the completeness calculation; the corresponding area loss is accounted for separately through the unmasked area fraction listed in Table~\ref{tab:hosts}. 

As an example, the top-right panel of Figure~\ref{fig:det_demo} shows the completeness as a function of absolute magnitude $M_g$ and central surface brightness $\mu_0(g)$ for one representative host, NGC~59. Six of the ten detected satellite candidates fall within the displayed completeness grid and are overplotted using the same symbols as in the top-left panel; the other four fall outside the plotted $M_g$--$\mu_0(g)$ range. We also show the average mass--size relation of satellites from the ELVES survey \citep{ELVES-I}, together with its $1\sigma$ scatter. The completeness declines for both very faint compact objects and very diffuse systems, but remains relatively uniform along the observed satellite relation except at the faintest luminosities.

The completeness for each host in our sample is shown in Figure~\ref{fig:host_completeness} in Appendix~\ref{ap:footprint}. The completeness varies from host to host because of differences in distance, imaging depth, and data quality in the Legacy Surveys. For hosts at $D<8$~Mpc, we achieve $>50\%$ completeness to $M_g\approx-8.5$, corresponding to $M_\star\approx10^{5.4}\,M_\odot$ for a representative color of $g-r=0.4$. For more distant hosts at $D>8$~Mpc, the 50\% completeness limit is $M_g\approx-9.5$, corresponding to $M_\star\approx10^{5.7}\,M_\odot$. These limits are comparable to those of the ID-MAGE survey and slightly deeper than those reported by ELVES for hosts at similar distances, primarily because we adopt a larger smoothing scale and a lower detection threshold.

Although the completeness declines rapidly below $M_g\approx-9$~mag, the deeper Legacy Surveys imaging available for some hosts allows us to identify several lower-luminosity satellite candidates. For example, dw0312m6643 around NGC~1313 has $M_g\approx-8$~mag. We retain these candidates in the sample and attempt to obtain distance measurements for them, as described in \S\ref{sec:sbf}.


\subsection{Photometry}\label{sec:photometry}

We measure the structural and photometric properties of the satellite candidates by fitting single-\sersic models to the Legacy Surveys $g$ and $r$-band (also $i$-band where available) coadded images, following a procedure similar to that used in \citet{ELVES-I}, \citet{CarlstenELVES2022}, and \citet{Li2024}. Most candidates appear smooth in the Legacy Surveys data, show no detectable SBF signal at this depth, and are well described by a single \sersic component. Using \code{IMFIT} \citep{imfit}, we first fit a \sersic model to the $g$-band data, allowing the central R.A. and decl., total magnitude ($m_g$), effective radius ($r_e$, defined as the half-light radius along the semi-major axis), \sersic index ($n$), and ellipticity ($\varepsilon$) to vary. We then fit the $r$-band data using the best-fit $g$-band structural parameters and allowing only the total magnitude ($m_r$) to vary. The resulting photometric properties are listed in Table~\ref{tab:sats}. 

Photometric uncertainties are estimated from injection-and-recovery tests following Appendix~E of \citet{ELVES-I}. Assuming each candidate lies at the distance of its host, we derive stellar masses using the best-fit \sersic photometry and the color--$M_\star/L$ relation of \citet{Into2013}. We adopt the \citet{Into2013} calibration for consistency with previous ELVES and \elvesdwarf analyses; this relation is also in good agreement with \citet{delosReyes2025} at $M_\star\lesssim10^8\,M_\odot$. As a sanity check, we cross-match our candidate list with the Systematically Measuring Ultra-Diffuse Galaxies catalog \citep[SMUDGes;][]{Zaritsky2023} and find good agreement for objects in common. For the few bright candidates that are not well fit by a single \sersic model, we adopt photometric measurements from the SGA \citep{Moustakas2023}.

An integrated color (e.g., $g-r$ or $g-i$) is required to measure the SBF distance, and the uncertainty in color often dominates the resulting SBF distance uncertainty. For candidates that are well described by a single \sersic model, we adopt the $g-r$ color from the \sersic fit. For a small number of brighter dwarfs whose light distributions are not well captured by a single \sersic model, we instead measure surface-brightness profiles through isophotal fitting with \code{photutils}\footnote{\url{https://photutils.readthedocs.io/en/stable/reference/isophote\_api.html}}, following an approach similar to that used in the SGA \citep{Moustakas2023}. In these cases, we measure the color profile over the same radial range used for the SBF analysis and adopt that value in the SBF measurement.

\section{Archival Data and Follow-up Observations}\label{sec:followup}

After identifying satellite candidates around each host, the next step is to determine which candidates are physically associated with the host as real satellites and which are background galaxies. For most candidates, no direct distance measurements or spectroscopic velocities are available in the literature. We therefore rely on the SBF method to measure distances systematically across the whole sample. Because SBF measurements at $D>3$~Mpc require imaging that is substantially deeper and of higher angular resolution than the Legacy Surveys data, we use suitable archival data where available and otherwise obtain new deep $i$-band follow-up imaging with ground-based facilities.

For targets in the northern sky, we first search for archival Hyper Suprime-Cam \citep[HSC;][]{Miyazaki2018,Tanaka2021,Aihara2022} imaging data. These data are well suited for SBF measurements because of their excellent depth and image quality, typically reaching $i>26.2$ mag at $5\sigma$ for point sources with median $i$-band seeing of $\sim0\farcs6$. In \citet{Li2025}, we identified eight such hosts with overlapping HSC data and obtained SBF distances for satellite candidates around them. For northern hosts without HSC coverage, we obtained follow-up imaging with the Gemini Multi-Object Spectrographs \citep[GMOS;][]{GMOS} on the 8.1 m Gemini-North telescope through programs GN-2024B-Q-115 and GN-2026A-Q-307. Although GMOS has a relatively small field of view ($5\arcmin\times5\arcmin$), it provides superb image quality for SBF work. We obtained $i$-band imaging for the candidates with total exposure times of 30--40 minutes under median seeing of $0\farcs8$.

For targets in the southern sky, we use the Inamori Magellan Areal Camera and Spectrograph \citep[IMACS;][]{Dressler2011} on the 6.5 m Magellan Baade telescope. IMACS provides a much larger field of view than GMOS, allowing multiple candidates to be observed simultaneously in many cases. Most observations were obtained with the F2 camera, which has a field of view of $27\arcmin\times27\arcmin$, although the F4 camera was used for a small number of targets. We took deep $i$-band imaging with typical total exposure times of 30--40 minutes and average seeing of $0\farcs7$--$0\farcs8$. The Magellan campaign was carried out between February 2024 and May 2026.

The reduction of the Magellan and Gemini data follows the procedures described in \citet{Li2024} and \citet{Li2026_DDO161}, and full details are provided in Appendix~\ref{sec:data_reduction}.


\begin{figure*}
    \centering
    \includegraphics[width=1\linewidth]{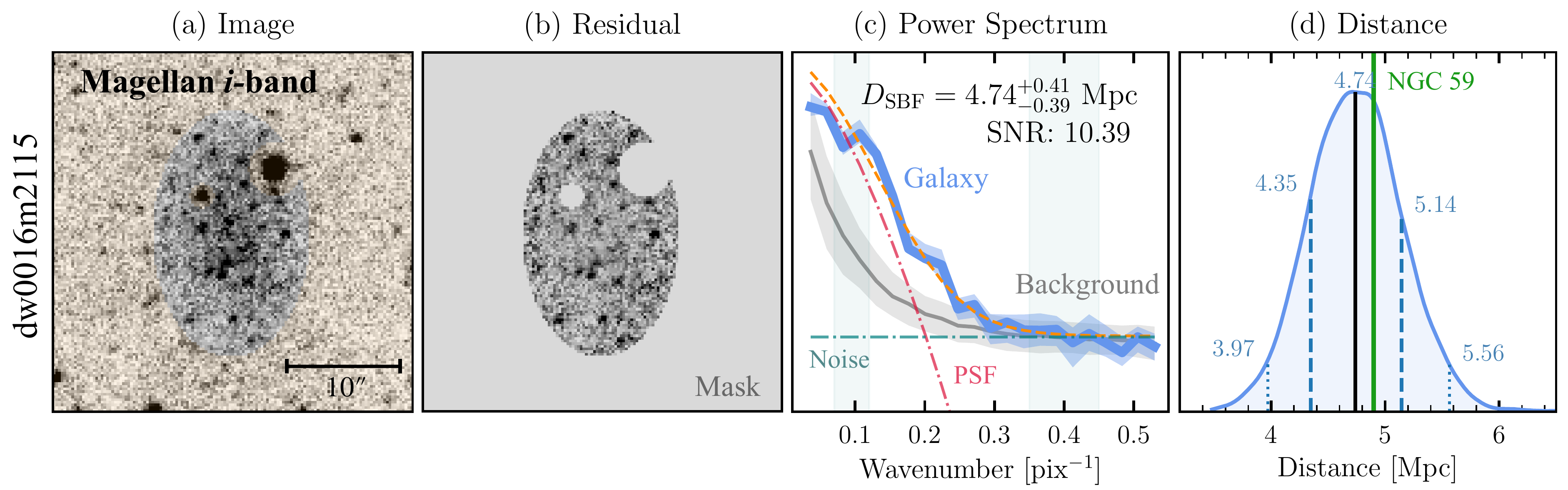}
    \vspace{-2em}
    \caption{Example surface brightness fluctuation (SBF) distance measurement for dw0016m2115, a confirmed satellite of NGC~59 ($D=4.90$~Mpc). (a) Magellan $i$-band image. (b) Residual image after subtracting a smooth galaxy model and masking bright sources; the SBF signal is measured from this image. (c) Azimuthally averaged power spectrum of the residual image (blue), fit with the PSF component (red) and the white-noise component (teal). The best-fit model is shown in orange, and the gray shaded region indicates the contribution from unmasked background sources. The light-green vertical bands mark the allowed ranges for the lower and upper wavenumber limits used in the fit. (d) Measured SBF distance distribution, with the median, $1\sigma$ interval (16th--84th percentiles), and $2\sigma$ interval (2.5th--97.5th percentiles) marked. The SBF distance is consistent with the host distance of 4.90~Mpc, confirming dw0016m2115 as a satellite of NGC~59.}
    \label{fig:sbf_demo}
\end{figure*}

\section{SBF Distances to Satellite Candidates}\label{sec:sbf}

In this section, we describe how we measure SBF distances for the satellite candidates using the imaging data presented in Section~\ref{sec:followup}, and how we use these distances to determine their association with the host galaxies. Our analysis follows the methodology of \citet{CarlstenELVES2022}, \citet{Li2024}, \citet{Li2025}, and \citet{Li2026_DDO161}; we refer readers to these works for a detailed discussion of the SBF measurement procedure.

\subsection{SBF Measurement}\label{sec:sbf_meas}

SBF quantifies the pixel-to-pixel fluctuations in the light of a semi-resolved galaxy caused by Poisson fluctuations in the number of luminous stars within each resolution element, i.e., on the scale of the PSF (\citealt{Tonry1988}; see \citealt{cantiello2023review} for a review). At larger distances, each resolution element contains more stars, and the fractional fluctuation amplitude decreases. Therefore, the SBF signal can be used as a distance indicator, although the absolute SBF magnitude also depends on the underlying stellar population of the galaxy \citep{Greco2021}. In optical, SBF measurements are commonly performed in the $r$ and $i$ bands, where the intrinsic fluctuation signal is stronger than in bluer bands, and optical colors are used as empirical proxies for the stellar population when calibrating the absolute SBF magnitude. Such SBF--color relations are well-calibrated in the optical, including the $r$ and $i$ bands \citep[e.g.,][]{Cantiello2018,Carlsten2019,Kim2021}. For dwarf galaxies in the Local Volume, SBF distances typically achieve uncertainties of $\sim10\%$--$15\%$ \citep{Carlsten2019,Li2026_DDO161}. In contrast, SBF distances to luminous early-type galaxies can reach $\lesssim5\%$ precision with HST imaging \citep{cantiello2023review}. In this work, we measure SBF in the $r$ band for satellite candidates around the five hosts with HSC imaging from \citet{Li2025}, and in the $i$ band for the remaining systems. We adopt the $i$-band SBF calibration of \citet{Carlsten2019} and the $r$-band calibration of \citet{Li2025}.

An example SBF measurement is shown in Figure \ref{fig:sbf_demo}. We first subtract the smooth galaxy light and construct a fluctuation image,
\begin{equation}
    {\rm Residual} = \frac{{\rm Galaxy} - {\rm Model}}{\sqrt{\rm Model}} \times {\rm Mask},
\end{equation}
where ``Galaxy'' is the observed image, ``Model'' describes the smooth light distribution of the galaxy, and ``Mask'' excludes sources whose fluctuations are not associated with the smooth stellar population of the galaxy itself, such as foreground stars, background galaxies, compact star-forming regions, and globular clusters. The division by $\sqrt{\rm Model}$ normalizes the fluctuation amplitude by the local galaxy light, producing a residual image from which the SBF signal is measured (see panel (b) of Figure~\ref{fig:sbf_demo} as an example, with masked regions shown in gray).

We measure the SBF amplitude from the azimuthally averaged power spectrum of the residual image. Following the standard SBF formalism, the power spectrum can be modeled as
\begin{equation}\label{eq:sbf}
    P(k) = f_{\rm SBF} \cdot E(k) + P_{\rm noise}(k),
\end{equation}
where $f_{\rm SBF}$ is the SBF amplitude, $E(k)$ is the power spectrum given by the PSF power spectrum convolved with the mask power spectrum, and $P_{\rm noise}(k)$ is the power spectrum of the noise in the image. We refer readers to \citet{Liu2000} for a more detailed derivation. An example of the measured power spectrum is shown as the blue line in the panel (c) of Figure \ref{fig:sbf_demo}. We then determine $f_{\rm SBF}$ by fitting the observed power spectrum with a combination of the PSF's power spectrum (red dash-dotted line) and the noise power spectrum (purple dash-dotted line). Although coadded images can contain correlated noise because of the interpolation kernel used in coaddition \citep{Mei2005}, the noise power spectrum is often flat over a wavenumber range (e.g., $k_{\rm min} < k < k_{\rm max}$). In practice, we treat $P_{\rm noise}(k)$ as a constant noise floor and select the fitting range in $k$ to avoid scales dominated by correlated noise or large-scale subtraction residuals.

The observed SBF magnitude is therefore $\overline{m}_{\rm SBF} = -2.5\log_{10}(f_{\rm SBF}) + \mathrm{ZP}$, and the distance modulus is $\mu_{\mathrm{DM}} = \overline{m}_{\rm SBF} - \overline{M}_{\rm SBF}$, where $\overline{M}_{\rm SBF}$ is the absolute SBF magnitude derived from the calibrated SBF--color relation \citep{Carlsten2019,Li2025}. For the imaging data used in this work, we set the photometric zeropoint to $\mathrm{ZP}=27.0$~mag (see Appendix \ref{sec:data_reduction}). Below, we describe each step of the measurement procedure in more detail.

\textit{(a) Smooth model.} We construct the smooth galaxy model in two ways, depending on the morphology of the target galaxy. For galaxies with regular morphologies, we fit a single \sersic model with \code{IMFIT} \citep{imfit}. For galaxies with more irregular or non-\sersic morphologies, the smooth model is built by median-filtering the image, following \citet{Kim2021}. The filter scale is about 5--10 times the seeing, and is adjusted case by case such that it removes large-scale structure while preserving the small-scale SBF signal. In practice, imperfect smooth models can leave residual large-scale power in the measured power spectrum. We therefore impose a lower bound on the wavenumber range, $k>k_{\rm min}$, to exclude these large-scale residuals from the SBF fit.

\textit{(b) Mask.} We mask globular clusters, star-forming regions, background galaxies, and other compact contaminants using an absolute-magnitude threshold, $M_{\rm thresh}$, following \citet{Carlsten2019} and \citet{Li2025}. The threshold is defined assuming that the candidate lies at the distance of its potential host. It is chosen to remove sources brighter than the brightest RGB stars, which have typical magnitudes of $M_i \approx -4.5$ and $M_r \approx -4$, while retaining the galaxy's RGB population. In practice, we tune $M_{\rm thresh}$ around these values for each target to ensure that bright compact non-SBF sources are masked. The adopted thresholds typically lie in the range $-5 \lesssim M_{\rm thresh} < -4$ in both the $r$ and $i$ bands, with the $r$-band threshold generally set slightly fainter because RGB stars are intrinsically fainter in $r$.

To improve the SBF signal-to-noise ratio (S/N), we restrict the SBF measurement to regions with sufficient galaxy light by masking pixels where the surface brightness of the smooth model falls below $f_{\rm mask}$ times the central surface brightness, with $f_{\rm mask}$ typically set to 0.2--0.3. For galaxies with central star-forming regions or nuclear star clusters, we additionally mask the central regions. For bright galaxies (such as the two host galaxies NGC~4700 and NGC~1800), we measure the SBF signal in the outskirts where the fluctuations are less affected by star-forming regions and strong color gradients. After constructing both the smooth model and the mask, we generate the residual image. The second panel of Figure~\ref{fig:sbf_demo} shows an example, with masked regions shown in gray.

\textit{(c) Background contamination.} Although sources brighter than $M_{\rm thresh}$ are masked, fainter unresolved objects, primarily background galaxies, can still contribute to the measured fluctuation signal. To correct for this contamination, we select relatively empty regions near the target galaxy and analyze them in the same way as the target by constructing the residual image as $\rm Background / \sqrt{Model} \times Mask$, where `Model' is the same as that of the target galaxy, and `Mask' is generated using the same $M_{\rm thresh}$. We then measure the fluctuation signal in these background fields, $f_{\rm bkg}$, using the same power-spectrum fitting procedure. The background power spectrum is shown as the gray shaded regions in panel (c) of Figure~\ref{fig:sbf_demo}. The resulting background contribution is subtracted from the SBF signal measured from the target galaxy. 

\textit{(d) Uncertainty.} The uncertainty of the measured SBF signal is estimated through Monte Carlo resampling over several choices in the measurement procedure, including the background field, the wavenumber range for fitting, and the surface brightness masking threshold \citep{Cohen2018}. In each realization, we select one of the background fields and draw $k_{\rm min}\sim \mathcal{U}(0.05,0.10)\ \mathrm{pixel}^{-1}$, $k_{\rm max}\sim \mathcal{U}(0.3,0.4)\ \mathrm{pixel}^{-1}$, and $f_{\rm mask}\sim \mathcal{U}(0.2,0.3)$. We typically perform 100 Monte Carlo realizations for each target galaxy. The shaded regions around the power spectra in panel (c) of Figure~\ref{fig:sbf_demo} show the resulting range of the Monte Carlo results, and the light-green vertical bands indicate the allowed ranges of $k_{\rm min}$ and $k_{\rm max}$. For each Monte Carlo realization, we compute the background-corrected SBF amplitude, $f_{\rm SBF}$; the ensemble of realizations then provides the distribution of $f_{\rm SBF}$ used to estimate the measurement uncertainty.

We define the SBF S/N as the median background-corrected SBF amplitude divided by its uncertainty. The example in Figure~\ref{fig:sbf_demo} has a high S/N of 10.39. We generally find an SBF measurement of S/N$>5$ to be reliable \citep{CarlstenELVES2022}, which has been validated against HST TRGB distances \citep{Bennet2019}. Finally, we convert $f_{\rm SBF}$ and its uncertainty into a distance using the adopted SBF--color relations from \citet{Carlsten2019} for $i$ band and \citet{Li2025} for $r$ band, including the color uncertainty and the intrinsic scatter of the SBF calibration. For bright galaxies with significant color gradients, we use the mean color measured over the same region used for the SBF measurement. Panel (d) of Figure~\ref{fig:sbf_demo} shows the inferred SBF distance distribution, with the median and the $1\sigma$ (68\%) and $2\sigma$ (95\%) intervals marked by blue vertical lines; the host distance is marked in green. 

For galaxies with a significant SBF detection, such as the example shown in Figure~\ref{fig:sbf_demo}, we measure an SBF distance directly. For galaxies without a significant SBF detection, the absence of a measurable fluctuation signal instead provides a lower limit on the distance.

The depth of the imaging data is reflected self-consistently in the SBF S/N and distance uncertainty: shallower images have higher noise floors and larger variation in the background corrections, which lower the SBF S/N and broaden the inferred distance distribution.

\subsection{Satellite--Host Association}\label{sec:psat}

We use the available distance and velocity information to determine the association of each satellite candidate with its potential host. Distance information comes either from literature TRGB measurements or from the SBF measurements presented in this work, while velocity information is compiled from existing spectroscopic measurements such as DESI \citep{DESI-DR1} and \ion{H}{1} measurements using SIMBAD and NED. Based on these constraints, we classify candidates into three categories: confirmed satellites, rejected foreground/background galaxies, and unconfirmed candidates. The unconfirmed category includes both candidates with inconclusive distance or velocity constraints and candidates that have not yet been observed in our follow-up campaign.

For candidates with distance measurements, we compare the candidate distance to that of the host, accounting for the uncertainties in both distances. A candidate is classified as a confirmed satellite if its $2\sigma$ distance interval overlaps the $2\sigma$ distance interval of the host. If the distance is from SBF, we further require the SBF measurement to have a S/N$>5$ to ensure that the distance is reliable \citep{Carlsten2021}. Conversely, if the candidate's $2\sigma$ interval lies entirely beyond or in front of the host distance interval, we reject it as a background or foreground galaxy, respectively. 

The SBF S/N threshold helps ensure that the measured fluctuation signal is dominated by SBF rather than contaminants such as unresolved background galaxies or star-forming regions. If a candidate has a low-S/N SBF measurement but its $2\sigma$ distance interval remains consistent with the host distance, we classify it as unconfirmed, since deeper or higher-resolution imaging is required to definitively determine its membership.

We also use radial-velocity information, when available, to assign satellite candidates to hosts \citep[e.g.,][]{SAGA-III}. For candidates with existing radial velocities but no independent distance measurements, we classify the candidate as a confirmed satellite if its line-of-sight velocity offset from the host satisfies $|\Delta v_h| < \Delta v_{\rm max}(M_\star^{\rm host})$, where $\Delta v_{\rm max}$ is a host-mass-dependent threshold calibrated using cosmological simulation TNG50 (Appendix~\ref{ap:vel_thresh}). This threshold corresponds to the 99th percentile of the velocity-offset distribution of simulated satellites within the projected virial radius, and therefore retains nearly all true satellites while rejecting candidates with substantially larger velocity offsets. We describe this calibration in detail in Appendix~\ref{ap:vel_thresh}. We note that our follow-up campaign generally prioritizes candidates without existing membership constraints, and therefore often does not include objects that already have existing velocity measurements.

For candidates with both distance and velocity information, we assess their memberships on a case-by-case basis, particularly when the two diagnostics are in tension. For example, LEDA~18431, a satellite candidate of NGC~2188, has a very small velocity offset from the host, $|\Delta v_{\rm h}|\sim20\ \mathrm{km\ s^{-1}}$. However, its TRGB distance, $9.64\pm0.10$~Mpc \citep{Tully2013}, is inconsistent with the TRGB distance of NGC~2188, $8.39\pm0.10$~Mpc. Because the color-magnitude diagram of LEDA~18431 is sparsely populated, the TRGB distance may be less secure than implied by the quoted formal uncertainty. We therefore classify LEDA~18431 as \textit{unconfirmed} rather than rejecting it as a background galaxy. We discuss three more such cases in Appendix \ref{ap:satcat}, all classified as unconfirmed.

Candidates without conclusive distance or velocity information are classified as unconfirmed. This includes candidates that are not followed up in our observing campaign, as well as objects for which the image quality is not sufficient to yield a reliable distance constraint. For these unconfirmed candidates, we assign a satellite probability, $P_{\rm sat}$, based on their absolute magnitude and surface brightness, following the empirical model of \citet{CarlstenELVES2022}. This model is motivated by the fact that confirmed satellites and rejected background galaxies occupy different regions of the absolute magnitude--surface brightness plane, with rejected candidates typically being more compact and having higher surface brightness. The probability is calibrated from the confirmed and rejected candidates in the ELVES survey \citep[see Section~5.6 of][]{CarlstenELVES2022}. Confirmed and rejected candidates are assigned $P_{\rm sat}=1$ and $P_{\rm sat}=0$, respectively.

The fraction of rejected background galaxies varies substantially across hosts because of line-of-sight structure. The most extreme case is NGC~5068 ($D=5.15$~Mpc, $v_h=668\,\kms$), whose field lies in front of two background groups associated with NGC~5084 ($v_h\approx1700\,\kms$) and IC~4237 ($v_h\approx2600\,\kms$; \citealt{Firth2006}). Consequently, many visually selected candidates in this field belong to the background galaxy groups and are rejected. The field also contains a foreground dwarf, the Hedgehog galaxy at $D\approx2.4$~Mpc, an isolated quiescent dwarf identified in \citet{Li2024}.

In Appendix~\ref{ap:purity}, we estimate the purity of our SBF-based satellite association using the cosmological simulation TNG50. For a host at $D_{\rm host}=6$~Mpc, comparable to the median distance of our sample, and a $1\sigma$ SBF distance uncertainty of $7\%$, we find that foreground and background interlopers contribute $13\pm2\%$ of the SBF-selected satellite sample. The contamination fraction increases to $20\pm2\%$ for hosts at $D_{\rm host}=8$~Mpc when adopting a larger SBF uncertainty of $10\%$. Given the relatively modest impact of this effect, and the absence of a strong host-mass dependence, we do not apply an explicit contamination correction in this work. Instead, we interpret our SBF-selected satellite abundances as potentially biased high at the $\sim10$--$20\%$ level, depending on host distance and SBF precision.


\begin{figure*}
    \centering
    \includegraphics[width=1\linewidth]{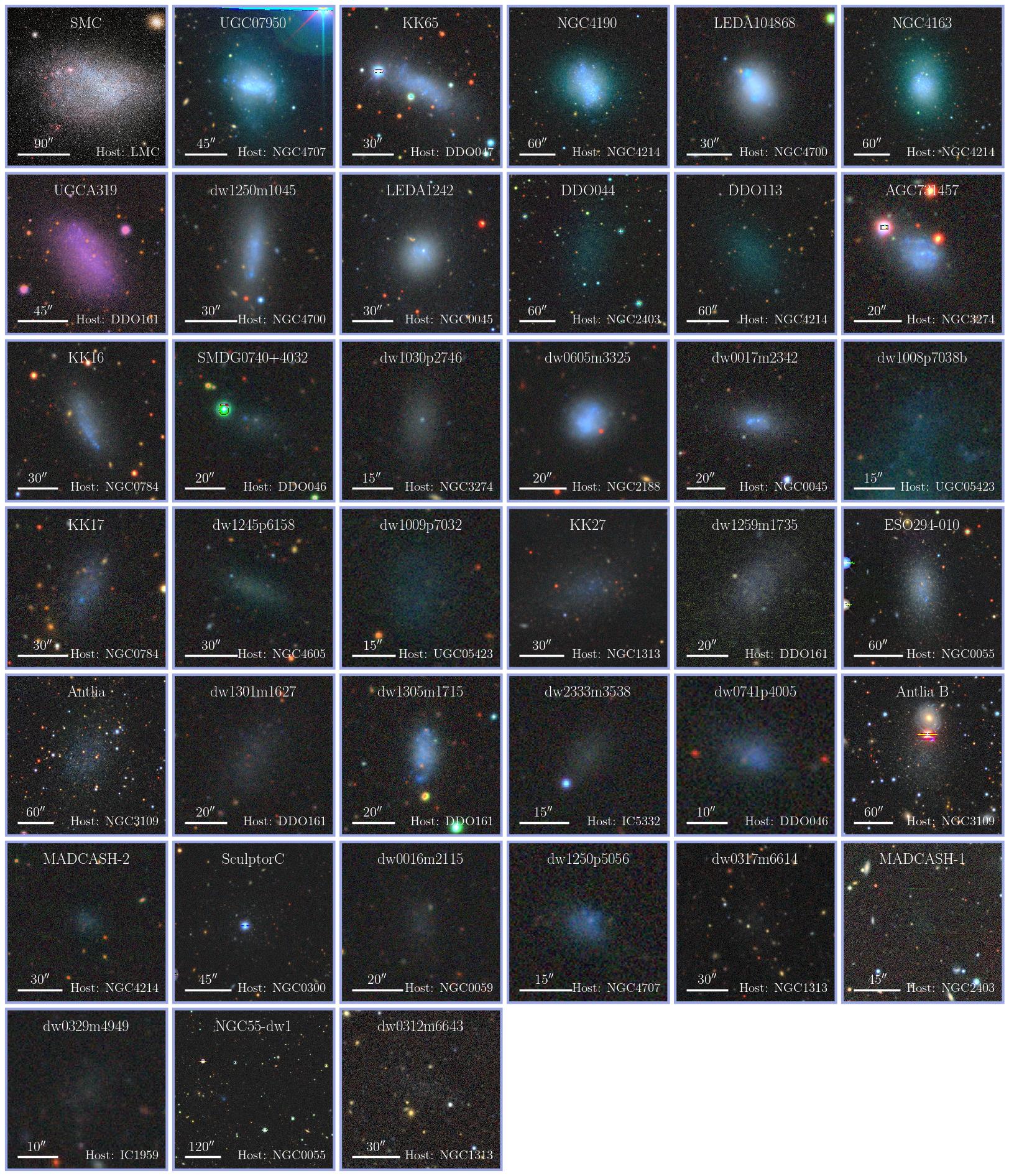}
    \caption{Color-composite cutouts of all \numsats confirmed satellites in \elvesdwarf, ordered by decreasing stellar mass from left to right and top to bottom. Each panel is labeled with the satellite name and includes an angular scale bar. Most images are from the Legacy Surveys, while the SMC image is taken from the SMASH survey \citep{Nidever2021}. This gallery highlights the broad diversity of the confirmed sample, including luminosity, color, and structure.}
    \label{fig:confirmed_gallery}
\end{figure*}

\section{Theoretical Predictions}\label{sec:model}\label{sec:tng}

To compare our observed satellite populations with theoretical expectations, we construct model satellite populations around dwarf hosts using the TNG50 cosmological hydrodynamical simulation. The predictions from TNG50 provide a useful benchmark for the measured satellite abundance, stellar mass function, satellite-to-host mass-ratio function, and projected radial distribution in this work. In Appendices~\ref{ap:vel_thresh} and \ref{ap:purity}, we also use TNG50 to calibrate the velocity-based satellite-association criterion and estimate contamination in our SBF-selected sample.

Our analysis is based on the TNG50-1 run of the Illustris-TNG simulation suite \citep{Pillepich2018,Pillepich2019,Nelson2019}. TNG50-1 has a box size of 50~Mpc and dark-matter and baryonic mass resolutions of $4.5\times10^5\,M_\odot$ and $8.5\times10^4\,M_\odot$, respectively. Halos and subhalos are identified with \textsc{FoF} and \textsc{Subfind} \citep{Davis1985,Springel2001}; we take the most massive subhalo in each FoF group to be the central galaxy and define the halo virial mass following the \citet{Bryan1998} definition. 

To construct a host sample analogous to \elvesdwarf in TNG50, we select central galaxies at $z=0$ with $10^{10.3}<M_{\rm vir}/M_\odot<10^{12}$ and with no more massive halos within 1~Mpc. We then compute the tidal index, $\Theta_1$, following \citet{Karachentsev2013} and retain hosts with $\Theta_1\leqslant0$, yielding 3437 TNG50 systems. The central galaxies are well resolved at these masses, so we use their native TNG50 stellar masses, measured within twice the stellar half-mass radius \citep[e.g.,][]{Shi2020,Engler2021}.

For satellites, we do not use the native TNG50 stellar masses because the stellar components of the low-mass systems considered in this work are poorly resolved. Their dark-matter subhalos, however, are sufficiently well resolved to trace their mass assembly histories. We therefore use the peak dark-matter mass of each satellite subhalo along its merger tree, $M_{\rm peak}$, and assign its stellar mass using the empirical stellar-to-halo mass relation of \citet{Nadler2020}, a galaxy--halo connection model calibrated on satellites of the Milky Way. We generate 100 Monte Carlo realizations to incorporate the intrinsic scatter in this relation.

To mimic the satellite selection of the observations, we randomly project each TNG50 satellite system onto a two-dimensional plane and select satellites within the projected virial radius. This procedure resembles the satellite selection in \S\ref{sec:detection} and naturally contains a small number of systems that lie beyond the three-dimensional virial radius but project within the two-dimensional virial radius, including recently accreted satellites or objects in merging pairs. Relative to a strict three-dimensional virial-radius cut, the projected selection increases the mean satellite abundance by approximately 10\% (see Appendix \ref{ap:purity}).

\section{Results}\label{sec:results}

In this section, we first summarize the satellite systems identified around the \elvesdwarf hosts, then measure the satellite abundances and stellar-mass functions, and compare these measurements with theoretical predictions. We also examine the spatial and color distributions of the confirmed satellites. We do not correct for the small area lost to bright-star masks or for detection incompleteness, since most confirmed satellites used in the analysis have high completeness ($>90\%$; Table~\ref{tab:sats}). A more complete treatment of these effects, including corrections for area loss, detection incompleteness, and interloper contamination, will be presented in future work.

\subsection{Overview of the Satellite Systems}

Our host sample, listed in Table~\ref{tab:hosts}, consists of \numhosts galaxies, including 32 surveyed systematically for the first time in this work and 7 taken from the literature. Of these, 31 satisfy our isolation criteria. The host-by-host survey footprints, candidate locations, and detection completeness are shown in Appendix~\ref{ap:footprint} in Figures~\ref{fig:host_coverage} and~\ref{fig:host_completeness}. In total, we identify \numcands satellite candidates around the \numhosts hosts.

Among these candidates, we confirm \numsats satellites with $M_\star>10^5\,M_\odot$, reject 131 as non-satellites of their assigned hosts, and classify 37 as unconfirmed. The confirmed sample includes 18 satellites confirmed using SBF distances, 17 using literature TRGB distances, three using radial velocities, and the SMC using a Cepheid distance and kinematics. The unconfirmed sample comprises 20 candidates with inconclusive distance or velocity constraints and 17 candidates that have not yet been observed in our SBF follow-up campaign. Table~\ref{tab:candidate_summary} summarizes these membership classifications and the methods used to confirm the satellites. The distance measurements and photometric properties of all satellite candidates considered in this work are presented in Table~\ref{tab:sats}; a machine-readable version is available online\footnote{\catalogurl}.

Figure~\ref{fig:confirmed_gallery} presents color-composite images of all \numsats confirmed satellites, ranked by their stellar masses. The sample spans from bright irregular galaxies to faint, diffuse systems, visually highlighting the diversity of satellites in \elvesdwarf.

\begin{deluxetable}{lc}
\tablecaption{Satellite Candidate Membership Summary\label{tab:candidate_summary}}
\tablehead{
    \colhead{Classification or confirmation method} & \colhead{Number}
}
\startdata
Confirmed satellites & \numsats \\
\quad \quad Cepheid distance & 1 \\
\quad \quad SBF distance & 18 \\
\quad \quad TRGB distance & 17 \\
\quad \quad Radial velocity & 3 \\
Rejected candidates & 131 \\
Unconfirmed candidates & 37 \\
\quad \quad Inconclusive distance or velocity constraints & 20 \\
\quad \quad Not observed in the follow-up campaign & 17 \\
\hline
All satellite candidates & \numcands
\enddata
\end{deluxetable}

\begin{figure*}[t!]
    \centering
    \includegraphics[width=0.65\linewidth]{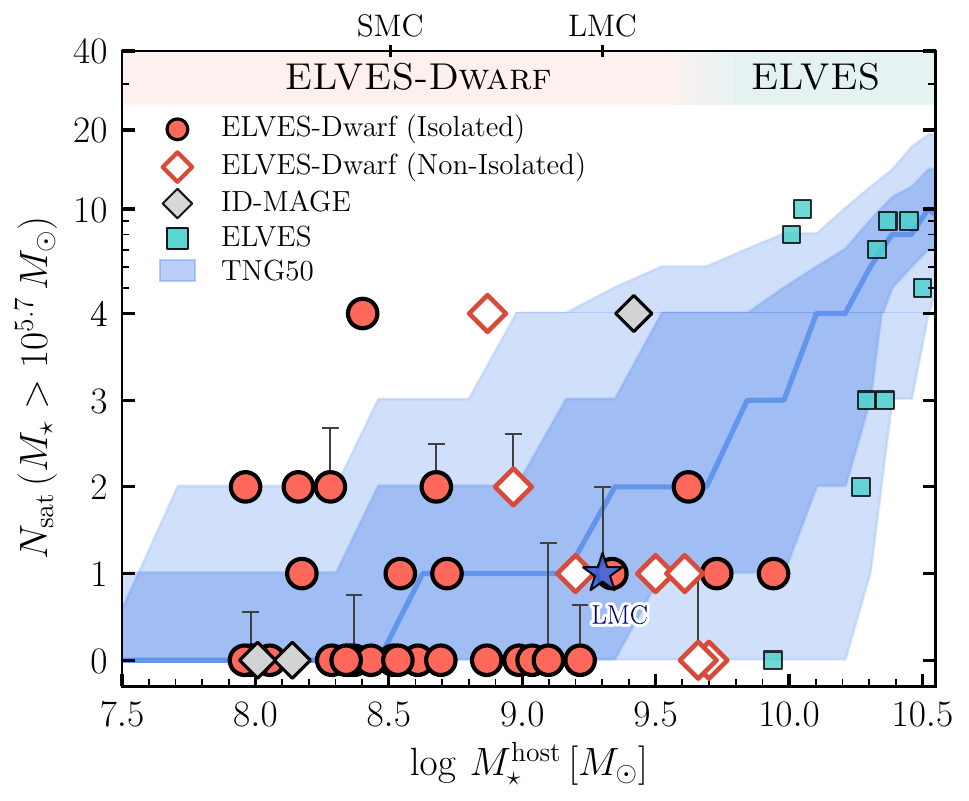}
    \vspace{-1em}
    \caption{Satellite abundance as a function of host stellar mass, counting satellites with $M_\star \gtrsim 10^{5.7}\,M_\odot$ within the projected virial radius. This threshold approximately matches the average survey completeness in \elvesdwarf and is conservative for more nearby hosts. The symbols show the number of confirmed satellites, while capped vertical lines show the possible additional contribution from unconfirmed candidates. Isolated hosts, non-isolated hosts, and massive hosts taken from the ELVES survey are shown as filled circles, open diamonds, and turquoise squares, respectively. The LMC is shown as a blue star, with the SMC counted as a confirmed satellite and Carina as a possible satellite. The gray diamonds show the three fully searched hosts from the ID-MAGE survey \citep{Hunter2026}. Blue shaded regions denote the $1\sigma$ (16th--84th percentile) and $2\sigma$ (2.5th--97.5th percentile) intervals of the TNG50 predictions, assuming the stellar-to-halo mass relation of \citet{Nadler2020}. The satellite populations of dwarf hosts show substantial host-to-host diversity. The overall satellite abundances are broadly consistent with TNG50, although a few systems appear as outliers, having more satellites than expected.}
    \label{fig:nsat}
\end{figure*}

\subsection{Satellite Abundance}\label{sec:nsat}

Here we present the satellite abundance, \nsat, for the \numhosts hosts in \elvesdwarf. We count satellites within the projected virial radius of each host and above our fiducial stellar-mass threshold of $M_\star \gtrsim 10^{5.7}\,M_\odot$. This mass threshold approximately corresponds to the characteristic completeness of the survey and is therefore conservative for more nearby systems. Some hosts, such as NGC~1313, have confirmed satellites below our fiducial mass threshold. These hosts remain in the sample, but their satellites with $M_\star<10^{5.7}\,M_\odot$ are not counted in the results presented here.

For the fiducial \nsat measurement, we count only confirmed satellites. To estimate the possible contribution from unconfirmed candidates, we weight each candidate by $P_{\rm sat}/C$, where $P_{\rm sat}$ is the satellite probability described in Section~\ref{sec:psat}, and $C$ is the detection completeness evaluated at the candidate's magnitude and central surface brightness (Section~\ref{sec:completeness}). The resulting upper limits are shown by the capped vertical lines in Figure~\ref{fig:nsat}. Summing $P_{\rm sat}/C$ over all candidates gives a probability- and completeness-weighted total of approximately 58 satellites; we regard this as an upper limit, implying that up to roughly 20 additional satellites may remain unconfirmed. 

Figure~\ref{fig:nsat} shows \nsat as a function of host stellar mass. The \elvesdwarf systems are shown as red symbols, with filled circles denoting isolated systems and open diamonds denoting non-isolated systems. To place \elvesdwarf in a broader context, we also show the ELVES hosts that have been surveyed to their full virial radii, as well as the three ID-MAGE hosts for which satellite searches have been completed \citep{Hunter2026}: DDO~052 (UGC~04426), DDO~165 (UGC~08201), and NGC~3432. The LMC is shown for comparison, with the SMC counted as a confirmed satellite and Carina shown as a possible additional satellite \citep[e.g.,][]{Pardy2020,Patel2020,Santos-Santos2021,Battaglia2022}. The top axis indicates the approximate stellar masses of the SMC and LMC, highlighting that \elvesdwarf extends satellite-abundance measurements below the MW regime for the first time.

Above this fiducial mass limit, 21 hosts have no confirmed satellites, 10 hosts have one, 6 hosts have two, no hosts have three, and two hosts have four. This distribution highlights the substantial host-to-host diversity in satellite abundance. The two richest systems in \elvesdwarf are DDO~161 and NGC~4214, each with four confirmed satellites above the fiducial mass limit. \citet{Li2026_DDO161} discuss the DDO~161 system in depth and argue that it may introduce a ``too-many-satellites'' problem. NGC~3432 (gray diamond) from ID-MAGE likewise has four confirmed satellites above our mass limit.

This diversity is not uniform across the full host-mass range. In particular, the LMC-mass hosts in our sample ($10^{9.3} < M_\star < 10^{10}\, M_\odot$) span a relatively narrow range in \nsat compared with the lower-mass systems. This may partly reflect limited sampling of hosts in the LMC-mass regime: because only a small number of \elvesdwarf hosts lie at these masses, rare systems in the high occupation tails are unlikely to be present. A larger sample of LMC-mass hosts is therefore needed to determine whether satellite abundance is intrinsically less stochastic at these host masses, or whether the apparent uniformity is a consequence of small-number statistics.

We next compare the observed satellite abundances with the TNG50 predictions described in Section~\ref{sec:tng}. In Figure~\ref{fig:nsat}, the solid blue line shows the median predicted \nsat, while the blue shaded regions show the $1\sigma$ (16th--84th percentile) and $2\sigma$ (2.5th--97.5th percentile) intervals of the predicted distribution. These predictions are broadly consistent with other theoretical expectations for satellite abundances of dwarf hosts, including the semi-analytic models of \citet{Dooley2017a} and \citet{Dooley2017b}, and the FIRE hydrodynamical simulations \citep{Jahn2019}. Overall, the \elvesdwarf measurements agree well with TNG50 across the host-mass range probed here. However, the most satellite-rich systems, DDO~161 and NGC~4214, lie above the $2\sigma$ range of the TNG50 prediction. These satellite-rich systems may therefore be particularly informative for constraining the SHMR of satellites in low-density environments \citep{Li2026_DDO161}. Although each of these systems is unusual when considered individually, their presence does not necessarily imply tension between \elvesdwarf and the theoretical models. A quantitative comparison requires constructing mock surveys from TNG50 with the same host-mass distribution as \elvesdwarf and determining how often they contain satellite systems as rich as those observed. We leave this survey-level comparison to future work.

Finally, by comparing the isolated and non-isolated hosts, we do not find a clear dependence of satellite abundance on host isolation in the current \elvesdwarf sample. A larger sample of non-isolated dwarf hosts, such as in the ID-MAGE survey \citep{Hunter2025,Hunter2026}, will be needed to test this environmental dependence more robustly.

\begin{figure*}
    \centering
    \includegraphics[width=1\linewidth]{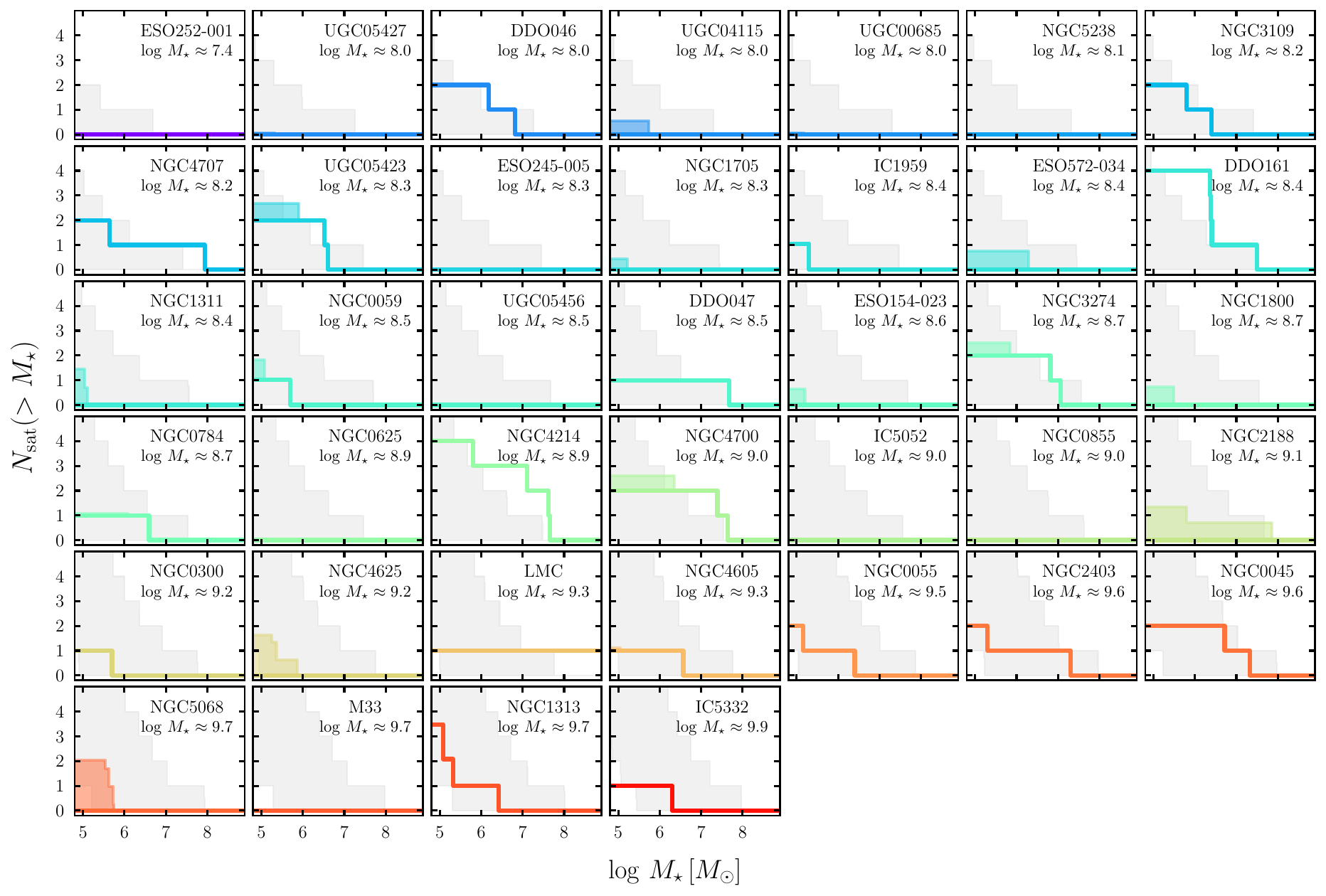}
    \vspace{-2em}
    \caption{Cumulative satellite stellar mass functions for the dwarf hosts in our sample, arranged by increasing host stellar mass. Only satellite candidates within the projected virial radius are included. Solid colored lines show the mass functions for confirmed satellites, while the colored shaded regions show the possible contribution from unconfirmed satellite candidates. For each host, the gray shaded region shows the $2\sigma$ range (2.5th--97.5th percentile) of the TNG50 predictions, after selecting simulated hosts within a 0.4 dex bin in host stellar mass and assigning satellite stellar masses using the \citet{Nadler2020} SHMR. The wide gray regions highlight the large host-to-host scatter expected in satellite populations at fixed host stellar mass. The observed satellite mass functions are broadly consistent with the TNG50 predictions for most hosts, although several satellite-rich systems lie outside the predicted $2\sigma$ range.}
    \label{fig:sat_mass_function}
\end{figure*}

\subsection{Satellite Stellar Mass Function}

We present the cumulative satellite stellar mass functions for each host in Figure \ref{fig:sat_mass_function}. As in Section \ref{sec:nsat}, we include only satellite candidates within the projected \rvir. Solid lines show the mass functions for confirmed satellites only, color-coded by host stellar mass, while the colored shaded regions include the possible contribution from unconfirmed candidates, each contributing $P_{\rm sat} / C$. For each host, we construct the corresponding TNG50 prediction by selecting simulated hosts within a 0.4 dex stellar-mass bin, consistent with our adopted 0.2 dex uncertainty in host stellar mass, and assigning satellite stellar masses using the SHMR of \citet{Nadler2020}. For clarity, we show only the $2\sigma$ range (2.5th--97.5th percentiles) of the predicted cumulative satellite mass functions as gray shaded regions. These gray regions illustrate the substantial host-to-host scatter expected in satellite populations even at fixed host stellar mass.

As shown in Figure \ref{fig:sat_mass_function}, the observed satellite stellar mass functions broadly agree with the TNG50 predictions for most hosts in our sample. However, several systems, including DDO~161, NGC~4214, NGC~4700, and UGC~05423, lie outside the predicted $2\sigma$ range. The LMC is also an outlier in this comparison, primarily because its most massive satellite, the SMC, is more massive than expected for satellites of LMC-mass hosts in the model. We leave a more detailed comparison between the observed satellite stellar mass functions and simulations to future work, where this host-by-host comparison can be used to constrain the satellite SHMR in the dwarf-host regime \citep[e.g.,][]{Danieli2023}.

\subsection{Satellite-to-Host Mass Ratio}

\begin{figure}
    \centering
    \includegraphics[width=\linewidth]{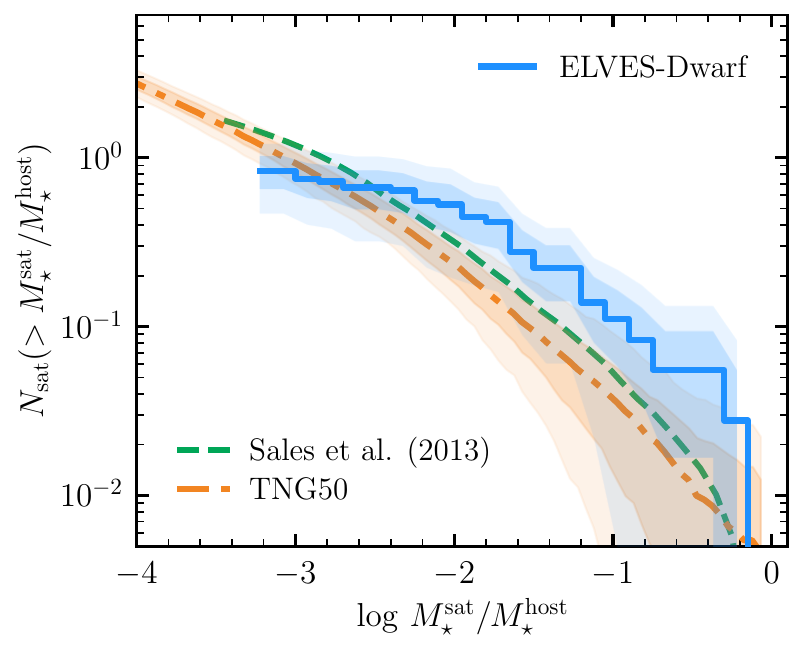}
    \vspace{-2em}
    \caption{Cumulative number of satellites per host as a function of satellite-to-host stellar mass ratio, $M_{\star}^{\rm sat}/M_{\star}^{\rm host}$. The blue line shows the measurement from \elvesdwarf using all confirmed satellites, with the light and dark blue shaded regions denoting the $1\sigma$ and $2\sigma$ jackknife uncertainties. The green dashed line shows the dwarf-host prediction from \citet{Sales2013}. The orange line shows the TNG50 prediction, with the shaded regions showing its $1\sigma$ and $2\sigma$ scatter. The observed mass-ratio function is broadly consistent with both models. The low-mass-ratio flattening in our measurement likely arises because satellites at these mass ratios fall below the detection limit for the lowest-mass hosts.}
    \label{fig:sat_mass_ratio}
\end{figure}

We also characterize the satellite populations in terms of the satellite-to-host stellar mass ratio, $M_{\star}^{\rm sat}/M_{\star}^{\rm host}$. In $\Lambda$CDM, the subhalo mass function is nearly scale-free when expressed as a function of subhalo-to-host halo mass ratio \citep{Wang2012}. If the SHMR is approximately a power law in the dwarf-galaxy regime, this scale-free behavior should translate into similar satellite abundances at fixed stellar mass ratio across dwarf hosts \citep{Sales2013}. \citet{Sales2013} showed that this expectation holds for hosts with $M_{\star}^{\rm host}\lesssim10^{9.75}\,M_\odot$, which closely matches the mass range probed by our sample.

Figure~\ref{fig:sat_mass_ratio} shows the cumulative mass-ratio function per host by taking all confirmed satellites in \elvesdwarf. We estimate the uncertainty using a jackknife procedure, recomputing the mass-ratio function after removing one host at a time and then measuring the standard deviation in each bin. The blue shaded regions show the resulting $1\sigma$ and $2\sigma$ jackknife uncertainties. The additional uncertainty from including unconfirmed satellite candidates is much smaller than the jackknife uncertainty and is therefore not shown separately.

We compare these measurements with two theoretical models. The green dashed line shows the prediction from \citet{Sales2013} for dwarf hosts, based on the semi-analytic galaxy catalog constructed from the Millennium-II simulation \citep{Guo2011}. We also make a similar prediction from TNG50 by randomly selecting \numhosts dwarf hosts with $10^8 < M_{\star}^{\rm host}/M_\odot < 10^{8.5}$, where the host galaxies are safely in the scale-free regime, and measuring the cumulative mass-ratio function of their satellites. We repeat this procedure 1000 times; the orange dashed line shows the median TNG50 prediction, while the orange shaded regions show the corresponding $1\sigma$ and $2\sigma$ ranges.

Overall, the observed mass-ratio function agrees well with both the \citet{Sales2013} expectation and the TNG50 prediction. This agreement supports the conclusion that the SHMR is approximately a power-law in the low-mass regime. The observed function falls below the models at $M_{\star,\rm sat}/M_{\star,\rm host}\lesssim10^{-2.5}$, likely because the satellite stellar masses approach our detection limit for the lower-mass hosts.

\subsection{Radial distribution}\label{sec:radial}

\begin{figure}
    \centering
    \includegraphics[width=\linewidth]{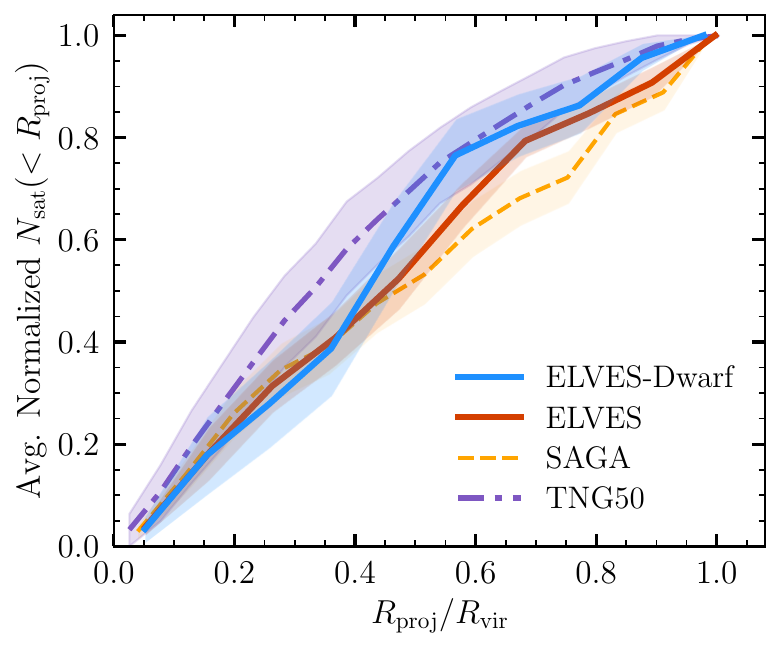}
    \caption{Average radial distribution of satellites, normalized by the total number of satellites within the virial radius. The blue and red solid lines show the \elvesdwarf and ELVES radial distributions, respectively, with shaded regions showing jackknife uncertainties. The orange dashed line and shaded region show the radial distribution of satellites in the SAGA gold sample ($M_\star > 10^{7.5}\,M_\odot$) \citep{SAGA-III}. For a consistent comparison, we include only hosts with satellite searches extending to the full virial radius; this leaves 53 hosts in SAGA and 21 hosts in ELVES. The purple dashed line shows the TNG50 prediction for satellites around dwarf hosts, with shaded regions denoting the $1\sigma$ range. The \elvesdwarf radial distribution is broadly consistent with those measured for ELVES and SAGA and with the TNG50 prediction, although it appears slightly more centrally concentrated than ELVES and SAGA.}
    \label{fig:radial}
\end{figure}

The radial distribution of satellites provides a complementary test of the connection between galaxies and their dark matter subhalos. Beyond tracing the underlying subhalo distribution, satellite spatial distributions are sensitive to tidal stripping and disruption, the baryonic influence of the central galaxy, and the nature of dark matter itself \citep[e.g.,][]{Nagai2005,Samuel2020,McDonough2022,Nadler2021,Lovell2021}. Recent measurements around MW-mass hosts provide a useful benchmark for comparison with \elvesdwarf \citep[e.g.,][]{elves-radial,SAGA-III}. Here, we present the radial distribution of satellites in \elvesdwarf and compare with that of MW-mass hosts and simulations.

We measure the projected radial distribution of confirmed satellites in \elvesdwarf as follows. For each host with at least one satellite above our fiducial mass threshold of $M_\star=10^{5.7}\,M_\odot$, we compute the cumulative radial distribution normalized by the total number of satellites within the virial radius, $N_i(<R_{\rm proj})/N_i(<R_{\rm vir})$, as a function of $R_{\rm proj}/R_{\rm vir}$. We then average these normalized profiles over all contributing hosts. This per-host normalization isolates the shape of the radial distribution from host-to-host differences in satellite abundance. Hosts with no satellites above the mass threshold do not contribute to this calculation. We only include confirmed satellites in this analysis.

The resulting \elvesdwarf radial distribution is shown as the blue line in Figure~\ref{fig:radial}. The blue shaded region shows the jackknife uncertainty, estimated by recomputing the average radial distribution after excluding each host in turn. For comparison, we show the corresponding radial distribution for the more massive hosts in ELVES and SAGA \citep{SAGA-III}, computed using the same procedure. To ensure a consistent comparison, we include only hosts for which the satellite search extends to the full virial radius, leaving 21 hosts in ELVES and 53 hosts in SAGA. ELVES reaches a satellite stellar-mass limit comparable to that of \elvesdwarf, whereas SAGA has a higher threshold of $M_\star>10^{7.5}\,M_\odot$ (i.e., the SAGA gold sample). As shown in Figure \ref{fig:radial}, the \elvesdwarf radial distribution is broadly consistent with those of ELVES and SAGA, although it appears somewhat more centrally concentrated than them both.

We also compare the observed radial distribution of \elvesdwarf with TNG50. We randomly select \numhosts dwarf hosts from TNG50 with $10^8 < M_{\star}^{\rm host}/M_\odot < 10^{10}$, compute the average normalized satellite radial distribution using the same procedure as for \elvesdwarf, and repeat this process 1000 times. The purple dashed line in Figure~\ref{fig:radial} shows the median TNG50 prediction, while the shaded regions show the corresponding $1\sigma$ (16th--84th percentiles) range. The \elvesdwarf radial distribution is consistent with the TNG50 prediction over the full radial range within uncertainty, although TNG50 is slightly more centrally concentrated.


\subsection{Satellite Quenching}

\begin{figure}
    \centering
    \includegraphics[width=1\linewidth]{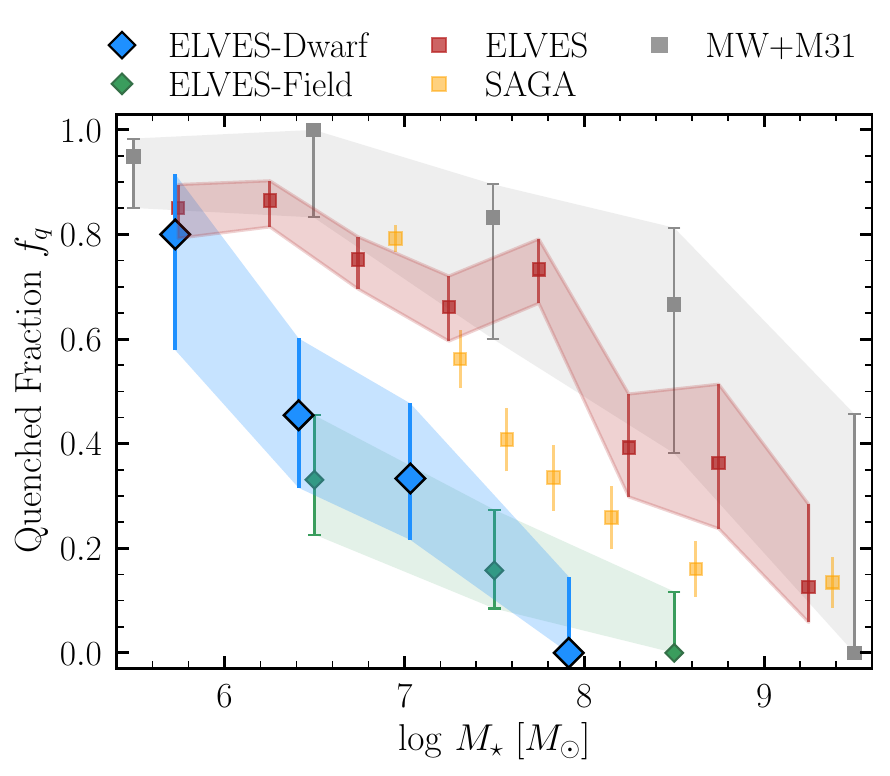}
    \caption{Quenched fraction as a function of satellite stellar mass for confirmed \elvesdwarf satellites (blue squares). Points show the quenched fraction in bins of stellar mass, with error bars indicating the statistical uncertainties. For comparison, we also show satellites around MW-mass hosts from ELVES (red squares; \citealt{CarlstenELVES2022,Greene2023}), SAGA (orange squares; \citealt{SAGA-III}), satellites of the MW and M31 (gray squares; \citealt{McConnachie2012}), and isolated field dwarfs from ELVES-Field (green diamonds; \citealt{ELVES_Field_II}). 
    Satellites in \elvesdwarf have a substantially lower quenched fraction than satellites around MW-mass hosts, approaching that of isolated field dwarfs.}
    \label{fig:quenched_frac}
\end{figure}

Star formation in dwarf galaxies is regulated by both internal and environmental processes. Stellar feedback, gas accretion, and reionization can strongly affect the gas reservoirs of low-mass galaxies, while external mechanisms such as ram-pressure stripping, tidal stripping, tidal heating, and starvation can further suppress star formation after infall into a larger halo. Observationally, the star-forming properties of dwarf galaxies exhibit a strong environmental dependence. Nearly all isolated field dwarfs with $M_\star\gtrsim10^7\,M_\odot$ are star-forming \citep{Geha2012}, although recent studies indicate that the field quenched fraction rises to $\sim30\%$ below $M_\star\sim10^7\,M_\odot$ \citep{ELVES_Field_I,ELVES_Field_II}. In contrast, quenched satellites are commonly found around massive hosts, with quenched fractions of $>50-60\%$ \citep{Greene2023,SAGA-IV}. Dwarf hosts therefore provide an important intermediate environment for exploring how satellite star formation varies with host mass. Here, we use the integrated colors of \elvesdwarf satellites to characterize their star-formation properties and compare them with satellites around more massive hosts.

We classify satellites as blue/star-forming or red/quenched using the empirical color--magnitude division $g-i=-0.067\,M_V-0.23$ calibrated using the ELVES satellites by \citet{ELVES-I}. This luminosity-dependent color cut provides an effective separation between star-forming and quiescent satellites \citep{Font2022,SAGA-III,Li2023}. Because not all satellites have $i$-band coverage in the Legacy Surveys, we convert $g-r$ to $g-i$ using the relation $g-i=1.53 \cdot (g-r)-0.032$ from \citet{ELVES-I} when necessary. Satellites redder than the dividing relation are classified as quenched, while those bluer than the relation are classified as star-forming. Robust integrated colors are not available for a few very extended, well-resolved satellites; we therefore classify these systems using their known star-formation properties and morphology: the SMC is classified as star-forming, and Antlia \citep{McQuinn2010} and Sculptor~C \citep{Sand2024} are classified as quenched.

Among the 34 confirmed \elvesdwarf satellites with $M_V<-9$~mag, we classify 13 as red/quenched and 21 as blue/star-forming, corresponding to a quenched fraction of $f_{\rm q}=38\%$ for the full compilation. Restricting the analysis to satellites of isolated hosts gives $f_{\rm q}=33\%$, with 8 quenched satellites out of 24, while considering only the uniformly surveyed systems and excluding hosts taken from the literature gives $f_{\rm q}=25\%$, with 6 quenched satellites out of 24. Thus, across these selections, we infer a fiducial quenched fraction of $\sim25$--$40\%$ for \elvesdwarf satellites.

We also find very tentative evidence that the quenched fraction increases with host stellar mass even within the mass range probed by \elvesdwarf. Focusing only on the 33 confirmed satellites with $M_V<-9$~mag, 5 out of 10 satellites around the 13 hosts with $M_{\star}^{\rm host}>10^9\,M_\odot$ are classified as quenched, corresponding to $f_{\rm q}=50\%\pm16\%$, where the uncertainty is estimated assuming binomial statistics. In contrast, 7 out of 23 satellites around the 26 hosts with $M_{\star}^{\rm host}<10^9\,M_\odot$ are quenched, corresponding to $f_{\rm q}=30\%\pm10\%$. Given the small sample sizes, this trend is not statistically significant, but it is qualitatively consistent with the host-mass dependence of satellite quenching observed around more massive hosts \citep{Greene2023}. 

Furthermore, Figure~\ref{fig:quenched_frac} shows the quenched fraction as a function of satellite stellar mass. The quenched fraction of \elvesdwarf satellites increases steeply toward lower satellite masses, consistent with the mass dependence of satellite quenching observed around MW-mass hosts \citep{Pan2022,Greene2023,SAGA-IV,Putman2021,Mercado2026}. For comparison, we also show isolated field dwarfs from the ELVES-Field survey \citep{ELVES_Field_II}, satellites around MW-mass hosts from ELVES and SAGA \citep{SAGA-III}, and the satellite populations of the MW and M31 \citep{McConnachie2012}. We apply the same luminosity-dependent optical color cut to the \elvesdwarf, ELVES, ELVES-Field, and MW+M31 samples. SAGA classifies satellites as quenched based on the specific SFR ($\mathrm{sSFR}<10^{-11}\,\mathrm{yr}^{-1}$) derived from a combination of H$\alpha$ and NUV measurements \citep{SAGA-III,SAGA-IV}. Across these samples, we find a qualitative environmental sequence over the overlapping stellar-mass range: the quenched fraction of \elvesdwarf satellites is modestly higher than that of isolated field dwarfs, but substantially lower than that of satellites around MW-mass hosts. For comparison, the overall quenched fraction of ELVES satellites is $f_{\rm q}\sim65\%$ \citep{Greene2023}, considerably higher than that inferred for \elvesdwarf. This suggests that environmental quenching becomes progressively more efficient from the field to dwarf-host halos and then to MW-mass halos.

Our results are broadly consistent with other recent studies for satellites around dwarf hosts. The inferred quenched fraction is slightly higher than that from the initial \elvesdwarf results of \citet{Li2025}, where only one quiescent dwarf was identified among six satellites ($f_{\rm q}=17\%$). For more massive satellites with $M_\star>10^7\,M_\odot$, the ID-MAGE survey measured a quenched fraction of $\sim10$--$25\%$ using UV-derived star-formation rates \citep{Hunter2026}; applying the same stellar mass threshold to \elvesdwarf gives $f_{\rm q}=17\%$ (2/12), consistent with the ID-MAGE measurement. The abundance of star-forming satellites is also consistent with recent \ion{H}{1} work: \elvesdwarf contains 21 star-forming satellites around 39 hosts, corresponding to $\langle N_{\rm sat,SF}\rangle\simeq0.5$, broadly in line with the gas-rich satellite census from \citet{Zhu2025}.

We do not attempt to compare the observed quenched fractions with TNG50 because its baryonic mass resolution is insufficient to robustly resolve star formation and quenching in satellites at these low stellar masses; such a comparison requires higher-resolution zoom-in simulations \citep[e.g.,][]{Hopkins2018,Jahn2022,Cruz2026}. The FIRE simulations of LMC-mass hosts by \citet{Jahn2022} predict that satellites of LMC-mass hosts have quenched fractions similar to those of satellites around MW-mass hosts at fixed satellite stellar mass. The \elvesdwarf quenched fraction is generally lower than this prediction. However, this comparison should be interpreted cautiously because of differences in host and satellite selection and in the definitions of quenching used in the observations and simulations.

A more complete characterization of the star-formation properties and quenching of \elvesdwarf satellites will require multi-wavelength data beyond the optical colors used here. In future work, we will combine optical photometry with UV measurements and \ion{H}{1} observations to obtain more direct constraints on recent star formation and neutral gas content. In addition, we have secured JWST imaging for seven quenched \elvesdwarf satellites, which will provide precise TRGB distances and detailed star formation histories. Together, these data will provide a more detailed view of how satellite quenching proceeds in the dwarf-host regime.

\section{Discussion and Summary}\label{sec:discussion}

In this work, we present the full \elvesdwarf survey, a systematic search for satellite systems around primarily isolated dwarf galaxies in the Local Volume. We combine wide-field satellite searches in the Legacy Surveys with membership classifications based on SBF distances from our follow-up and archival imaging, literature TRGB distances, and radial velocities. In total, we have \numhosts dwarf hosts with well-characterized completeness and \numsats confirmed satellites with $M_\star>10^5\,M_\odot$. This sample provides the first large, homogeneous census of satellites of dwarf hosts, complementing targeted searches such as MADCASH \citep{Carlin2016,Carlin2019,Hargis2020,Carlin2024}, LBT-SONG \citep{Garling2021,Davis2021,Davis2024}, and DELVE-DEEP \citep{Doliva-Dolinsky2025,Medoff2025}, as well as the ID-MAGE survey \citep{Hunter2025,Hunter2026}, which probes a similar host-mass and distance range across a broader range of host environments. The distance measurements and photometric properties of all satellite candidates are provided in Table~\ref{tab:sats}. The data products are available online, accompanied by interactive sky viewers and tutorial notebooks\footnote{\catalogurl}.

The satellite abundances of dwarf hosts, shown in Figure~\ref{fig:nsat}, are broadly consistent with the theoretical predictions from TNG50. We find no strong evidence for either a \textit{global} ``missing-satellite'' problem or a \textit{global} ``too-many-satellites'' problem in the dwarf-host regime. Compared with the first results of \elvesdwarf from \citet{Li2025}, the larger sample presented here reveals substantial host-to-host diversity. Most hosts contain zero or one confirmed satellite above our fiducial mass threshold of $M_\star\gtrsim10^{5.7}\,M_\odot$, while a small number of systems are much richer. In particular, DDO~161 and NGC~4214 lie above the $2\sigma$ range of the TNG50 prediction in Figure~\ref{fig:nsat}. The host-by-host satellite stellar mass functions in Figure~\ref{fig:sat_mass_function} similarly highlight systems such as the LMC, NGC~4700, and UGC~05423, whose most massive satellites are unusually massive compared with the model expectations. These individual systems do not by themselves rule out the adopted galaxy--halo connection model from \citet{Nadler2020}, but they provide important leverage on the slope, scatter, and possible environmental dependence of the SHMR in the low-mass regime \citep[e.g.,][]{Danieli2023,Li2026_DDO161,Kado-Fong2025}.

An equally important constraint comes from the low-\nsat end of the \elvesdwarf sample, which is less visually emphasized in Figure~\ref{fig:nsat}. A complete interpretation of the sample should therefore model the full host-by-host satellite abundance, rather than focusing only on the satellite-rich outliers. In practice, this requires fitting the joint likelihood of the observed satellite abundances and stellar mass functions across the full sample, while accounting for host stellar-mass uncertainties, which are non-negligible (Table~\ref{tab:hosts} and Appendix~\ref{ap:mstar}), detection completeness, unconfirmed candidates, and projection effects (Appendix~\ref{ap:purity}). We will build such a framework in future work, similar in spirit to \citet{Danieli2023}. Applied to \elvesdwarf, this approach will use the full range of satellite populations, from hosts with no detected satellites to satellite-rich outliers, to constrain the low-mass SHMR and test whether the galaxy--halo connection depends on environment.

The improved statistics of the full \elvesdwarf sample also allow us to move beyond simple satellite counts. The cumulative satellite-to-host mass-ratio function (Figure~\ref{fig:sat_mass_ratio}) agrees well with both the expectation from \citet{Sales2013} and the TNG50 prediction, supporting the idea that satellite populations remain approximately scale-free when expressed in terms of stellar mass ratio. The radial distribution of \elvesdwarf satellites (Figure \ref{fig:radial}) is also broadly consistent with TNG50. Although the \elvesdwarf radial distribution appears slightly more centrally concentrated than those of ELVES and SAGA, the current statistics are insufficient to determine whether these differences are significant. 

This similarity among satellite radial distributions across host mass is qualitatively expected if satellites trace the underlying dark matter halo, since halo concentration changes only slightly between dwarf-host halos ($M_h\approx10^{11}\,M_\odot$, $c\approx12$) and Milky Way-mass halos ($M_h\approx10^{12}\,M_\odot$, $c\approx10$) \citep[e.g.,][]{Bullock2001,Diemer2019}. If real, the more concentrated \elvesdwarf profile may indicate that satellite disruption is less efficient around dwarf hosts, consistent with their weaker tidal fields and the absence of a massive stellar and gaseous disk \citep[e.g.,][]{Nagai2005,Jahn2019,Samuel2020,Jiang2021,Green2022}.

The star-forming properties of the satellites provide a complementary view of how the environment regulates satellite evolution. We find a quenched fraction of only $\sim25$--$40\%$ for \elvesdwarf satellites, substantially lower than the $\sim50$--$70\%$ measured for satellites in the same stellar mass range around MW-mass hosts in ELVES and SAGA \citep{Greene2023,SAGA-III,SAGA-IV}. This lower quenched fraction suggests that environmental quenching is less efficient in dwarf-host halos, consistent with their weaker tidal fields and less dense circumgalactic media relative to MW-mass hosts \citep{Bordoloi2014,Zheng2024}. At the same time, quenching is not absent: the quenched fraction appears to increase with host stellar mass even within the \elvesdwarf sample, although the trend is not statistically significant. Combined with ELVES-Field, where nearly all isolated dwarfs above $M_\star\sim10^7\,M_\odot$ are star-forming and the field quenched fraction rises to $\sim30\%$ only below this mass \citep{ELVES_Field_II}, these results demonstrate an environmental sequence in quenching, from isolated field dwarfs to satellites of dwarf hosts to satellites of MW-mass hosts. A more complete physical interpretation will require star-formation diagnostics beyond optical colors. In future work, we will use UV photometry, \ion{H}{1} measurements, and JWST-resolved stellar populations to map out this environmental sequence of galaxy quenching.

At present, extending \elvesdwarf to more hosts still requires follow-up imaging to confirm satellite membership, which limits our ability to build a substantially larger sample. As shown by \citet{ELVES_Field_I}, sufficiently deep wide-field imaging surveys including HSC, Rubin/LSST \citep{LSST2019}, Euclid \citep{Euclid_Overview}, and Roman \citep{Spergel2015} can instead enable SBF distances directly from the discovery data, greatly reducing this observational cost. With these upcoming surveys, dwarf galaxy searches can be extended over much larger areas and volumes, increasing the number of dwarf hosts with well-characterized satellite systems by orders of magnitude. Together with deeper resolved-star studies from space-based facilities (e.g., HST, JWST), these future samples will turn satellite populations of dwarf galaxies into a powerful statistical probe of how low-mass galaxies form and evolve across different cosmic environments. 

\appendix

\section{Survey footprint and detection completeness of individual hosts}\label{ap:footprint}

In this appendix, we present host-by-host summaries of the survey footprint and detection completeness for the \elvesdwarf sample. Figure~\ref{fig:host_coverage} shows the sky coverage for satellite candidate detection around each host. For each host, we show the Legacy Surveys imaging footprint adopted in our search, the projected virial radius of the host, and the locations of satellite candidates classified according to their membership status. For most of the hosts, our search covers almost the entire virial area. The survey footprint files are also available on our survey website. 

Figure~\ref{fig:host_completeness} shows the detection completeness for each host as a function of central surface brightness, $\mu_0(g)$, and absolute magnitude, $M_g$, as described in \S\ref{sec:completeness}. The average satellite mass--size relation from \citet{ELVES-I}, together with its $1\sigma$ scatter, is overplotted to indicate the region of parameter space occupied by typical dwarf satellites.

\begin{figure*}
    \centering
    \includegraphics[width=1\linewidth]{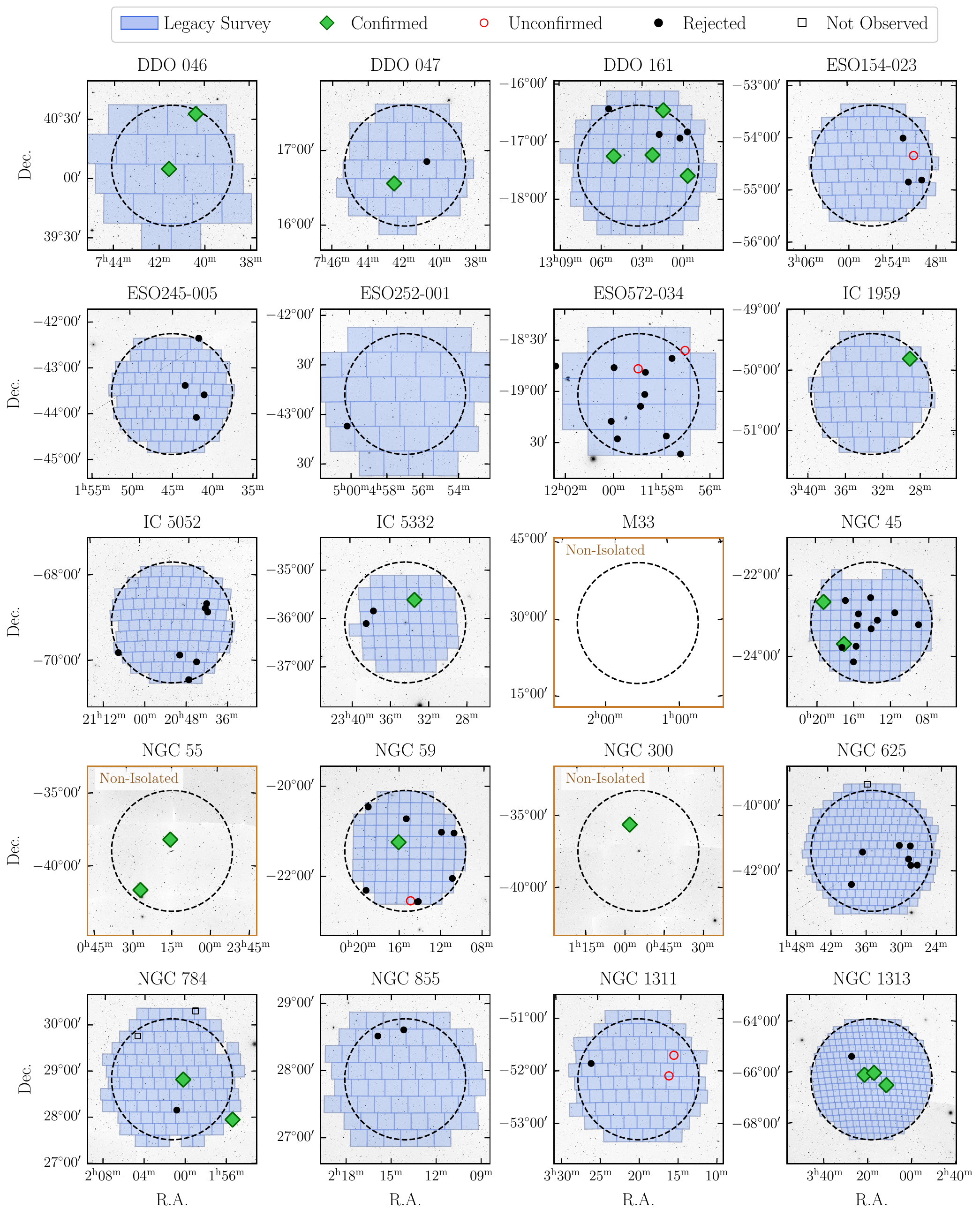}
    \caption{Survey footprint of the dwarf hosts in our sample. For each host, the coverage of the Legacy Survey (used for satellite candidate detection) is shown in blue, with the projected virial radius \rvir highlighted by a black dashed circle. The background image is taken from DSS. We mark the positions of the confirmed satellites (green diamonds), unconfirmed satellite candidates (red open circles), rejected candidates (black-filled circles), and candidates without follow-up observations (open squares). We only show satellites that are more massive than $M_\star \geqslant 10^5\,M_\odot$. The non-isolated hosts are highlighted in dark orange. }
    \label{fig:host_coverage}
\end{figure*}

\begin{figure*}
    \centering
    \includegraphics[width=1\linewidth]{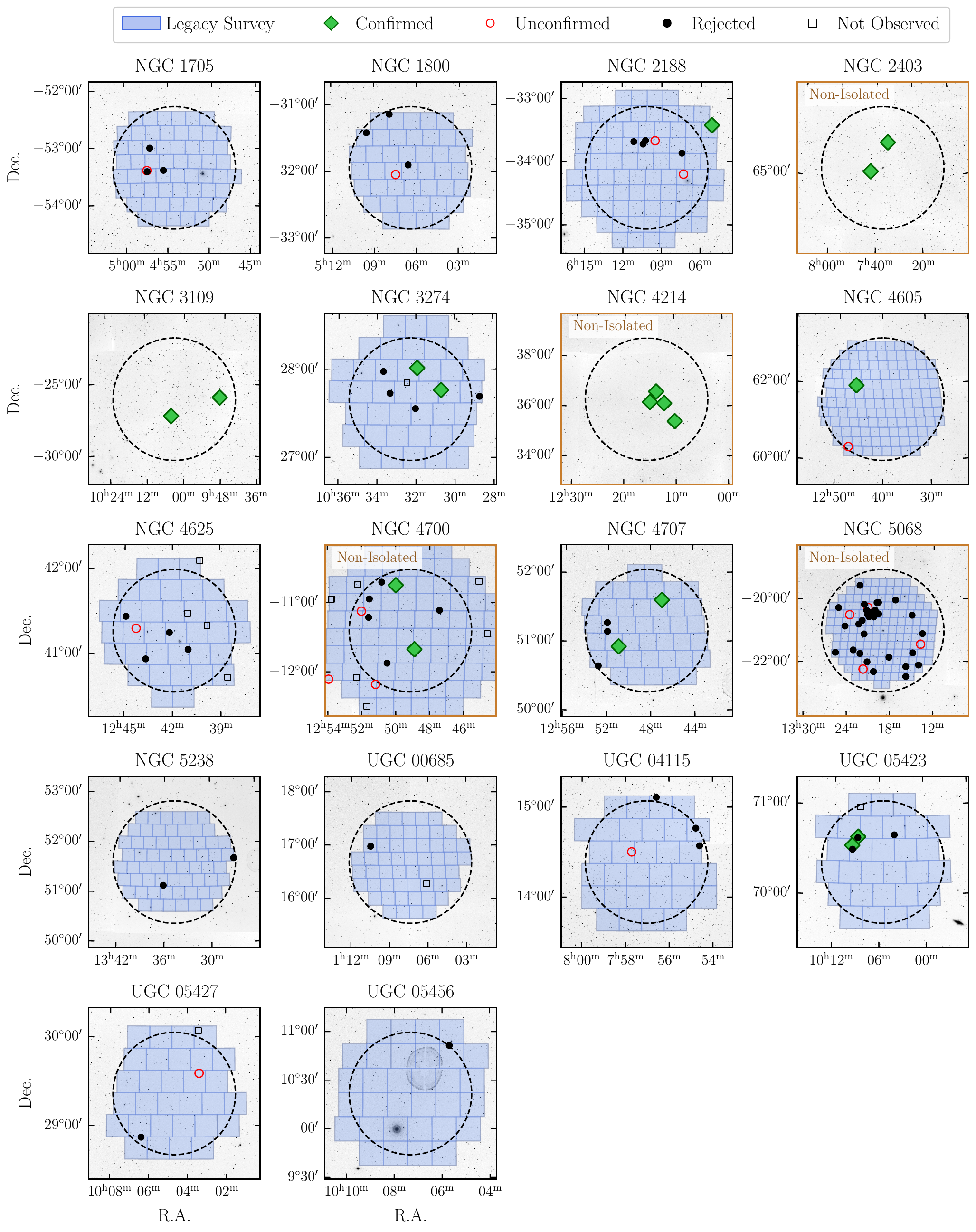}
    \addtocounter{figure}{-1}
    \caption{(Continued.)}
\end{figure*}

\begin{figure*}
    \centering
    \includegraphics[width=0.95\linewidth]{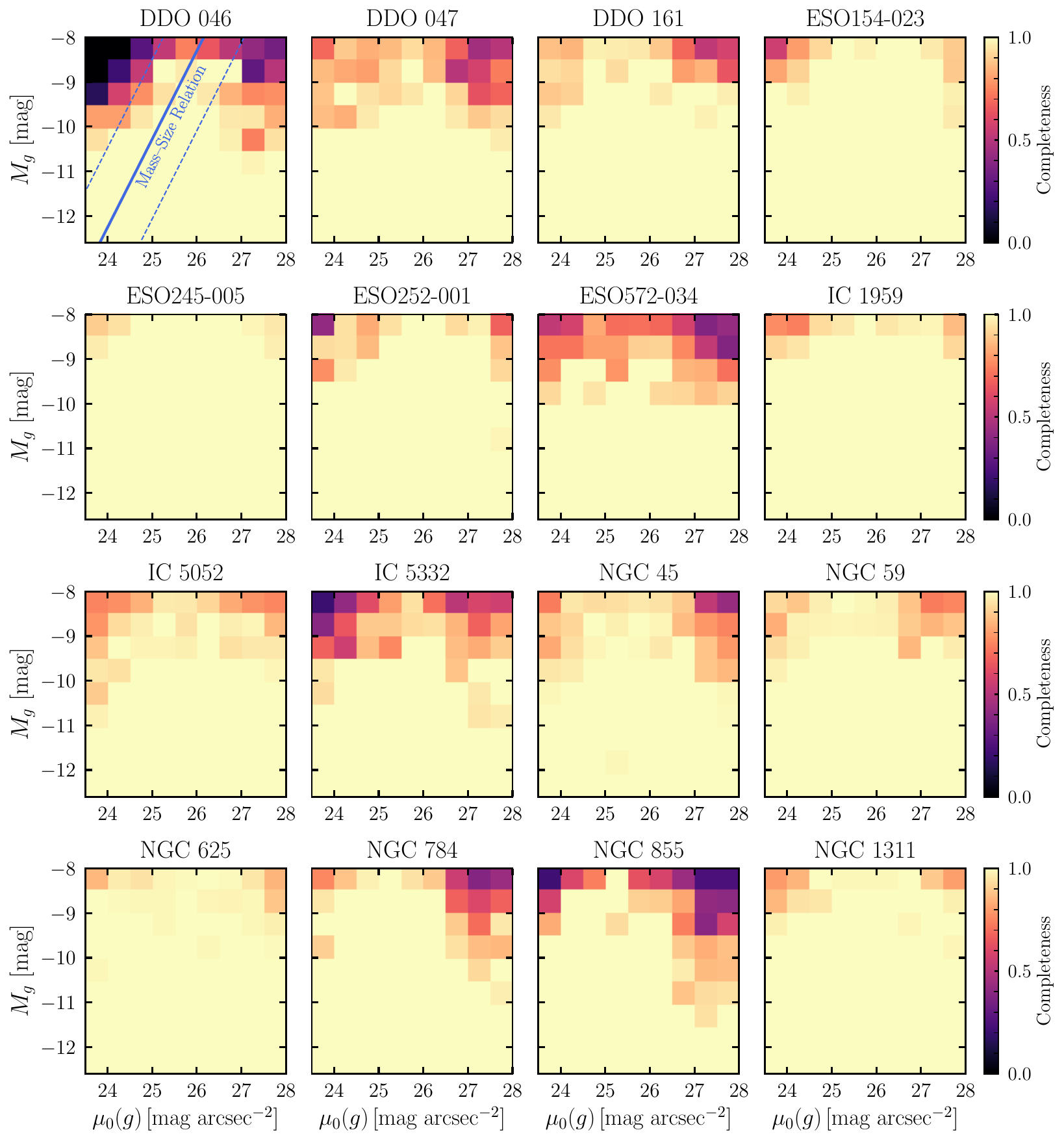}
    \caption{Detection completeness of each host as a function of central surface brightness $\mu_0(g)$ and absolute magnitude $M_g$. The blue solid line shows the average mass--size relation for dwarf satellites from \citet{ELVES-I}, and the blue dashed lines show $1\sigma$ scatter in the mass--size relation. We only show the hosts searched in this work, excluding the hosts taken from the literature.}
    \label{fig:host_completeness}
\end{figure*}

\begin{figure*}
    \centering
    \includegraphics[width=0.95\linewidth]{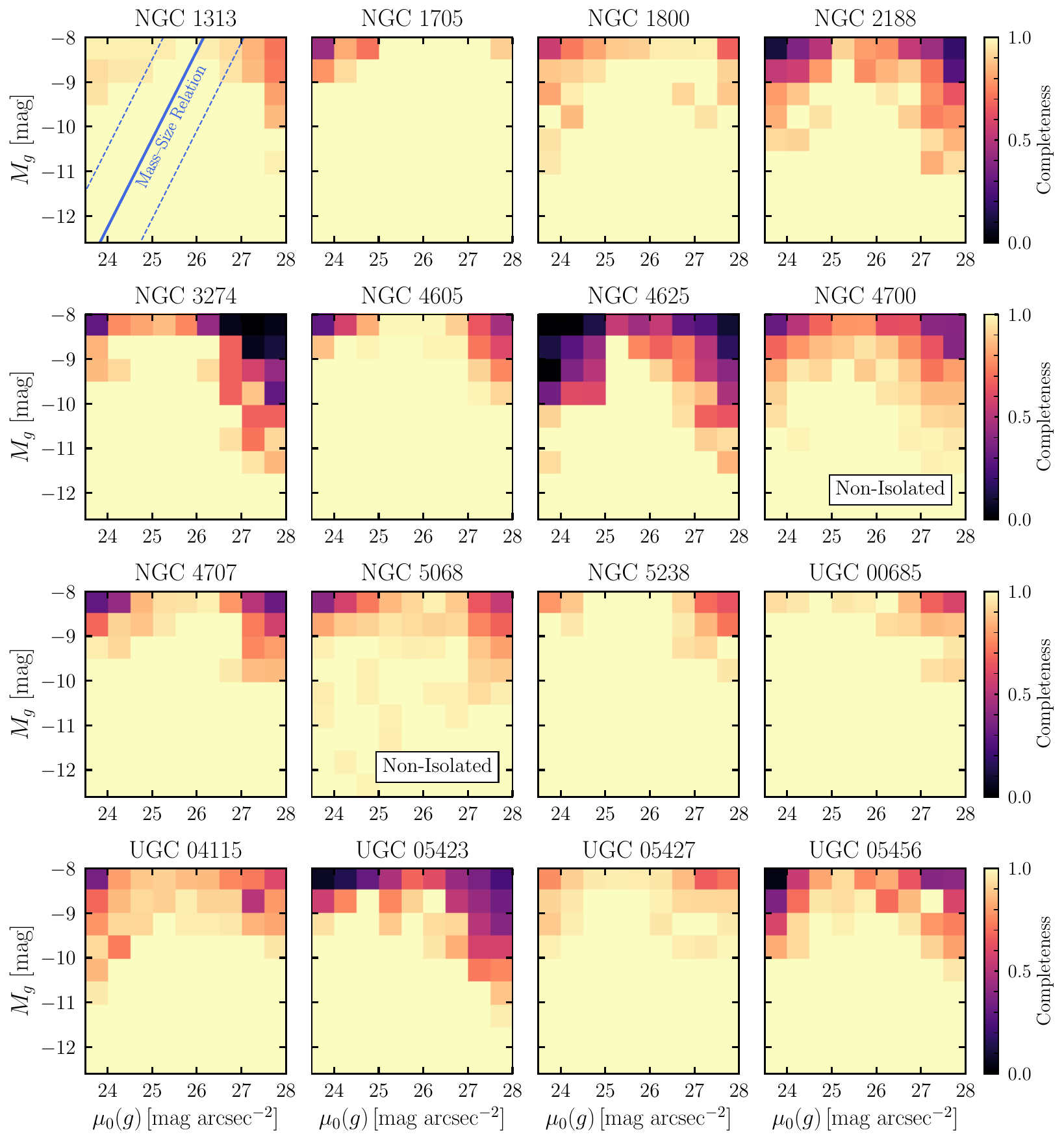}
    \addtocounter{figure}{-1}
    \caption{(Continued.)}
\end{figure*}

\section{Imaging Data Reduction}\label{sec:data_reduction}

In Section~\ref{sec:followup}, we described the imaging data obtained in our follow-up campaign to measure SBF distances for the satellite candidates, using Subaru/HSC, Magellan/IMACS, and Gemini/GMOS. In this section, we summarize the data reduction procedures for each instrument. Our reductions are optimized for accurate sky-background subtraction, photometric calibration, and PSF characterization, all of which are critical for reliable SBF measurements. The reduction of the Subaru/HSC data is described in detail in \citet{Li2025}; we therefore do not repeat those procedures here.

\subsection{Magellan/IMACS}
We reduce the Magellan/IMACS imaging data using a custom pipeline. The observations analyzed in this work were obtained either with the IMACS F2 camera or with the IMACS F4 camera, with the latter read out in $2\times2$ binning mode. The resulting native pixel scale is approximately $0.2\arcsec\ \mathrm{pixel}^{-1}$ in both configurations.

Each exposure is first bias-subtracted and flat-fielded on a chip-by-chip basis. The flat fields are built from twilight flat exposures obtained on the same night as the science data. Cosmic rays are removed from the individual calibrated exposures using \code{Astro-SCRAPPY} \citep{astroscrappy,LACosmic}. Astrometric calibration is performed in two stages: we first derive an initial solution with \code{astrometry.net}\footnote{\url{https://astrometry.net/}} \citep{Lang2010}, and then refine it relative to Gaia DR3 reference stars \citep{Gaia-dr3} using \code{SCAMP}\footnote{\url{https://www.astromatic.net/software/scamp/}} \citep{scamp}.

The treatment of the sky background depends on the IMACS camera. For data taken with the F2 camera, we subtract the sky background directly from each exposure using \code{photutils} \citep{photutils} with a mesh size of $20\arcsec$ to capture large-scale sky variations without over-subtracting the local structure. For large and bright galaxies (e.g., NGC~1800 and NGC~4700), we increase the mesh size to avoid over-subtracting the extended outskirts of the galaxy. The $i$-band exposures taken with the F4 camera exhibit strong interference fringes arising from sky-line emission, which, if not corrected, can bias the SBF measurements. We therefore construct a fringe frame from a subset of science exposures with matched exposure times but different pointings. After aggressively masking astrophysical sources, we subtract a global sky level from each image and median-combine the masked frames to produce the fringe template. We visually verify that the resulting fringe frame is free of significant contamination from residual flux associated with bright or saturated sources, and then subtract it from each science exposure. After this fringing correction, the smooth sky background is removed using \code{photutils} with the same $20\arcsec$ mesh size. No fringe correction is applied to the F2 data.

For photometric calibration, we compare instrumental PSF magnitudes measured from each exposure against external reference catalogs. We first construct a PSF model using \code{PSFEx} \citep{psfex} from non-saturated point sources in each exposure, and perform PSF-fitting photometry using \code{SExtractor} \citep{SExtractor}. For targets in the southern sky, we calibrate against DELVE DR2 \citep{DELVE-DR2} where available; otherwise we use Pan-STARRS1 DR2 \citep{Chambers2016}. Zeropoints are derived separately for each chip and exposure, and the calibrated images are then rescaled to a common zeropoint of $\mathrm{ZP}=27.0$~mag. We tested for the presence of a color term in the calibration and found it negligible, so no color-term correction is applied.

After astrometric and photometric calibration, the individual exposures are coadded with \code{SWarp} \citep{swarp}. To minimize the impact of correlated noise on the SBF analysis, we adopt the \code{NEAREST} interpolation kernel when resampling the input images \citep{Mei2005}. The coadd images are designed to have a pixel scale of $0.2\arcsec\, \mathrm{pixel}^{-1}$ for IMACS F2 and $0.23\arcsec\,\mathrm{pixel}^{-1}$ for IMACS F4. We then derive PSF models for the coadded images using \code{PSFEx}.

\subsection{Gemini/GMOS}

We reduce the Gemini/GMOS imaging data using the \code{DRAGONS} data reduction package \citep{DRAGONS_paper,DRAGONS_software}. We first construct a master bias and normalized flat field from calibration exposures taken during the same observing run, and apply these calibrations to individual science frames. The \code{DRAGONS} pipeline then performs astrometric calibration, matches sky levels across exposures, and combines the images into a final coadded image.

Because \code{DRAGONS} does not perform background subtraction or photometric calibration by default, we manually carry out these steps on the coadded images. We model and subtract the sky background using \code{photutils} \citep{photutils}, with a mesh size of $20\arcsec$, similar to the IMACS reduction. Photometric calibration is then performed using the same PSF-based procedure described for the IMACS data, with Pan-STARRS1 DR2 as the reference catalog. The photometric zeropoint is set to $\mathrm{ZP}=27.0$~mag. Finally, we construct a PSF model for each coadded image using \code{PSFEx} \citep{psfex}.

\section{SBF Distances to Two Host Galaxies: NGC~1800 and NGC~4700}\label{ap:ngc4700}

\begin{figure*}
    \centering
    \includegraphics[width=1\linewidth]{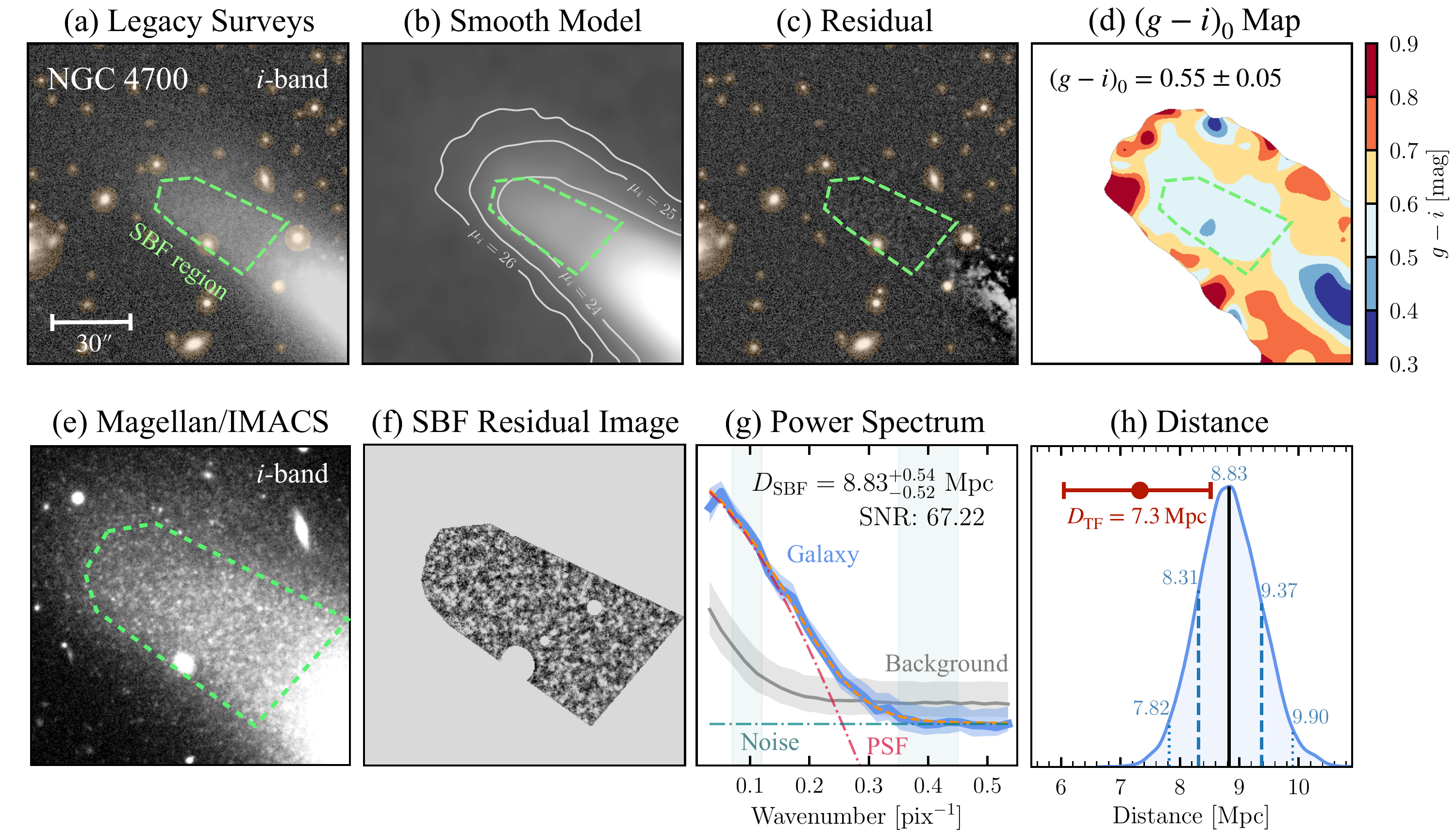}
    \caption{SBF distance measurement for NGC~4700. \textit{Top panels}: Local color measurement for converting SBF amplitude to distance. (a) Legacy Surveys $i$-band image of the northeastern outskirts of NGC~4700; the green dashed polygon marks the SBF measurement region. (b) Smooth $i$-band model from the Legacy Surveys imaging, with contours showing constant surface brightness. (c) Residual after subtracting the smooth model, showing that the large-scale galaxy light is well modeled. (d) Extinction-corrected $(g-i)_0$ map derived from the smooth $g$- and $i$-band models. The SBF region is chosen to minimize color gradients and has a color of $(g-i)_0=0.55\pm0.05$~mag. \textit{Bottom panels}: SBF measurement from the Magellan/IMACS data. (e) IMACS $i$-band image of the same northeastern region, showing strong SBF signals. (f) SBF residual image after subtracting a smooth galaxy model and applying the mask. (g) Power spectrum of the residual image. The measured spectrum is shown in blue, the best-fit model in orange, the PSF-shaped SBF component in red, and the white-noise component in teal. The gray shaded region shows the contribution from unmasked background sources. (h) Inferred SBF distance posterior, giving $D_{\rm SBF}=8.83\pm0.54$~Mpc. The Tully--Fisher distance from \citetalias{Karachentsev2013} is shown in red.}
    \label{fig:NGC4700_sbf}
\end{figure*}

\begin{figure*}
    \centering
    \includegraphics[width=1\linewidth]{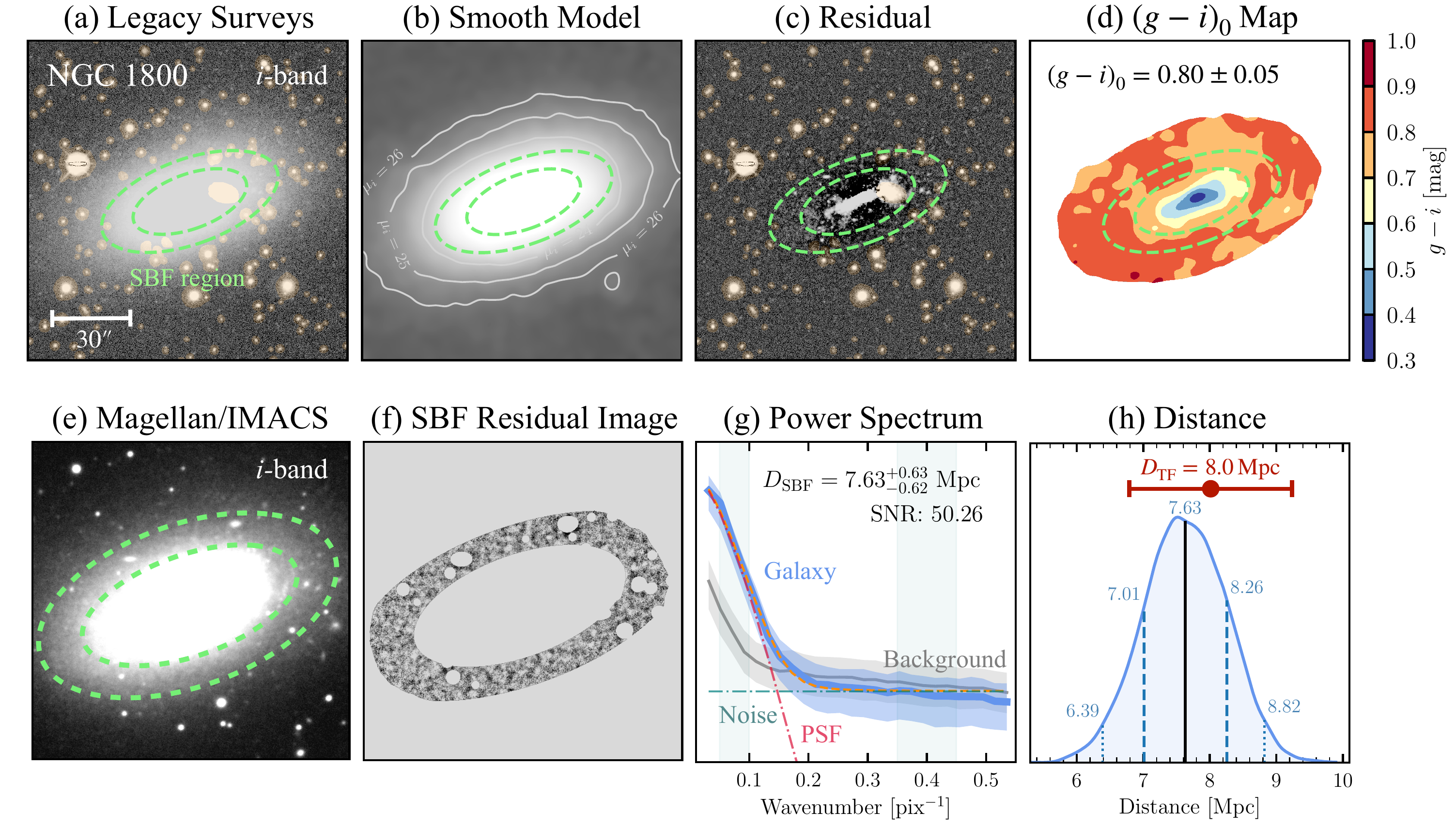}
    \caption{SBF distance measurement for NGC~1800 using Magellan/IMACS imaging, shown in the same format as Figure~\ref{fig:NGC4700_sbf}. The inferred distance is $D_{\rm SBF}=7.63\pm0.63$~Mpc, consistent with the Tully--Fisher distance from \citetalias{Karachentsev2013}, shown in red.}
    \label{fig:NGC1800_sbf}
\end{figure*}

Two galaxies in our sample, NGC~1800 and NGC~4700, do not have published TRGB distances. Instead, \citet{Karachentsev2013} report Tully--Fisher distances of $D_{\rm TF}=8.0\pm1.4$~Mpc for NGC~1800 and $D_{\rm TF}=7.3\pm1.3$~Mpc for NGC~4700. These measurements have large uncertainties and are not always reliable. We therefore estimate distances to these two hosts using SBF measurements from our Magellan/IMACS imaging.

These measurements require exquisite treatment because both galaxies have large angular sizes and complex stellar populations. NGC~4700 is a star-forming edge-on disk galaxy, while NGC~1800 is a dwarf galaxy with a compact blue star-forming central region. A straightforward SBF measurement over the main galaxy body is therefore not appropriate: the bright inner regions contain young stellar populations, dust, and strong stellar-population gradients, all of which complicate the interpretation of the SBF amplitude. We instead restrict the measurements to low-surface-brightness outskirts, where the stellar populations are expected to be older and more spatially homogeneous, and where the local color can be estimated with minimal contamination from strong color gradients.

We first attempt to estimate the local $g-i$ color using the multi-band Legacy Surveys imaging. Figure~\ref{fig:NGC4700_sbf} illustrates this procedure for NGC~4700. The northeastern outskirts of NGC~4700 contain a relatively smooth, low-surface-brightness extension suitable for the SBF measurement. To build a $g-i$ color map, we construct smooth two-dimensional models of the galaxy light in the $g$ and $i$ bands by first building an aggressive mask to exclude foreground stars, background galaxies, saturated pixels, and other artifacts (Panel (a)). We then median-filter the masked image with a kernel size of $\sim6\arcsec$ to obtain a smooth model in each band. The resulting flux models are additionally smoothed with a Gaussian kernel of $\sigma=2\arcsec$ to suppress small-scale artifacts before constructing the color map. Panel (b) shows the smooth $i$-band model, with contours corresponding to $\mu_i=24$, 25, and 26~mag~arcsec$^{-2}$. Panel (c) shows the residual image after subtracting the smooth model, demonstrating that the model captures the large-scale galaxy light. Since the relevant color structures are much larger than the seeing scale, we do not PSF-match the two bands. 

With the smooth models in $g$ and $i$ bands, we build the color map by taking their ratios. Panel (d) shows the extinction-corrected $(g-i)_0$ map. A color gradient is present in this region of the galaxy, and we choose a region where the gradient is minimal, highlighted by the green polygon, for SBF measurement. Within this aperture, we measure the color by summing the unmasked pixels in the $g$- and $i$-band models, obtaining $(g-i)_0=0.55\pm0.05$~mag. The quoted uncertainty accounts for both changes under modest aperture variations and systematic uncertainty in the smooth-model construction.

We then measure the SBF signal from the Magellan/IMACS data following the methodology described in Section~\ref{sec:sbf_meas} and in \citet{Li2024}. Panel (e) of Figure~\ref{fig:NGC4700_sbf} shows the IMACS $i$-band image, obtained in $0\farcs5$ seeing with a total exposure time of 30 minutes. A strong SBF signal is visible in the selected outer region. We construct the smooth galaxy model using the median-filtering approach as described in \S\ref{sec:sbf_meas}, with a kernel size of ten times the PSF FWHM. Panel (f) shows the residual image used for the SBF measurement, and panel (g) shows the azimuthally averaged power spectrum together with the fitted SBF, PSF, and white-noise components. The SBF signal is detected at high significance, with ${\rm S/N}\simeq67$. The resulting SBF distance distribution is shown in panel (h), with $D_{\rm SBF}=8.83^{+0.54}_{-0.52}~{\rm Mpc}$. This distance is broadly consistent with the Tully--Fisher estimate of $D_{\rm TF}=7.3\pm1.3$~Mpc from \citet{Karachentsev2013}, although the SBF measurement places NGC~4700 slightly farther away and has a smaller uncertainty.

We apply the same procedure to NGC~1800, as shown in Figure \ref{fig:NGC1800_sbf}. Its IMACS observations were obtained in $0\farcs9$ seeing with a total exposure time of 30 minutes. We again avoid the bright central region and define an outer annular aperture for the SBF measurement. The local color in this annulus is $(g-i)_0=0.80\pm0.05$~mag. The resulting SBF distance is $D_{\rm SBF}=7.63\pm0.63~{\rm Mpc}$, with a high detection significance of ${\rm S/N}\simeq50$. This distance agrees well with the published Tully--Fisher estimate of $D_{\rm TF}=8.0\pm1.4$~Mpc \citep{Karachentsev2013}. We adopt the SBF distances measured here for NGC~4700 and NGC~1800 in this paper. 

We recompute the tidal indices of NGC~4700 and NGC~1800 using the new SBF distances, following the prescription of \citet{Karachentsev2013}. We obtain $\Theta_1=0.60$ and $\Theta_5=0.60$ for NGC~4700, and $\Theta_1=-1.70$ and $\Theta_5=-1.50$ for NGC~1800. With the revised distance, NGC~4700 is therefore classified as non-isolated, with the M94 group as its dominant background perturber.

\section{Comparison among Different Host Stellar-Mass Estimates}\label{ap:mstar}

\begin{figure*}
    \centering
    \includegraphics[width=1\linewidth]{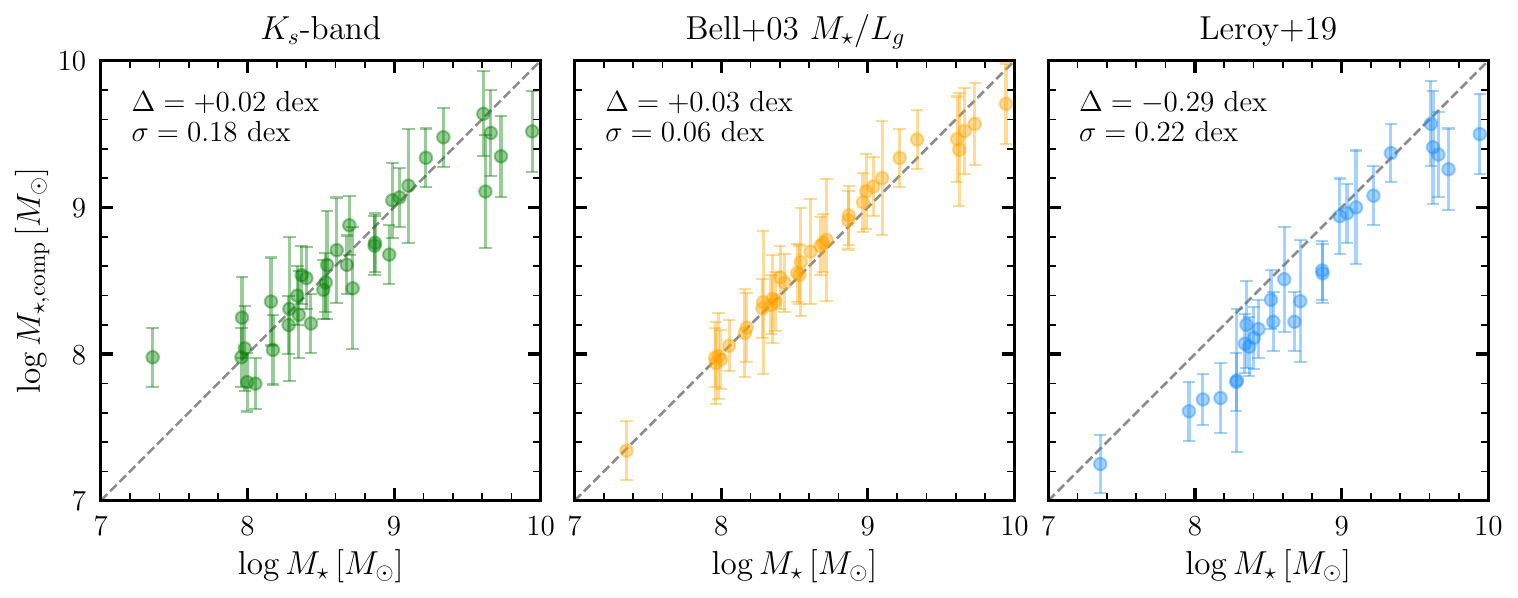}
    \caption{Comparison of host stellar-mass estimates used in this work. The $x$-axis shows our fiducial stellar masses, derived from Siena Galaxy Atlas photometry using the optical color--$M_\star/L_g$ calibration of \citet{delosReyes2025}. The $y$-axis shows alternative estimates based on $K_s$-band luminosities assuming $M_\star /L_{\rm K_s} = 0.6$ (left panel), the optical color--$M_\star/L_g$ relation from \citet{Bell2003} applied to the same SGA photometry (middle panel), and the WISE-based measurements from \citet{Leroy2019} (right panel). The dashed gray line indicates one-to-one agreement. These three comparisons show broad consistency among the different stellar-mass estimators, with a typical scatter of $\sim0.2$ dex. This scatter reflects systematic uncertainties associated with photometric measurements and color--$M_\star/L$ calibrations in different wavelengths. We note that the \citet{Leroy2019} estimates are systematically lower than our fiducial stellar masses.
    }
    \label{fig:host_mstar_comparison}
\end{figure*}

In this appendix, we compare our fiducial host stellar masses with several alternative stellar mass estimates. Our fiducial values are derived from Siena Galaxy Atlas optical photometry \citep{Moustakas2023,Moustakas2025_inprep} using the optical color--$M_\star/L_g$ calibration of \citet{delosReyes2025}. As a consistency check, we compare these values against estimates based on (1) $K_s$-band luminosities from \citet{Karachentsev2013}, assuming a fixed $M_\star/L_{K_s}=0.6$; (2) the same SGA photometry, but using the color--$M_\star/L_g$ relation of \citet{Bell2003}; and (3) WISE-based stellar masses from \citet{Leroy2019}. The $K_s$-band and WISE-based estimates provide useful external comparisons because both are widely used stellar-mass estimators for nearby galaxies. As shown in Figure~\ref{fig:host_mstar_comparison}, the different estimators are broadly consistent, with a typical scatter of $\sim0.2$ dex, although the \citet{Leroy2019} masses are systematically lower than our fiducial values \citep{Li2025}. We take the 0.2 dex uncertainty as the minimum host stellar mass uncertainty in Table \ref{tab:hosts}, which reflects systematic uncertainties associated with photometric measurements and color--$M_\star/L$ calibrations.

\section{Velocity-Based Satellite Association Criterion}\label{ap:vel_thresh}

In this work, we use radial velocity measurements, when available, as an additional criterion for associating satellite candidates with their host galaxies. To define an appropriate velocity threshold, we calibrate the expected line-of-sight velocity offsets of true satellites using the TNG50 hydrodynamical cosmological simulation. For the TNG50 host sample described in \S\ref{sec:tng}, we select satellites with $M_\star > 10^{5.6}\,M_\odot$ that lie within the projected virial radius of their host, and calculate the absolute line-of-sight velocity offset $\Delta v = |v_{\rm sat} - v_{\rm host}|$. 

\begin{figure}
    \centering
    \includegraphics[width=1\linewidth]{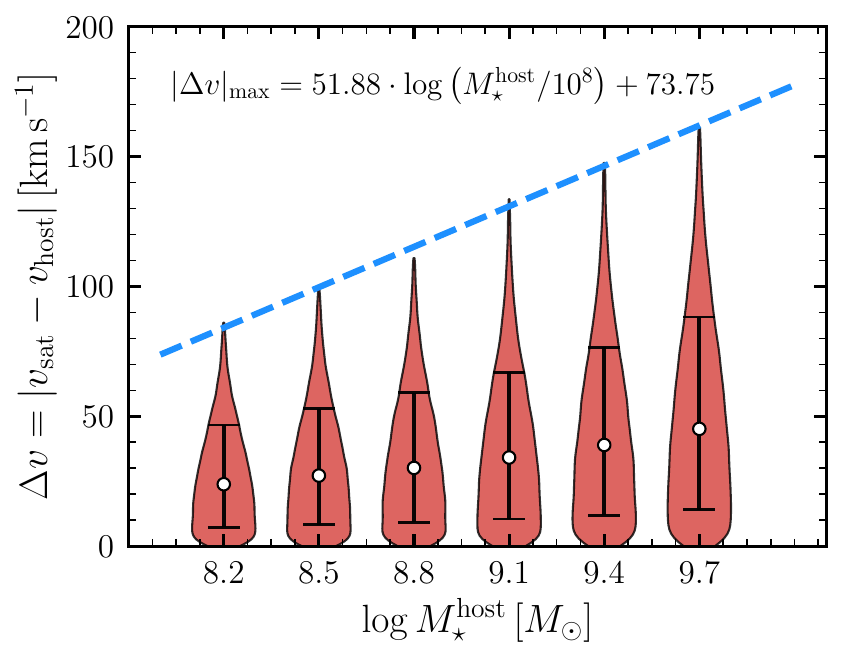}
    \caption{Distribution of line-of-sight velocity offsets for satellites in TNG50, defined as $\Delta v = |v_{\rm sat} - v_{\rm host}|$. Satellites are selected to have $M_\star > 10^{5.6}\,M_\odot$ and to lie within the projected virial radius of their host. The violin distributions are truncated at the 99th percentile for visualization. The vertical black bars indicate the 16th--84th percentile range, with white circles marking the median. We fit the 99th percentile of the velocity-offset distribution as a function of host stellar mass; the best-fit relation is shown by the blue dashed line and is used to define the velocity-based association threshold in \S\ref{sec:psat}.}
    \label{fig:tng_sat_vel}
\end{figure}

Figure~\ref{fig:tng_sat_vel} shows the resulting distribution of $\Delta v$ as a function of host stellar mass. As expected, both the median velocity offset and the width of the distribution increase with host mass, reflecting the deeper potential wells of more massive systems. To define a velocity threshold that is conservative in the sense of maximizing satellite completeness, we characterize the upper envelope of the simulated satellite velocity distribution using the 99th percentile in each host-mass bin. We then fit this 99th-percentile velocity offset as a function of host stellar mass, and get
\begin{equation}
    (\Delta v)_{\rm max}\ [\kms] = 51.88 \cdot \log \left(M_\star^{\rm host}/10^8\right) + 73.75.
\end{equation}

This best-fit relation is shown as the blue dashed line in Figure~\ref{fig:tng_sat_vel}. We adopt this mass-dependent threshold when classifying the satellite candidates using velocity information in \S\ref{sec:psat}. This choice retains nearly all simulated satellites while rejecting candidates with velocity offsets that are unlikely to be dynamically associated with the host.

\section{Purity of SBF-based Satellite Association}\label{ap:purity}

SBF distances provide an efficient way to determine the membership of satellite candidates in the Local Volume. As we described in \S\ref{sec:psat}, we classify a candidate as a confirmed satellite if its SBF distance is consistent with the host distance within $2\sigma$. This criterion is designed to be inclusive: it includes the majority of true satellites. However, ground-based SBF distances for dwarf galaxies within $\sim 10$~Mpc typically have uncertainties of order $\sim 10\%$ \citep{Carlsten2019,Greco2021,CarlstenELVES2022}. As a result, foreground and background galaxies can be scattered into the host distance, and consequently classified as real satellites in our observation. For example, a $10\%$ distance uncertainty at $D=8$~Mpc corresponds to $\sim 0.8$~Mpc, substantially larger than the virial radii of the dwarf hosts considered here. It is therefore important to quantify the expected contamination fraction from galaxies that are projected near a host but are not physically associated with it.

Similar to \citet{Li2026_DDO161}, we estimate this contamination fraction using TNG50, which provides a cosmological environment such that we can select both true satellites and interlopers. We select isolated TNG50 hosts as described in \S\ref{sec:tng} and identify all candidate galaxies projected within the virial radius of each host, regardless of the line-of-sight distance. We then apply a stellar mass threshold of $M_\star > 10^{5.6}\,M_\odot$, matched to the approximate detection limit of our survey. This sample contains both true satellites and interlopers along the line of sight. To mimic our observational SBF selection, we place each host at a fiducial distance of $D_{\rm host}=6$~Mpc, approximately the median distance of the observed host sample, and perturb the line-of-sight distances of the candidate galaxies by a Gaussian uncertainty corresponding to the assumed SBF distance error. Candidates whose perturbed distances agree with the host within $2\sigma$ are then classified as satellites. We repeat this procedure in Monte Carlo realizations, including both the scatter in the assigned stellar masses and the SBF distance perturbations.

For a fiducial SBF uncertainty of $\sigma_D=7\%$, we find a contamination fraction of $13\pm2\%$ for hosts at $D_{\rm host}=6$~Mpc. This contamination level is consistent with previous estimates for nearby satellite searches, including the $10$--$15\%$ contamination fraction reported for Milky-Way-mass hosts by \citet{Carlsten2021}, as well as the estimate for the DDO~161 satellite system in \citet{Li2026_DDO161}. If instead we place the hosts at $D_{\rm host}=8$~Mpc and assume a larger SBF uncertainty of $\sigma_D=10\%$, the contamination fraction increases to $20\pm2\%$. 

We also investigate the dependence of the contamination fraction on host and satellite stellar mass. Over the host-mass range considered here, the interloper fraction shows no strong dependence on host mass. It increases mildly toward lower satellite masses. If the satellite mass limit is $M_\star > 10^{5}\,M_\odot$, the contamination fraction is $17\pm3\%$.


\section{Satellite Distances and Photometric Properties}
\label{ap:satcat}

Table~\ref{tab:sats} presents the distance measurements and photometric properties of all satellite candidates considered in this work. The table includes the candidate positions, adopted distances or distance constraints, membership classifications, integrated magnitudes, stellar masses, structural parameters, projected separations, and completeness. Membership classifications (`Status' in the table) follow the criteria described in Section~\ref{sec:psat}. The distance-dependent quantities, such as stellar mass and effective radius, are all computed assuming the distance of the host. A machine-readable version of this table is available online\footnote{\catalogurl}.

Here, we note several special cases in the membership classification:
\begin{itemize}

    \item The satellite system of NGC~4214: We use previous studies to define the satellite system of NGC~4214 ($D=3.04$~Mpc; \citealt{Dalcanton2009}). We include its two previously reported satellites, DDO~113 and MADCASH-2 \citep{Garling2020,Carlin2021}, together with NGC~4190 and NGC~4163, whose association with NGC~4214 has been discussed in the literature \citep{Allsopp1979a,Allsopp1979b,Karachentsev2013,Annibali2020}. NGC~4190 and NGC~4163 lie close to NGC~4214 in projection, and both have TRGB distances of $D=2.95\pm0.05$~Mpc \citep{McQuinn2015starburst,Tully2013}, supporting their inclusion in the NGC~4214 satellite system.

    \item LEDA18431: This satellite candidate of NGC~2188 has a small velocity offset from the host, $|\Delta v_{\rm h}|\sim20\ \kms$. However, its TRGB distance, $9.64\pm0.10$~Mpc \citep{Tully2013}, is inconsistent with the TRGB distance of NGC~2188, $8.39\pm0.10$~Mpc. Because the CMD of LEDA~18431 is sparsely populated, its TRGB distance may be less secure than the quoted formal uncertainty. We therefore classify it as \textit{unconfirmed}.

    \item dw0457m5324: This satellite candidate of NGC~1705 has a blue color and shows a possible SBF signal in our IMACS data. However, the apparent fluctuations may be affected, or dominated, by star-forming regions. We therefore do not regard the SBF measurement as a secure distance confirmation and retain the object as \textit{unconfirmed}.

    \item DDO~153: Although its TRGB distance, $9.04\pm0.16$~Mpc, is consistent with the SBF distance of NGC~4700, $8.8\pm0.53$~Mpc, its heliocentric radial velocity differs from that of NGC~4700 by $|\Delta v_{\rm h}|\simeq590\,\kms$, which is too large for the two galaxies to be dynamically associated. We therefore classify DDO~153 as \textit{rejected} as a satellite of NGC~4700 and consider it more likely to be associated with the M104 group \citep{Karachentsev2020}.

    \item dw0251m5421: This satellite candidate of ESO154-023 has very low surface brightness ($\mu_{\mathrm{eff},g}\approx 28\,\sbunit$) and low luminosity ($M_V\approx-7.8$~mag) with tentative evidence for an SBF signal. The measurement is not sufficiently secure for confirmation, and we retain it as \textit{unconfirmed}. Space-based imaging would be needed for a definitive distance measurement.
\end{itemize}

\clearpage
\setlength{\tabcolsep}{2pt}
\begin{longrotatetable}
\begin{deluxetable*}{lCClccCcccCCCCCCccc}\label{tab:sats}
\tabletypesize{\tiny}
\tablecaption{Distances and Photometric Properties of Satellite Candidates}
\tablewidth{0pt}
\tablehead{
    \colhead{Name} & \colhead{R.A.} & \colhead{Dec.} & \colhead{Host} & \colhead{$D$} & \colhead{Meth.} & \colhead{S/N} & \colhead{$v_{h}$} & \colhead{Status} & \colhead{$P_{\rm sat}$} & \colhead{$R_{\rm proj}$} & \colhead{$m_g$} & \colhead{$g-r$} & \colhead{$g-i$} & \colhead{$M_V$} & \colhead{$\log M_\star$} & \colhead{$r_{\rm eff}$} & \colhead{Comp.} & \colhead{Ref.} \\
    \colhead{} & \colhead{(deg)} & \colhead{(deg)} & \colhead{} & \colhead{(Mpc)} & \colhead{} & \colhead{} & \colhead{(km/s)} & \colhead{} & \colhead{} & \colhead{(kpc)} & \colhead{(mag)} & \colhead{(mag)} & \colhead{(mag)} & \colhead{(mag)} & \colhead{($M_\odot$)} & \colhead{(kpc)} & \colhead{} & \colhead{}
}
\startdata
SMC & 13.1580 & -72.8003 & LMC & $0.06^{+0.00,0.00}_{-0.00,0.00}$ & Ceph & \nodata & \nodata & C & 1.0 & 18 & \nodata & \nodata & \nodata & $-16.79 \pm 0.20$ & $8.95 \pm 0.10$ & $1.08 \pm 0.10$ & 1.00 & \citetalias{Pace2024} \\
Antlia & 151.0171 & -27.3311 & NGC3109 & $1.35^{+0.15,0.30}_{-0.15,0.30}$ & TRGB & \nodata & 362 & C & 1.0 & 28 & \nodata & \nodata & \nodata & $-10.40 \pm 0.20$ & $6.40 \pm 0.10$ & $0.46 \pm 0.10$ & 1.00 & \citetalias{McConnachie2012}, \citetalias{Doliva-Dolinsky2025} \\
Antlia B & 147.2337 & -25.9900 & NGC3109 & $1.35^{+0.06,0.12}_{-0.06,0.12}$ & TRGB & \nodata & 376 & C & 1.0 & 75 & $16.23 \pm 0.40$ & $0.50 \pm 0.40$ & \nodata & $-9.70 \pm 0.40$ & $5.80 \pm 0.10$ & $0.27 \pm 0.10$ & 1.00 & \citetalias{Sand2015}, \citetalias{Doliva-Dolinsky2025} \\
SculptorC & 14.4675 & -35.8523 & NGC0300 & $2.04^{+0.13,0.26}_{-0.13,0.26}$ & TRGB & \nodata & \nodata & C & 1.0 & 67 & \nodata & \nodata & \nodata & $-9.10 \pm 0.10$ & $5.70 \pm 0.10$ & $0.36 \pm 0.10$ & 1.00 & \citetalias{Sand2024} \\
NGC55-dw1 & 3.8740 & -38.4190 & NGC0055 & $2.20^{+0.10,0.20}_{-0.10,0.20}$ & TRGB & \nodata & \nodata & C & 1.0 & 32 & \nodata & \nodata & \nodata & $-8.00 \pm 0.30$ & $5.15 \pm 0.10$ & $2.20 \pm 0.10$ & 1.00 & \citetalias{McNanna2024} \\
ESO294-010 & 6.6391 & -41.8561 & NGC0055 & $2.03^{+0.05,0.10}_{-0.05,0.10}$ & TRGB & \nodata & 107 & C & 1.0 & 141 & $14.84 \pm 0.20$ & $0.47 \pm 0.20$ & $0.65 \pm 0.20$ & $-11.30 \pm 0.20$ & $6.40 \pm 0.10$ & $0.25 \pm 0.10$ & 1.00 & \citetalias{Medoff2025} \\
MADCASH-1 & 115.6620 & 65.4172 & NGC2403 & $3.41^{+0.23,0.46}_{-0.23,0.46}$ & TRGB & \nodata & \nodata & C & 1.0 & 35 & $20.12 \pm 0.40$ & \nodata & $0.70 \pm 0.40$ & $-7.81 \pm 0.40$ & $5.30 \pm 0.10$ & $0.18 \pm 0.10$ & 1.00 & \citetalias{Carlin2016}, \citetalias{Carlin2024} \\
\enddata
\end{deluxetable*}
\end{longrotatetable}
\clearpage

\begin{acknowledgments}
J.L. is grateful for helpful discussions with Andrey Kravtsov, Risa Wechsler, Zhiwei Shao, Alex Drlica-Wagner, Alex Ji, Burçin Mutlu-Pakdil, Laura Hunter, Jingyao Zhu, Stacy Kim, Ethan Nadler, Rachael Beaton, Alyson Brooks, Annika Peter, Yifei Luo, Xiaojing Lin, Dustin Lang, Dylan Folsom, Sebastian Monzon, Lee Kelvin, and Yilun Ma. We thank the staff at Las Campanas Observatory and Gemini Observatory, in particular Matias Diaz and Clara Martinez, for their assistance with our observations.

This research was supported by Grant No. 2024777 from the United States-Israel Binational Science Foundation (BSF) and by Grant No. 2506292 from the United States National Science Foundation (NSF).

This paper includes data gathered with the 6.5-meter Magellan Telescopes at the Las Campanas Observatory in Chile. 

The Gemini Observatory is operated by the Association of Universities for Research in Astronomy, Inc., under a cooperative agreement with the NSF on behalf of the Gemini partnership: the National Science Foundation (United States), the National Research Council (Canada), CONICYT (Chile), Ministerio de Ciencia, Tecnología e Innovación Productiva (Argentina), Ministério da Ciência, Tecnologia e Inovação (Brazil), and Korea Astronomy and Space Science Institute (Republic of Korea).

The Hyper Suprime-Cam (HSC) collaboration includes the astronomical communities of Japan and Taiwan, and Princeton University. The HSC instrumentation and software were developed by the National Astronomical Observatory of Japan (NAOJ), the Kavli Institute for the Physics and Mathematics of the Universe (Kavli IPMU), the University of Tokyo, the High Energy Accelerator Research Organization (KEK), the Academia Sinica Institute for Astronomy and Astrophysics in Taiwan (ASIAA), and Princeton University. Funding was contributed by the FIRST program from the Japanese Cabinet Office, the Ministry of Education, Culture, Sports, Science and Technology (MEXT), the Japan Society for the Promotion of Science (JSPS), Japan Science and Technology Agency (JST), the Toray Science Foundation, NAOJ, Kavli IPMU, KEK, ASIAA, and Princeton University. 

This paper is based on data from the Hyper Suprime-Cam Legacy Archive (HSCLA), and the data collected at the Subaru Telescope and retrieved from the HSC data archive system, which is operated by the Subaru Telescope and Astronomy Data Center (ADC) at the National Astronomical Observatory of Japan. Data analysis was in part carried out with the cooperation of the Center for Computational Astrophysics (CfCA), National Astronomical Observatory of Japan. The Subaru Telescope is honored and grateful for the opportunity to observe the Universe from Maunakea, which has cultural, historical, and natural significance in Hawaii. 

This paper makes use of software developed for the Vera C. Rubin Observatory. We thank the observatory for making their code available as free software at http://dm.lsst.org.

The Legacy Surveys consist of three individual and complementary projects: the Dark Energy Camera Legacy Survey (DECaLS; Proposal ID \#2014B-0404; PIs: David Schlegel and Arjun Dey), the Beijing-Arizona Sky Survey (BASS; NOAO Prop. ID \#2015A-0801; PIs: Zhou Xu and Xiaohui Fan), and the Mayall z-band Legacy Survey (MzLS; Prop. ID \#2016A-0453; PI: Arjun Dey). DECaLS, BASS and MzLS together include data obtained, respectively, at the Blanco telescope, Cerro Tololo Inter-American Observatory, NSF’s NOIRLab; the Bok telescope, Steward Observatory, University of Arizona; and the Mayall telescope, Kitt Peak National Observatory, NOIRLab. Pipeline processing and analyses of the data were supported by NOIRLab and the Lawrence Berkeley National Laboratory (LBNL). The Legacy Surveys project is honored to be permitted to conduct astronomical research on Iolkam Du’ag (Kitt Peak), a mountain with particular significance to the Tohono O’odham Nation. The Legacy Surveys imaging of the DESI footprint is supported by the Director, Office of Science, Office of High Energy Physics of the U.S. Department of Energy under Contract No. DE-AC02-05CH1123, by the National Energy Research Scientific Computing Center, a DOE Office of Science User Facility under the same contract; and by the U.S. National Science Foundation, Division of Astronomical Sciences under Contract No. AST-0950945 to NOAO.

NOIRLab is operated by the Association of Universities for Research in Astronomy (AURA) under a cooperative agreement with the National Science Foundation. LBNL is managed by the Regents of the University of California under contract to the U.S. Department of Energy.

This project used data obtained with the Dark Energy Camera (DECam), which was constructed by the Dark Energy Survey (DES) collaboration. Funding for the DES Projects has been provided by the U.S. Department of Energy, the U.S. National Science Foundation, the Ministry of Science and Education of Spain, the Science and Technology Facilities Council of the United Kingdom, the Higher Education Funding Council for England, the National Center for Supercomputing Applications at the University of Illinois at Urbana-Champaign, the Kavli Institute of Cosmological Physics at the University of Chicago, Center for Cosmology and Astro-Particle Physics at the Ohio State University, the Mitchell Institute for Fundamental Physics and Astronomy at Texas A\&M University, Financiadora de Estudos e Projetos, Fundacao Carlos Chagas Filho de Amparo, Financiadora de Estudos e Projetos, Fundacao Carlos Chagas Filho de Amparo a Pesquisa do Estado do Rio de Janeiro, Conselho Nacional de Desenvolvimento Cientifico e Tecnologico and the Ministerio da Ciencia, Tecnologia e Inovacao, the Deutsche Forschungsgemeinschaft and the Collaborating Institutions in the Dark Energy Survey. The Collaborating Institutions are Argonne National Laboratory, the University of California at Santa Cruz, the University of Cambridge, Centro de Investigaciones Energeticas, Medioambientales y Tecnologicas-Madrid, the University of Chicago, University College London, the DES-Brazil Consortium, the University of Edinburgh, the Eidgenossische Technische Hochschule (ETH) Zurich, Fermi National Accelerator Laboratory, the University of Illinois at Urbana-Champaign, the Institut de Ciencies de l’Espai (IEEC/CSIC), the Institut de Fisica d’Altes Energies, Lawrence Berkeley National Laboratory, the Ludwig Maximilians Universitat Munchen and the associated Excellence Cluster Universe, the University of Michigan, NSF’s NOIRLab, the University of Nottingham, the Ohio State University, the University of Pennsylvania, the University of Portsmouth, SLAC National Accelerator Laboratory, Stanford University, the University of Sussex, and Texas A\&M University.

BASS is a key project of the Telescope Access Program (TAP), which has been funded by the National Astronomical Observatories of China, the Chinese Academy of Sciences (the Strategic Priority Research Program “The Emergence of Cosmological Structures” Grant \# XDB09000000), and the Special Fund for Astronomy from the Ministry of Finance. The BASS is also supported by the External Cooperation Program of Chinese Academy of Sciences (Grant \# 114A11KYSB20160057), and Chinese National Natural Science Foundation (Grant \# 12120101003, \# 11433005).

The Siena Galaxy Atlas was made possible by funding support from the U.S. Department of Energy, Office of Science, Office of High Energy Physics under Award Number DE-SC0020086 and from the National Science Foundation under grant AST-1616414.

This research used data obtained with the Dark Energy Spectroscopic Instrument (DESI). DESI construction and operations are managed by the Lawrence Berkeley National Laboratory. This material is based upon work supported by the U.S. Department of Energy, Office of Science, Office of High-Energy Physics, under Contract No. DE-AC02-05CH11231, and by the National Energy Research Scientific Computing Center, a DOE Office of Science User Facility under the same contract. Additional support for DESI was provided by the U.S. National Science Foundation (NSF), Division of Astronomical Sciences under Contract No. AST-0950945 to the NSF’s National Optical-Infrared Astronomy Research Laboratory; the Science and Technology Facilities Council of the United Kingdom; the Gordon and Betty Moore Foundation; the Heising-Simons Foundation; the French Alternative Energies and Atomic Energy Commission (CEA); the National Council of Humanities, Science and Technology of Mexico (CONAHCYT); the Ministry of Science and Innovation of Spain (MICINN), and by the DESI Member Institutions: www.desi.lbl.gov/collaborating-institutions. The DESI collaboration is honored to be permitted to conduct scientific research on I’oligam Du’ag (Kitt Peak), a mountain with particular significance to the Tohono O’odham Nation. Any opinions, findings, and conclusions or recommendations expressed in this material are those of the author(s) and do not necessarily reflect the views of the U.S. National Science Foundation, the U.S. Department of Energy, or any of the listed funding agencies.

The authors are pleased to acknowledge that the work reported in this paper was substantially performed using the Princeton Research Computing resources at Princeton University, a consortium of groups led by the Princeton Institute for Computational Science and Engineering (PICSciE) and the Office of Information Technology's Research Computing.

This research has made use of the SIMBAD database, operated at CDS, Strasbourg, France. The NASA/IPAC Extragalactic Database (NED) is funded by the National Aeronautics and Space Administration and operated by the California Institute of Technology. 

\end{acknowledgments}

\vspace{1em}
\facilities{Subaru (HSC), Blanco (DECam), CFHT (MegaCam), Magellan:Baade (IMACS), Gemini:Gillett (GMOS)}

\software{\href{http://www.numpy.org}{\code{NumPy}} \citep{Numpy},
          \href{https://www.astropy.org/}{\code{Astropy}} \citep{astropy}, \href{https://www.scipy.org}{\code{SciPy}} \citep{scipy}, \href{https://matplotlib.org}{\code{Matplotlib}} \citep{matplotlib},
          \href{https://artpop.readthedocs.io/en/latest/index.html}{\code{ArtPop}} \citep{artpop},
          \href{https://www.astromatic.net/software/sextractor/}{\code{SExtractor}} \citep{SExtractor},
          \href{https://www.astromatic.net/software/swarp/}{\code{SWarp}} \citep{swarp},
          \href{https://www.astromatic.net/software/scamp/}{\code{SCAMP}} \citep{scamp},
          \href{https://www.astromatic.net/software/psfex/}{\code{PSFEx}} \citep{psfex},
          \href{https://dragons.readthedocs.io/projects/gmosimg-drtutorial/en/stable/}{\code{DRAGONS}} \citep{DRAGONS_paper},
          \href{https://sep.readthedocs.io/en/v1.1.x/}{\code{sep}} \citep{Barbary2016},
          \href{https://www.mpe.mpg.de/~erwin/code/imfit/}{\code{imfit}} \citep{imfit},
          \href{https://github.com/kbarbary/sfdmap}{\code{sfdmap}},
          \href{https://github.com/dr-guangtou/unagi/}{\code{unagi}},
          \href{https://bdiemer.bitbucket.io/colossus/index.html}{\code{colossus}} \citep{Colossus}
          }

\bibliography{master,software}

\begin{thebibliography}{}
\expandafter\ifx\csname natexlab\endcsname\relax\def\natexlab#1{#1}\fi
\providecommand{\url}[1]{\href{#1}{#1}}
\providecommand{\dodoi}[1]{doi:~\href{http://doi.org/#1}{\nolinkurl{#1}}}
\providecommand{\doeprint}[1]{\href{http://ascl.net/#1}{\nolinkurl{http://ascl.net/#1}}}
\providecommand{\doarXiv}[1]{\href{https://arxiv.org/abs/#1}{\nolinkurl{https://arxiv.org/abs/#1}}}

\bibitem[{{Abbott} {et~al.}(2021){Abbott}, {Adam{\'o}w}, {Aguena}, {Allam}, {Amon}, {Annis}, {Avila}, {Bacon}, {Banerji}, {Bechtol}, {Becker}, {Bernstein}, {Bertin}, {Bhargava}, {Bridle}, {Brooks}, {Burke}, {Carnero Rosell}, {Carrasco Kind}, {Carretero}, {Castander}, {Cawthon}, {Chang}, {Choi}, {Conselice}, {Costanzi}, {Crocce}, {da Costa}, {Davis}, {De Vicente}, {DeRose}, {Desai}, {Diehl}, {Dietrich}, {Drlica-Wagner}, {Eckert}, {Elvin-Poole}, {Everett}, {Evrard}, {Ferrero}, {Fert{\'e}}, {Flaugher}, {Fosalba}, {Friedel}, {Frieman}, {Garc{\'\i}a-Bellido}, {Gaztanaga}, {Gelman}, {Gerdes}, {Giannantonio}, {Gill}, {Gruen}, {Gruendl}, {Gschwend}, {Gutierrez}, {Hartley}, {Hinton}, {Hollowood}, {Honscheid}, {Huterer}, {James}, {Jeltema}, {Johnson}, {Kent}, {Kron}, {Kuehn}, {Kuropatkin}, {Lahav}, {Li}, {Lidman}, {Lin}, {MacCrann}, {Maia}, {Manning}, {Maloney}, {March}, {Marshall}, {Martini}, {Melchior}, {Menanteau}, {Miquel}, {Morgan}, {Myles}, {Neilsen}, {Ogando}, {Palmese}, {Paz-Chinch{\'o}n}, {Petravick},
  {Pieres}, {Plazas}, {Pond}, {Rodriguez-Monroy}, {Romer}, {Roodman}, {Rykoff}, {Sako}, {Sanchez}, {Santiago}, {Scarpine}, {Serrano}, {Sevilla-Noarbe}, {Smith}, {Smith}, {Soares-Santos}, {Suchyta}, {Swanson}, {Tarle}, {Thomas}, {To}, {Tremblay}, {Troxel}, {Tucker}, {Turner}, {Varga}, {Walker}, {Wechsler}, {Weller}, {Wester}, {Wilkinson}, {Yanny}, {Zhang}, {Nikutta}, {Fitzpatrick}, {Jacques}, {Scott}, {Olsen}, {Huang}, {Herrera}, {Juneau}, {Nidever}, {Weaver}, {Adean}, {Correia}, {de Freitas}, {Freitas}, {Singulani}, {Vila-Verde}, \& {Linea Science Server}}]{DES-DR2}
{Abbott}, T.~M.~C., {Adam{\'o}w}, M., {Aguena}, M., {et~al.} 2021, \apjs, 255, 20, \dodoi{10.3847/1538-4365/ac00b3}

\bibitem[{{Aihara} {et~al.}(2022){Aihara}, {AlSayyad}, {Ando}, {Armstrong}, {Bosch}, {Egami}, {Furusawa}, {Furusawa}, {Harasawa}, {Harikane}, {Hsieh}, {Ikeda}, {Ito}, {Iwata}, {Kodama}, {Koike}, {Kokubo}, {Komiyama}, {Li}, {Liang}, {Lin}, {Lupton}, {Lust}, {MacArthur}, {Mawatari}, {Mineo}, {Miyatake}, {Miyazaki}, {More}, {Morishima}, {Murayama}, {Nakajima}, {Nakata}, {Nishizawa}, {Oguri}, {Okabe}, {Okura}, {Ono}, {Osato}, {Ouchi}, {Pan}, {Plazas Malag{\'o}n}, {Price}, {Reed}, {Rykoff}, {Shibuya}, {Simunovic}, {Strauss}, {Sugimori}, {Suto}, {Suzuki}, {Takada}, {Takagi}, {Takata}, {Takita}, {Tanaka}, {Tang}, {Taranu}, {Terai}, {Toba}, {Turner}, {Uchiyama}, {Vijarnwannaluk}, {Waters}, {Yamada}, {Yamamoto}, \& {Yamashita}}]{Aihara2022}
{Aihara}, H., {AlSayyad}, Y., {Ando}, M., {et~al.} 2022, \pasj, 74, 247, \dodoi{10.1093/pasj/psab122}

\bibitem[{{Allsopp}(1979{\natexlab{a}})}]{Allsopp1979a}
{Allsopp}, N.~J. 1979{\natexlab{a}}, \mnras, 188, 371, \dodoi{10.1093/mnras/188.2.371}

\bibitem[{{Allsopp}(1979{\natexlab{b}})}]{Allsopp1979b}
---. 1979{\natexlab{b}}, \mnras, 188, 765, \dodoi{10.1093/mnras/188.4.765}

\bibitem[{{Anand} {et~al.}(2021{\natexlab{a}}){Anand}, {Rizzi}, {Tully}, {Shaya}, {Karachentsev}, {Makarov}, {Makarova}, {Wu}, {Dolphin}, \& {Kourkchi}}]{Anand2021-EDD}
{Anand}, G.~S., {Rizzi}, L., {Tully}, R.~B., {et~al.} 2021{\natexlab{a}}, \aj, 162, 80, \dodoi{10.3847/1538-3881/ac0440}

\bibitem[{{Anand} {et~al.}(2021{\natexlab{b}}){Anand}, {Lee}, {Van Dyk}, {Leroy}, {Rosolowsky}, {Schinnerer}, {Larson}, {Kourkchi}, {Kreckel}, {Scheuermann}, {Rizzi}, {Thilker}, {Tully}, {Bigiel}, {Blanc}, {Boquien}, {Chandar}, {Dale}, {Emsellem}, {Deger}, {Glover}, {Grasha}, {Groves}, {S. Klessen}, {Kruijssen}, {Querejeta}, {S{\'a}nchez-Bl{\'a}zquez}, {Schruba}, {Turner}, {Ubeda}, {Williams}, \& {Whitmore}}]{Anand2021}
{Anand}, G.~S., {Lee}, J.~C., {Van Dyk}, S.~D., {et~al.} 2021{\natexlab{b}}, \mnras, 501, 3621, \dodoi{10.1093/mnras/staa3668}

\bibitem[{{Annibali} {et~al.}(2020){Annibali}, {Beccari}, {Bellazzini}, {Tosi}, {Cusano}, {Paris}, {Cignoni}, {Ciotti}, {Nipoti}, \& {Sacchi}}]{Annibali2020}
{Annibali}, F., {Beccari}, G., {Bellazzini}, M., {et~al.} 2020, \mnras, 491, 5101, \dodoi{10.1093/mnras/stz3185}

\bibitem[{{Asali} {et~al.}(2025){Asali}, {Geha}, {Kado-Fong}, {Mao}, {Wechsler}, {de los Reyes}, {Pasha}, {Kallivayalil}, {Nadler}, {Tollerud}, {Wang}, {Weiner}, \& {Wu}}]{Asali2025}
{Asali}, Y., {Geha}, M., {Kado-Fong}, E., {et~al.} 2025, \apj, 995, 79, \dodoi{10.3847/1538-4357/ae147d}

\bibitem[{{Astropy Collaboration} {et~al.}(2013){Astropy Collaboration}, {Robitaille}, {Tollerud}, {Greenfield}, {Droettboom}, {Bray}, {Aldcroft}, {Davis}, {Ginsburg}, {Price-Whelan}, {Kerzendorf}, {Conley}, {Crighton}, {Barbary}, {Muna}, {Ferguson}, {Grollier}, {Parikh}, {Nair}, {Unther}, {Deil}, {Woillez}, {Conseil}, {Kramer}, {Turner}, {Singer}, {Fox}, {Weaver}, {Zabalza}, {Edwards}, {Azalee Bostroem}, {Burke}, {Casey}, {Crawford}, {Dencheva}, {Ely}, {Jenness}, {Labrie}, {Lim}, {Pierfederici}, {Pontzen}, {Ptak}, {Refsdal}, {Servillat}, \& {Streicher}}]{astropy}
{Astropy Collaboration}, {Robitaille}, T.~P., {Tollerud}, E.~J., {et~al.} 2013, \aap, 558, A33, \dodoi{10.1051/0004-6361/201322068}

\bibitem[{Barbary(2016)}]{Barbary2016}
Barbary, K. 2016, Journal of Open Source Software, 1(6), 58, \dodoi{10.21105/joss.0005}

\bibitem[{{Battaglia} {et~al.}(2022){Battaglia}, {Taibi}, {Thomas}, \& {Fritz}}]{Battaglia2022}
{Battaglia}, G., {Taibi}, S., {Thomas}, G.~F., \& {Fritz}, T.~K. 2022, \aap, 657, A54, \dodoi{10.1051/0004-6361/202141528}

\bibitem[{{Bell} {et~al.}(2003){Bell}, {McIntosh}, {Katz}, \& {Weinberg}}]{Bell2003}
{Bell}, E.~F., {McIntosh}, D.~H., {Katz}, N., \& {Weinberg}, M.~D. 2003, \apjs, 149, 289, \dodoi{10.1086/378847}

\bibitem[{{Bennet} {et~al.}(2019){Bennet}, {Sand}, {Crnojevi{\'c}}, {Spekkens}, {Karunakaran}, {Zaritsky}, \& {Mutlu-Pakdil}}]{Bennet2019}
{Bennet}, P., {Sand}, D.~J., {Crnojevi{\'c}}, D., {et~al.} 2019, \apj, 885, 153, \dodoi{10.3847/1538-4357/ab46ab}

\bibitem[{{Bertin}(2006)}]{scamp}
{Bertin}, E. 2006, in Astronomical Society of the Pacific Conference Series, Vol. 351, Astronomical Data Analysis Software and Systems XV, ed. C.~{Gabriel}, C.~{Arviset}, D.~{Ponz}, \& S.~{Enrique}, 112

\bibitem[{{Bertin}(2011)}]{psfex}
{Bertin}, E. 2011, in Astronomical Society of the Pacific Conference Series, Vol. 442, Astronomical Data Analysis Software and Systems XX, ed. I.~N. {Evans}, A.~{Accomazzi}, D.~J. {Mink}, \& A.~H. {Rots}, 435

\bibitem[{{Bertin} \& {Arnouts}(1996)}]{SExtractor}
{Bertin}, E., \& {Arnouts}, S. 1996, \aaps, 117, 393, \dodoi{10.1051/aas:1996164}

\bibitem[{{Bertin} {et~al.}(2002){Bertin}, {Mellier}, {Radovich}, {Missonnier}, {Didelon}, \& {Morin}}]{swarp}
{Bertin}, E., {Mellier}, Y., {Radovich}, M., {et~al.} 2002, in Astronomical Society of the Pacific Conference Series, Vol. 281, Astronomical Data Analysis Software and Systems XI, ed. D.~A. {Bohlender}, D.~{Durand}, \& T.~H. {Handley}, 228

\bibitem[{{Bhattacharyya} {et~al.}(2024){Bhattacharyya}, {Peter}, {Martini}, {Mutlu-Pakdil}, {Drlica-Wagner}, {Pace}, {Strigari}, {Cheng}, {Roberts}, {Tanoglidis}, {Aguena}, {Alves}, {Andrade-Oliveira}, {Bacon}, {Brooks}, {Carnero Rosell}, {Carretero}, {da Costa}, {Pereira}, {Davis}, {Desai}, {Doel}, {Ferrero}, {Frieman}, {Garc{\'\i}a-Bellido}, {Giannini}, {Gruen}, {Gruendl}, {Hinton}, {Hollowood}, {Honscheid}, {James}, {Kuehn}, {Marshall}, {Mena-Fern{\'a}ndez}, {Miquel}, {Palmese}, {Pieres}, {Plazas Malag{\'o}n}, {Sanchez}, {Santiago}, {Schubnell}, {Sevilla-Noarbe}, {Smith}, {Suchyta}, {Swanson}, {Tarle}, {Vincenzi}, {Walker}, {Weaverdyck}, \& {Wiseman}}]{Bhattacharyya2024}
{Bhattacharyya}, J., {Peter}, A.~H.~G., {Martini}, P., {et~al.} 2024, \apj, 975, 244, \dodoi{10.3847/1538-4357/ad79fe}

\bibitem[{{Bordoloi} {et~al.}(2014){Bordoloi}, {Tumlinson}, {Werk}, {Oppenheimer}, {Peeples}, {Prochaska}, {Tripp}, {Katz}, {Dav{\'e}}, {Fox}, {Thom}, {Ford}, {Weinberg}, {Burchett}, \& {Kollmeier}}]{Bordoloi2014}
{Bordoloi}, R., {Tumlinson}, J., {Werk}, J.~K., {et~al.} 2014, \apj, 796, 136, \dodoi{10.1088/0004-637X/796/2/136}

\bibitem[{Bradley {et~al.}(2025)Bradley, Sip{\H o}cz, Robitaille, Tollerud, Vin{\'{\i}}cius, Deil, Barbary, Wilson, Busko, Donath, G{\"u}nther, Cara, Lim, Me{\ss}linger, Burnett, Conseil, Droettboom, Bostroem, Bray, Bratholm, Jamieson, Ginsburg, Barentsen, Craig, Pascual, Rathi, Perrin, \& Morris}]{photutils}
Bradley, L., Sip{\H o}cz, B., Robitaille, T., {et~al.} 2025, astropy/photutils: 2.2.0, 2.2.0,  Zenodo, \dodoi{10.5281/zenodo.14889440}

\bibitem[{{Bryan} \& {Norman}(1998)}]{Bryan1998}
{Bryan}, G.~L., \& {Norman}, M.~L. 1998, \apj, 495, 80, \dodoi{10.1086/305262}

\bibitem[{{Bullock} \& {Boylan-Kolchin}(2017)}]{Bullock2017}
{Bullock}, J.~S., \& {Boylan-Kolchin}, M. 2017, \araa, 55, 343, \dodoi{10.1146/annurev-astro-091916-055313}

\bibitem[{{Bullock} {et~al.}(2001){Bullock}, {Kolatt}, {Sigad}, {Somerville}, {Kravtsov}, {Klypin}, {Primack}, \& {Dekel}}]{Bullock2001}
{Bullock}, J.~S., {Kolatt}, T.~S., {Sigad}, Y., {et~al.} 2001, \mnras, 321, 559, \dodoi{10.1046/j.1365-8711.2001.04068.x}

\bibitem[{{Bullock} {et~al.}(2000){Bullock}, {Kravtsov}, \& {Weinberg}}]{Bullock2000}
{Bullock}, J.~S., {Kravtsov}, A.~V., \& {Weinberg}, D.~H. 2000, \apj, 539, 517, \dodoi{10.1086/309279}

\bibitem[{{Cantiello} \& {Blakeslee}(2023)}]{cantiello2023review}
{Cantiello}, M., \& {Blakeslee}, J.~P. 2023, arXiv e-prints, arXiv:2307.03116, \dodoi{10.48550/arXiv.2307.03116}

\bibitem[{{Cantiello} {et~al.}(2018){Cantiello}, {Blakeslee}, {Ferrarese}, {C{\^o}t{\'e}}, {Roediger}, {Raimondo}, {Peng}, {Gwyn}, {Durrell}, \& {Cuillandre}}]{Cantiello2018}
{Cantiello}, M., {Blakeslee}, J.~P., {Ferrarese}, L., {et~al.} 2018, \apj, 856, 126, \dodoi{10.3847/1538-4357/aab043}

\bibitem[{{Carlin} {et~al.}(2016){Carlin}, {Sand}, {Price}, {Willman}, {Karunakaran}, {Spekkens}, {Bell}, {Brodie}, {Crnojevi{\'c}}, {Forbes}, {Hargis}, {Kirby}, {Lupton}, {Peter}, {Romanowsky}, \& {Strader}}]{Carlin2016}
{Carlin}, J.~L., {Sand}, D.~J., {Price}, P., {et~al.} 2016, \apjl, 828, L5, \dodoi{10.3847/2041-8205/828/1/L5}

\bibitem[{{Carlin} {et~al.}(2019){Carlin}, {Garling}, {Peter}, {Crnojevi{\'c}}, {Forbes}, {Hargis}, {Mutlu-Pakdil}, {Pucha}, {Romanowsky}, {Sand}, {Spekkens}, {Strader}, \& {Willman}}]{Carlin2019}
{Carlin}, J.~L., {Garling}, C.~T., {Peter}, A. H.~G., {et~al.} 2019, \apj, 886, 109, \dodoi{10.3847/1538-4357/ab4c32}

\bibitem[{{Carlin} {et~al.}(2021){Carlin}, {Mutlu-Pakdil}, {Crnojevi{\'c}}, {Garling}, {Karunakaran}, {Peter}, {Tollerud}, {Forbes}, {Hargis}, {Lim}, {Romanowsky}, {Sand}, {Spekkens}, \& {Strader}}]{Carlin2021}
{Carlin}, J.~L., {Mutlu-Pakdil}, B., {Crnojevi{\'c}}, D., {et~al.} 2021, \apj, 909, 211, \dodoi{10.3847/1538-4357/abe040}

\bibitem[{{Carlin} {et~al.}(2024){Carlin}, {Sand}, {Mutlu-Pakdil}, {Crnojevi{\'c}}, {Doliva-Dolinsky}, {Garling}, {Peter}, {Brodie}, {Forbes}, {Hargis}, {Romanowsky}, {Spekkens}, {Strader}, \& {Willman}}]{Carlin2024}
{Carlin}, J.~L., {Sand}, D.~J., {Mutlu-Pakdil}, B., {et~al.} 2024, \apj, 977, 112, \dodoi{10.3847/1538-4357/ad8dcd}

\bibitem[{{Carlsten} {et~al.}(2026{\natexlab{a}}){Carlsten}, {Li}, {Greene}, {Drlica-Wagner}, \& {Danieli}}]{ELVES_Field_I}
{Carlsten}, S., {Li}, J., {Greene}, J., {Drlica-Wagner}, A., \& {Danieli}, S. 2026{\natexlab{a}}, arXiv e-prints, arXiv:2602.16766, \dodoi{10.48550/arXiv.2602.16766}

\bibitem[{{Carlsten} {et~al.}(2026{\natexlab{b}}){Carlsten}, {Li}, {Greene}, {Drlica-Wagner}, \& {Danieli}}]{ELVES_Field_II}
---. 2026{\natexlab{b}}, arXiv e-prints, arXiv:2602.16778, \dodoi{10.48550/arXiv.2602.16778}

\bibitem[{{Carlsten} {et~al.}(2019){Carlsten}, {Beaton}, {Greco}, \& {Greene}}]{Carlsten2019}
{Carlsten}, S.~G., {Beaton}, R.~L., {Greco}, J.~P., \& {Greene}, J.~E. 2019, \apj, 879, 13, \dodoi{10.3847/1538-4357/ab22c1}

\bibitem[{{Carlsten} {et~al.}(2020{\natexlab{a}}){Carlsten}, {Greco}, {Beaton}, \& {Greene}}]{Carlsten2020}
{Carlsten}, S.~G., {Greco}, J.~P., {Beaton}, R.~L., \& {Greene}, J.~E. 2020{\natexlab{a}}, \apj, 891, 144, \dodoi{10.3847/1538-4357/ab7758}

\bibitem[{{Carlsten} {et~al.}(2022){Carlsten}, {Greene}, {Beaton}, {Danieli}, \& {Greco}}]{CarlstenELVES2022}
{Carlsten}, S.~G., {Greene}, J.~E., {Beaton}, R.~L., {Danieli}, S., \& {Greco}, J.~P. 2022, \apj, 933, 47, \dodoi{10.3847/1538-4357/ac6fd7}

\bibitem[{{Carlsten} {et~al.}(2021{\natexlab{a}}){Carlsten}, {Greene}, {Greco}, {Beaton}, \& {Kado-Fong}}]{ELVES-I}
{Carlsten}, S.~G., {Greene}, J.~E., {Greco}, J.~P., {Beaton}, R.~L., \& {Kado-Fong}, E. 2021{\natexlab{a}}, \apj, 922, 267, \dodoi{10.3847/1538-4357/ac2581}

\bibitem[{{Carlsten} {et~al.}(2021{\natexlab{b}}){Carlsten}, {Greene}, {Peter}, {Beaton}, \& {Greco}}]{Carlsten2021}
{Carlsten}, S.~G., {Greene}, J.~E., {Peter}, A. H.~G., {Beaton}, R.~L., \& {Greco}, J.~P. 2021{\natexlab{b}}, \apj, 908, 109, \dodoi{10.3847/1538-4357/abd039}

\bibitem[{{Carlsten} {et~al.}(2020{\natexlab{b}}){Carlsten}, {Greene}, {Peter}, {Greco}, \& {Beaton}}]{elves-radial}
{Carlsten}, S.~G., {Greene}, J.~E., {Peter}, A. H.~G., {Greco}, J.~P., \& {Beaton}, R.~L. 2020{\natexlab{b}}, \apj, 902, 124, \dodoi{10.3847/1538-4357/abb60b}

\bibitem[{{Chambers} {et~al.}(2016){Chambers}, {Magnier}, {Metcalfe}, {Flewelling}, {Huber}, {Waters}, {Denneau}, {Draper}, {Farrow}, {Finkbeiner}, {Holmberg}, {Koppenhoefer}, {Price}, {Rest}, {Saglia}, {Schlafly}, {Smartt}, {Sweeney}, {Wainscoat}, {Burgett}, {Chastel}, {Grav}, {Heasley}, {Hodapp}, {Jedicke}, {Kaiser}, {Kudritzki}, {Luppino}, {Lupton}, {Monet}, {Morgan}, {Onaka}, {Shiao}, {Stubbs}, {Tonry}, {White}, {Ba{\~n}ados}, {Bell}, {Bender}, {Bernard}, {Boegner}, {Boffi}, {Botticella}, {Calamida}, {Casertano}, {Chen}, {Chen}, {Cole}, {Deacon}, {Frenk}, {Fitzsimmons}, {Gezari}, {Gibbs}, {Goessl}, {Goggia}, {Gourgue}, {Goldman}, {Grant}, {Grebel}, {Hambly}, {Hasinger}, {Heavens}, {Heckman}, {Henderson}, {Henning}, {Holman}, {Hopp}, {Ip}, {Isani}, {Jackson}, {Keyes}, {Koekemoer}, {Kotak}, {Le}, {Liska}, {Long}, {Lucey}, {Liu}, {Martin}, {Masci}, {McLean}, {Mindel}, {Misra}, {Morganson}, {Murphy}, {Obaika}, {Narayan}, {Nieto-Santisteban}, {Norberg}, {Peacock}, {Pier}, {Postman}, {Primak}, {Rae}, {Rai},
  {Riess}, {Riffeser}, {Rix}, {R{\"o}ser}, {Russel}, {Rutz}, {Schilbach}, {Schultz}, {Scolnic}, {Strolger}, {Szalay}, {Seitz}, {Small}, {Smith}, {Soderblom}, {Taylor}, {Thomson}, {Taylor}, {Thakar}, {Thiel}, {Thilker}, {Unger}, {Urata}, {Valenti}, {Wagner}, {Walder}, {Walter}, {Watters}, {Werner}, {Wood-Vasey}, \& {Wyse}}]{Chambers2016}
{Chambers}, K.~C., {Magnier}, E.~A., {Metcalfe}, N., {et~al.} 2016, arXiv e-prints, arXiv:1612.05560, \dodoi{10.48550/arXiv.1612.05560}

\bibitem[{{Chazov} {et~al.}(2026){Chazov}, {Karachentsev}, \& {Kaisin}}]{Chazov2026}
{Chazov}, M.~I., {Karachentsev}, I.~D., \& {Kaisin}, S.~S. 2026, \mnras, 550, stag1276, \dodoi{10.1093/mnras/stag1276}

\bibitem[{{Cohen} {et~al.}(2018){Cohen}, {van Dokkum}, {Danieli}, {Romanowsky}, {Abraham}, {Merritt}, {Zhang}, {Mowla}, {Kruijssen}, {Conroy}, \& {Wasserman}}]{Cohen2018}
{Cohen}, Y., {van Dokkum}, P., {Danieli}, S., {et~al.} 2018, \apj, 868, 96, \dodoi{10.3847/1538-4357/aae7c8}

\bibitem[{{Crnojevi{\'c}} {et~al.}(2019){Crnojevi{\'c}}, {Sand}, {Bennet}, {Pasetto}, {Spekkens}, {Caldwell}, {Guhathakurta}, {McLeod}, {Seth}, {Simon}, {Strader}, \& {Toloba}}]{Crnojevic2019}
{Crnojevi{\'c}}, D., {Sand}, D.~J., {Bennet}, P., {et~al.} 2019, \apj, 872, 80, \dodoi{10.3847/1538-4357/aafbe7}

\bibitem[{{Cruz} {et~al.}(2026){Cruz}, {Brooks}, {Lisanti}, {Peter}, {Geda}, {Quinn}, {Tremmel}, {Munshi}, {Keller}, \& {Wadsley}}]{Cruz2026}
{Cruz}, A., {Brooks}, A.~M., {Lisanti}, M., {et~al.} 2026, \apj, 1005, 157, \dodoi{10.3847/1538-4357/ae69ce}

\bibitem[{{Dalcanton} {et~al.}(2009){Dalcanton}, {Williams}, {Seth}, {Dolphin}, {Holtzman}, {Rosema}, {Skillman}, {Cole}, {Girardi}, {Gogarten}, {Karachentsev}, {Olsen}, {Weisz}, {Christensen}, {Freeman}, {Gilbert}, {Gallart}, {Harris}, {Hodge}, {de Jong}, {Karachentseva}, {Mateo}, {Stetson}, {Tavarez}, {Zaritsky}, {Governato}, \& {Quinn}}]{Dalcanton2009}
{Dalcanton}, J.~J., {Williams}, B.~F., {Seth}, A.~C., {et~al.} 2009, \apjs, 183, 67, \dodoi{10.1088/0067-0049/183/1/67}

\bibitem[{{Danieli} {et~al.}(2023){Danieli}, {Greene}, {Carlsten}, {Jiang}, {Beaton}, \& {Goulding}}]{Danieli2023}
{Danieli}, S., {Greene}, J.~E., {Carlsten}, S., {et~al.} 2023, \apj, 956, 6, \dodoi{10.3847/1538-4357/acefbd}

\bibitem[{{Danieli} {et~al.}(2017){Danieli}, {van Dokkum}, {Merritt}, {Abraham}, {Zhang}, {Karachentsev}, \& {Makarova}}]{Danieli2017}
{Danieli}, S., {van Dokkum}, P., {Merritt}, A., {et~al.} 2017, \apj, 837, 136, \dodoi{10.3847/1538-4357/aa615b}

\bibitem[{{Davies} {et~al.}(1997){Davies}, {Allington-Smith}, {Bettess}, {Chadwick}, {Content}, {Dodsworth}, {Haynes}, {Lee}, {Lewis}, {Webster}, {Atad}, {Beard}, {Ellis}, {Hastings}, {Williams}, {Bond}, {Crampton}, {Davidge}, {Fletcher}, {Leckie}, {Morbey}, {Murowinski}, {Roberts}, {Saddlemyer}, {Sebesta}, {Stilburn}, \& {Szeto}}]{GMOS}
{Davies}, R.~L., {Allington-Smith}, J.~R., {Bettess}, P., {et~al.} 1997, in Society of Photo-Optical Instrumentation Engineers (SPIE) Conference Series, Vol. 2871, Optical Telescopes of Today and Tomorrow, ed. A.~L. {Ardeberg}, 1099--1106, \dodoi{10.1117/12.268996}

\bibitem[{{Davis} {et~al.}(2021){Davis}, {Nierenberg}, {Peter}, {Garling}, {Greco}, {Kochanek}, {Utomo}, {Casey}, {Pogge}, {Roberts}, {Sand}, \& {Sardone}}]{Davis2021}
{Davis}, A.~B., {Nierenberg}, A.~M., {Peter}, A. H.~G., {et~al.} 2021, \mnras, 500, 3854, \dodoi{10.1093/mnras/staa3246}

\bibitem[{{Davis} {et~al.}(2024){Davis}, {Garling}, {Nierenberg}, {Peter}, {Sardone}, {Kochanek}, {Leroy}, {Casey}, {Pogge}, {Roberts}, {Sand}, \& {Greco}}]{Davis2024}
{Davis}, A.~B., {Garling}, C.~T., {Nierenberg}, A.~M., {et~al.} 2024, arXiv e-prints, arXiv:2409.03999, \dodoi{10.48550/arXiv.2409.03999}

\bibitem[{{Davis} {et~al.}(1985){Davis}, {Efstathiou}, {Frenk}, \& {White}}]{Davis1985}
{Davis}, M., {Efstathiou}, G., {Frenk}, C.~S., \& {White}, S.~D.~M. 1985, \apj, 292, 371, \dodoi{10.1086/163168}

\bibitem[{{de los Reyes} {et~al.}(2025){de los Reyes}, {Asali}, {Wechsler}, {Geha}, {Mao}, {Kado-Fong}, {Pucha}, {Grant}, {Gandhi}, {Manwadkar}, {Engelhardt}, {Munshi}, \& {Wang}}]{delosReyes2025}
{de los Reyes}, M. A.~C., {Asali}, Y., {Wechsler}, R.~H., {et~al.} 2025, \apj, 989, 91, \dodoi{10.3847/1538-4357/ade4c5}

\bibitem[{{DESI Collaboration} {et~al.}(2026){DESI Collaboration}, {Abdul Karim}, {Adame}, {Aguado}, {Aguilar}, {Ahlen}, {Alam}, {Aldering}, {Alexander}, {Alfarsy}, {Allen}, {Allende Prieto}, {Alves}, {Anand}, {Andrade}, {Armengaud}, {Avila}, {Aviles}, {Awan}, {Bailey}, {Baleato Lizancos}, {Ballester}, {Bault}, {Bautista}, {Bean}, {Behera}, {BenZvi}, {Beraldo e Silva}, {Bermejo-Climent}, {Beutler}, {Bianchi}, {Blake}, {Blum}, {Bolton}, {Bonici}, {Brieden}, {Brodzeller}, {Brooks}, {Buckley-Geer}, {Burtin}, {Bystr{\"o}m}, {Canning}, {Carnero Rosell}, {Carr}, {Carrilho}, {Casas}, {Castander}, {Cereskaite}, {Cervantes-Cota}, {Chaussidon}, {Chaves-Montero}, {Chen}, {Chen}, {Circosta}, {Claybaugh}, {Cole}, {Cooper}, {Cousinou}, {Cuceu}, {Davis}, {Dawson}, {de Belsunce}, {de la Cruz}, {de la Macorra}, {de Mattia}, {Deiosso}, {Della Costa}, {Demina}, {Demirbozan}, {DeRose}, {Dey}, {Dey}, {Ding}, {Ding}, {Doel}, {Douglass}, {Dowicz}, {Ebina}, {Edelstein}, {Eisenstein}, {Elbers}, {Emas}, {Escoffier}, {Fagrelius},
  {Fan}, {Fanning}, {Favole}, {Fawcett}, {Fern{\'a}ndez-Garc{\'\i}a}, {Ferraro}, {Findlay}, {Font-Ribera}, {Forero-Romero}, {Forero-S{\'a}nchez}, {Frenk}, {G{\"a}nsicke}, {Galbany}, {Garc{\'\i}a-Bellido}, {Garcia-Quintero}, {Garrison}, {Gazta{\~n}aga}, {Gil-Mar{\'\i}n}, {Gloudemans}, {Gnedin}, {Gontcho A Gontcho}, {Gonzalez}, {Gonzalez-Morales}, {Gonzalez-Perez}, {Gordon}, {Graur}, {Green}, {Gruen}, {Gsponer}, {Guandalin}, {Gutierrez}, {Guy}, {Hahn}, {Han}, {Han}, {He}, {Herrera-Alcantar}, {Heydenreich}, {Honscheid}, {Hou}, {Howlett}, {Huterer}, {Ir{\v{s}}i{\v{c}}}, {Ishak}, {Jacques}, {Jiang}, {Jimenez}, {Jing}, {Joachimi}, {Joudaki}, {Joyce}, {Jullo}, {Juneau}, {Kara{\c{c}}ayl{\i}}, {Karim}, {Kehoe}, {Kent}, {Khederlarian}, {Kirkby}, {Kisner}, {Kitaura}, {Kizhuprakkat}, {Kong}, {Koposov}, {Kremin}, {Krolewski}, {Lahav}, {Lai}, {Lamman}, {Lan}, {Landriau}, {Lang}, {Lange}, {Lasker}, {Le Goff}, {Le Guillou}, {Leauthaud}, {Levi}, {Li}, {Li}, {Liu}, {Lodha}, {Lokken}, {Luo}, {Magneville}, {Manera}, {Manser},
  {Margala}, {Martini}, {Maus}, {McCullough}, {McDonald}, {Medina}, {Medina-Varela}, {Meisner}, {Mena-Fern{\'a}ndez}, {Menegas}, {Meneses-Rizo}, {Mezcua}, {Miquel}, {Montero-Camacho}, {Moon}, {Moustakas}, {Mu{\~n}oz-Guti{\'e}rrez}, {Mu noz-Santos}, {Myers}, {Myles}, {Nadathur}, {Najita}, {Napolitano}, {Newman}, {Nikakhtar}, {Nikutta}, {Niz}, {Noriega}, \& {Nugent}}]{DESI-DR1}
{DESI Collaboration}, {Abdul Karim}, M., {Adame}, A.~G., {et~al.} 2026, \aj, 171, 285, \dodoi{10.3847/1538-3881/ae4c43}

\bibitem[{{Dey} {et~al.}(2019){Dey}, {Schlegel}, {Lang}, {Blum}, {Burleigh}, {Fan}, {Findlay}, {Finkbeiner}, {Herrera}, {Juneau}, {Landriau}, {Levi}, {McGreer}, {Meisner}, {Myers}, {Moustakas}, {Nugent}, {Patej}, {Schlafly}, {Walker}, {Valdes}, {Weaver}, {Y{\`e}che}, {Zou}, {Zhou}, {Abareshi}, {Abbott}, {Abolfathi}, {Aguilera}, {Alam}, {Allen}, {Alvarez}, {Annis}, {Ansarinejad}, {Aubert}, {Beechert}, {Bell}, {BenZvi}, {Beutler}, {Bielby}, {Bolton}, {Brice{\~n}o}, {Buckley-Geer}, {Butler}, {Calamida}, {Carlberg}, {Carter}, {Casas}, {Castander}, {Choi}, {Comparat}, {Cukanovaite}, {Delubac}, {DeVries}, {Dey}, {Dhungana}, {Dickinson}, {Ding}, {Donaldson}, {Duan}, {Duckworth}, {Eftekharzadeh}, {Eisenstein}, {Etourneau}, {Fagrelius}, {Farihi}, {Fitzpatrick}, {Font-Ribera}, {Fulmer}, {G{\"a}nsicke}, {Gaztanaga}, {George}, {Gerdes}, {Gontcho}, {Gorgoni}, {Green}, {Guy}, {Harmer}, {Hernandez}, {Honscheid}, {Huang}, {James}, {Jannuzi}, {Jiang}, {Joyce}, {Karcher}, {Karkar}, {Kehoe}, {Kneib}, {Kueter-Young}, {Lan},
  {Lauer}, {Le Guillou}, {Le Van Suu}, {Lee}, {Lesser}, {Perreault Levasseur}, {Li}, {Mann}, {Marshall}, {Mart{\'\i}nez-V{\'a}zquez}, {Martini}, {du Mas des Bourboux}, {McManus}, {Meier}, {M{\'e}nard}, {Metcalfe}, {Mu{\~n}oz-Guti{\'e}rrez}, {Najita}, {Napier}, {Narayan}, {Newman}, {Nie}, {Nord}, {Norman}, {Olsen}, {Paat}, {Palanque-Delabrouille}, {Peng}, {Poppett}, {Poremba}, {Prakash}, {Rabinowitz}, {Raichoor}, {Rezaie}, {Robertson}, {Roe}, {Ross}, {Ross}, {Rudnick}, {Safonova}, {Saha}, {S{\'a}nchez}, {Savary}, {Schweiker}, {Scott}, {Seo}, {Shan}, {Silva}, {Slepian}, {Soto}, {Sprayberry}, {Staten}, {Stillman}, {Stupak}, {Summers}, {Sien Tie}, {Tirado}, {Vargas-Maga{\~n}a}, {Vivas}, {Wechsler}, {Williams}, {Yang}, {Yang}, {Yapici}, {Zaritsky}, {Zenteno}, {Zhang}, {Zhang}, {Zhou}, \& {Zhou}}]{Dey2019}
{Dey}, A., {Schlegel}, D.~J., {Lang}, D., {et~al.} 2019, \aj, 157, 168, \dodoi{10.3847/1538-3881/ab089d}

\bibitem[{{Diemer}(2018)}]{Colossus}
{Diemer}, B. 2018, \apjs, 239, 35, \dodoi{10.3847/1538-4365/aaee8c}

\bibitem[{{Diemer} \& {Joyce}(2019)}]{Diemer2019}
{Diemer}, B., \& {Joyce}, M. 2019, \apj, 871, 168, \dodoi{10.3847/1538-4357/aafad6}

\bibitem[{{Doliva-Dolinsky} {et~al.}(2025){Doliva-Dolinsky}, {Mutlu-Pakdil}, {Crnojevi{\'c}}, {Anbajagane}, {Carlin}, {Medoff}, {Sand}, {Tollerud}, {Lim}, {Bennet}, {Drlica-Wagner}, {Fielder}, {Hargis}, {Herron}, {Hunter}, {Jones}, {Karunakaran}, {Peter}, {Romanowsky}, {Spekkens}, {Strader}, {Willman}, {Carballo-Bello}, {Cerny}, {Chaturvedi}, {Kallivayalil}, {Mart{\'\i}nez-V{\'a}zquez}, {Medina}, {No{\"e}l}, {Pace}, {Riley}, {Sakowska}, {Smercina}, {Vivas}, {Adam{\'o}w}, {Bom}, {Choi}, {Ferguson}, {Navabi}, {Zenteno}, \& {Delve Collaboration}}]{Doliva-Dolinsky2025}
{Doliva-Dolinsky}, A., {Mutlu-Pakdil}, B., {Crnojevi{\'c}}, D., {et~al.} 2025, \apj, 989, 21, \dodoi{10.3847/1538-4357/ade9b8}

\bibitem[{{Dooley} {et~al.}(2017{\natexlab{a}}){Dooley}, {Peter}, {Carlin}, {Frebel}, {Bechtol}, \& {Willman}}]{Dooley2017b}
{Dooley}, G.~A., {Peter}, A. H.~G., {Carlin}, J.~L., {et~al.} 2017{\natexlab{a}}, \mnras, 472, 1060, \dodoi{10.1093/mnras/stx2001}

\bibitem[{{Dooley} {et~al.}(2017{\natexlab{b}}){Dooley}, {Peter}, {Yang}, {Willman}, {Griffen}, \& {Frebel}}]{Dooley2017a}
{Dooley}, G.~A., {Peter}, A. H.~G., {Yang}, T., {et~al.} 2017{\natexlab{b}}, \mnras, 471, 4894, \dodoi{10.1093/mnras/stx1900}

\bibitem[{{Dressler} {et~al.}(2011){Dressler}, {Bigelow}, {Hare}, {Sutin}, {Thompson}, {Burley}, {Epps}, {Oemler}, {Bagish}, {Birk}, {Clardy}, {Gunnels}, {Kelson}, {Shectman}, \& {Osip}}]{Dressler2011}
{Dressler}, A., {Bigelow}, B., {Hare}, T., {et~al.} 2011, \pasp, 123, 288, \dodoi{10.1086/658908}

\bibitem[{{Drlica-Wagner} {et~al.}(2022{\natexlab{a}}){Drlica-Wagner}, {Prescod-Weinstein}, {Yu}, {Albert}, {Amin}, {Banerjee}, {Baryakhtar}, {Bechtol}, {Bird}, {Birrer}, {Bringmann}, {Caputo}, {Chakrabarti}, {Chen}, {Croon}, {Cyr-Racine}, {Dawson}, {Dvorkin}, {Gluscevic}, {Gilman}, {Grin}, {Hlo{\v{z}}ek}, {Leane}, {Li}, {Mao}, {Meyers}, {Mishra-Sharma}, {Mu{\~n}oz}, {Munshi}, {Nadler}, {Parikh}, {Perez}, {Peter}, {Profumo}, {Schutz}, {Sehgal}, {Simon}, {Sinha}, {Valluri}, \& {Wechsler}}]{Drlica-Wagner2022}
{Drlica-Wagner}, A., {Prescod-Weinstein}, C., {Yu}, H.-B., {et~al.} 2022{\natexlab{a}}, arXiv e-prints, arXiv:2209.08215, \dodoi{10.48550/arXiv.2209.08215}

\bibitem[{{Drlica-Wagner} {et~al.}(2022{\natexlab{b}}){Drlica-Wagner}, {Ferguson}, {Adam{\'o}w}, {Aguena}, {Allam}, {Andrade-Oliveira}, {Bacon}, {Bechtol}, {Bell}, {Bertin}, {Bilaji}, {Bocquet}, {Bom}, {Brooks}, {Burke}, {Carballo-Bello}, {Carlin}, {Carnero Rosell}, {Carrasco Kind}, {Carretero}, {Castander}, {Cerny}, {Chang}, {Choi}, {Conselice}, {Costanzi}, {Crnojevi{\'c}}, {da Costa}, {de Vicente}, {Desai}, {Esteves}, {Everett}, {Ferrero}, {Fitzpatrick}, {Flaugher}, {Friedel}, {Frieman}, {Garc{\'\i}a-Bellido}, {Gatti}, {Gaztanaga}, {Gerdes}, {Gruen}, {Gruendl}, {Gschwend}, {Hartley}, {Hernandez-Lang}, {Hinton}, {Hollowood}, {Honscheid}, {Hughes}, {Jacques}, {James}, {Johnson}, {Kuehn}, {Kuropatkin}, {Lahav}, {Li}, {Lidman}, {Lin}, {March}, {Marshall}, {Mart{\'\i}nez-Delgado}, {Mart{\'\i}nez-V{\'a}zquez}, {Massana}, {Mau}, {McNanna}, {Melchior}, {Menanteau}, {Miller}, {Miquel}, {Mohr}, {Morgan}, {Mutlu-Pakdil}, {Mu{\~n}oz}, {Neilsen}, {Nidever}, {Nikutta}, {Nilo Castellon}, {No{\"e}l}, {Ogando}, {Olsen},
  {Pace}, {Palmese}, {Paz-Chinch{\'o}n}, {Pereira}, {Pieres}, {Plazas Malag{\'o}n}, {Prat}, {Riley}, {Rodriguez-Monroy}, {Romer}, {Roodman}, {Sako}, {Sakowska}, {Sanchez}, {S{\'a}nchez}, {Sand}, {Santana-Silva}, {Santiago}, {Schubnell}, {Serrano}, {Sevilla-Noarbe}, {Simon}, {Smith}, {Soares-Santos}, {Stringfellow}, {Suchyta}, {Suson}, {Tan}, {Tarle}, {Tavangar}, {Thomas}, {To}, {Tollerud}, {Troxel}, {Tucker}, {Varga}, {Vivas}, {Walker}, {Weller}, {Wilkinson}, {Wu}, {Yanny}, {Zaborowski}, {Zenteno}, {Delve Collaboration}, {Des Collaboration}, \& {Astro Data Lab}}]{DELVE-DR2}
{Drlica-Wagner}, A., {Ferguson}, P.~S., {Adam{\'o}w}, M., {et~al.} 2022{\natexlab{b}}, \apjs, 261, 38, \dodoi{10.3847/1538-4365/ac78eb}

\bibitem[{{Engler} {et~al.}(2021){Engler}, {Pillepich}, {Joshi}, {Nelson}, {Pasquali}, {Grebel}, {Lisker}, {Zinger}, {Donnari}, {Marinacci}, {Vogelsberger}, \& {Hernquist}}]{Engler2021}
{Engler}, C., {Pillepich}, A., {Joshi}, G.~D., {et~al.} 2021, \mnras, 500, 3957, \dodoi{10.1093/mnras/staa3505}

\bibitem[{{Erkal} \& {Belokurov}(2020)}]{Erkal2020}
{Erkal}, D., \& {Belokurov}, V.~A. 2020, \mnras, 495, 2554, \dodoi{10.1093/mnras/staa1238}

\bibitem[{{Erwin}(2015)}]{imfit}
{Erwin}, P. 2015, \apj, 799, 226, \dodoi{10.1088/0004-637X/799/2/226}

\bibitem[{{Euclid Collaboration} {et~al.}(2025){Euclid Collaboration}, {Mellier}, {Abdurro'uf}, {Acevedo Barroso}, {Ach{\'u}carro}, {Adamek}, {Adam}, {Addison}, {Aghanim}, {Aguena}, {Ajani}, {Akrami}, {Al-Bahlawan}, {Alavi}, {Albuquerque}, {Alestas}, {Alguero}, {Allaoui}, {Allen}, {Allevato}, {Alonso-Tetilla}, {Altieri}, {Alvarez-Candal}, {Alvi}, {Amara}, {Amendola}, {Amiaux}, {Andika}, {Andreon}, {Andrews}, {Angora}, {Angulo}, {Annibali}, {Anselmi}, {Anselmi}, {Arcari}, {Archidiacono}, {Aric{\`o}}, {Arnaud}, {Arnouts}, {Asgari}, {Asorey}, {Atayde}, {Atek}, {Atrio-Barandela}, {Aubert}, {Aubourg}, {Auphan}, {Auricchio}, {Aussel}, {Aussel}, {Avelino}, {Avgoustidis}, {Avila}, {Awan}, {Azzollini}, {Baccigalupi}, {Bachelet}, {Bacon}, {Baes}, {Bagley}, {Bahr-Kalus}, {Balaguera-Antolinez}, {Balbinot}, {Balcells}, {Baldi}, {Baldry}, {Balestra}, {Ballardini}, {Ballester}, {Balogh}, {Ba{\~n}ados}, {Barbier}, {Bardelli}, {Baron}, {Barreiro}, {Barrena}, {Barriere}, {Barros}, {Barthelemy}, {Bartolo}, {Basset},
  {Battaglia}, {Battisti}, {Baugh}, {Baumont}, {Bazzanini}, {Beaulieu}, {Beckmann}, {Belikov}, {Bel}, {Bellagamba}, {Bella}, {Bellini}, {Benabed}, {Bender}, {Benevento}, {Bennett}, {Benson}, {Bergamini}, {Bermejo-Climent}, {Bernardeau}, {Bertacca}, {Berthe}, {Berthier}, {Bethermin}, {Beutler}, {Bevillon}, {Bhargava}, {Bhatawdekar}, {Bianchi}, {Bisigello}, {Biviano}, {Blake}, {Blanchard}, {Blazek}, {Blot}, {Bosco}, {Bodendorf}, {Boenke}, {B{\"o}hringer}, {Boldrini}, {Bolzonella}, {Bonchi}, {Bonici}, {Bonino}, {Bonino}, {Bonvin}, {Bon}, {Booth}, {Borgani}, {Borlaff}, {Borsato}, {Bose}, {Botticella}, {Boucaud}, {Bouche}, {Boucher}, {Boutigny}, {Bouvard}, {Bouwens}, {Bouy}, {Bowler}, {Bozza}, {Bozzo}, {Branchini}, {Brando}, {Brau-Nogue}, {Brekke}, {Bremer}, {Brescia}, {Breton}, {Brinchmann}, {Brinckmann}, {Brockley-Blatt}, {Brodwin}, {Brouard}, {Brown}, {Bruton}, {Bucko}, {Buddelmeijer}, {Buenadicha}, {Buitrago}, {Burger}, {Burigana}, {Busillo}, {Busonero}, {Cabanac}, {Cabayol-Garcia}, {Cagliari}, {Caillat},
  {Caillat}, {Calabrese}, {Calabro}, {Calderone}, {Calura}, {Camacho Quevedo}, {Camera}, {Campos}, {Ca{\~n}as-Herrera}, {Candini}, {Cantiello}, {Capobianco}, {Cappellaro}, {Cappelluti}, {Cappi}, {Caputi}, {Cara}, {Carbone}, {Cardone}, {Carella}, {Carlberg}, {Carle}, {Carminati}, {Caro}, {Carrasco}, {Carretero}, {Carrilho}, {Carron Duque}, \& {Carry}}]{Euclid_Overview}
{Euclid Collaboration}, {Mellier}, Y., {Abdurro'uf}, {et~al.} 2025, \aap, 697, A1, \dodoi{10.1051/0004-6361/202450810}

\bibitem[{{Firth} {et~al.}(2006){Firth}, {Evstigneeva}, {Jones}, {Drinkwater}, {Phillipps}, \& {Gregg}}]{Firth2006}
{Firth}, P., {Evstigneeva}, E.~A., {Jones}, J.~B., {et~al.} 2006, \mnras, 372, 1856, \dodoi{10.1111/j.1365-2966.2006.10993.x}

\bibitem[{{Flaugher} {et~al.}(2015){Flaugher}, {Diehl}, {Honscheid}, {Abbott}, {Alvarez}, {Angstadt}, {Annis}, {Antonik}, {Ballester}, {Beaufore}, {Bernstein}, {Bernstein}, {Bigelow}, {Bonati}, {Boprie}, {Brooks}, {Buckley-Geer}, {Campa}, {Cardiel-Sas}, {Castander}, {Castilla}, {Cease}, {Cela-Ruiz}, {Chappa}, {Chi}, {Cooper}, {da Costa}, {Dede}, {Derylo}, {DePoy}, {de Vicente}, {Doel}, {Drlica-Wagner}, {Eiting}, {Elliott}, {Emes}, {Estrada}, {Fausti Neto}, {Finley}, {Flores}, {Frieman}, {Gerdes}, {Gladders}, {Gregory}, {Gutierrez}, {Hao}, {Holland}, {Holm}, {Huffman}, {Jackson}, {James}, {Jonas}, {Karcher}, {Karliner}, {Kent}, {Kessler}, {Kozlovsky}, {Kron}, {Kubik}, {Kuehn}, {Kuhlmann}, {Kuk}, {Lahav}, {Lathrop}, {Lee}, {Levi}, {Lewis}, {Li}, {Mandrichenko}, {Marshall}, {Martinez}, {Merritt}, {Miquel}, {Mu{\~n}oz}, {Neilsen}, {Nichol}, {Nord}, {Ogando}, {Olsen}, {Palaio}, {Patton}, {Peoples}, {Plazas}, {Rauch}, {Reil}, {Rheault}, {Roe}, {Rogers}, {Roodman}, {Sanchez}, {Scarpine}, {Schindler}, {Schmidt},
  {Schmitt}, {Schubnell}, {Schultz}, {Schurter}, {Scott}, {Serrano}, {Shaw}, {Smith}, {Soares-Santos}, {Stefanik}, {Stuermer}, {Suchyta}, {Sypniewski}, {Tarle}, {Thaler}, {Tighe}, {Tran}, {Tucker}, {Walker}, {Wang}, {Watson}, {Weaverdyck}, {Wester}, {Woods}, {Yanny}, \& {DES Collaboration}}]{DECam}
{Flaugher}, B., {Diehl}, H.~T., {Honscheid}, K., {et~al.} 2015, \aj, 150, 150, \dodoi{10.1088/0004-6256/150/5/150}

\bibitem[{{Font} {et~al.}(2022){Font}, {McCarthy}, {Belokurov}, {Brown}, \& {Stafford}}]{Font2022}
{Font}, A.~S., {McCarthy}, I.~G., {Belokurov}, V., {Brown}, S.~T., \& {Stafford}, S.~G. 2022, \mnras, 511, 1544, \dodoi{10.1093/mnras/stac183}

\bibitem[{{Gaia Collaboration} {et~al.}(2018){Gaia Collaboration}, {Brown}, {Vallenari}, {Prusti}, {de Bruijne}, {Babusiaux}, {Bailer-Jones}, {Biermann}, {Evans}, {Eyer}, {Jansen}, {Jordi}, {Klioner}, {Lammers}, {Lindegren}, {Luri}, {Mignard}, {Panem}, {Pourbaix}, {Randich}, {Sartoretti}, {Siddiqui}, {Soubiran}, {van Leeuwen}, {Walton}, {Arenou}, {Bastian}, {Cropper}, {Drimmel}, {Katz}, {Lattanzi}, {Bakker}, {Cacciari}, {Casta{\~n}eda}, {Chaoul}, {Cheek}, {De Angeli}, {Fabricius}, {Guerra}, {Holl}, {Masana}, {Messineo}, {Mowlavi}, {Nienartowicz}, {Panuzzo}, {Portell}, {Riello}, {Seabroke}, {Tanga}, {Th{\'e}venin}, {Gracia-Abril}, {Comoretto}, {Garcia-Reinaldos}, {Teyssier}, {Altmann}, {Andrae}, {Audard}, {Bellas-Velidis}, {Benson}, {Berthier}, {Blomme}, {Burgess}, {Busso}, {Carry}, {Cellino}, {Clementini}, {Clotet}, {Creevey}, {Davidson}, {De Ridder}, {Delchambre}, {Dell'Oro}, {Ducourant}, {Fern{\'a}ndez-Hern{\'a}ndez}, {Fouesneau}, {Fr{\'e}mat}, {Galluccio}, {Garc{\'\i}a-Torres},
  {Gonz{\'a}lez-N{\'u}{\~n}ez}, {Gonz{\'a}lez-Vidal}, {Gosset}, {Guy}, {Halbwachs}, {Hambly}, {Harrison}, {Hern{\'a}ndez}, {Hestroffer}, {Hodgkin}, {Hutton}, {Jasniewicz}, {Jean-Antoine-Piccolo}, {Jordan}, {Korn}, {Krone-Martins}, {Lanzafame}, {Lebzelter}, {L{\"o}ffler}, {Manteiga}, {Marrese}, {Mart{\'\i}n-Fleitas}, {Moitinho}, {Mora}, {Muinonen}, {Osinde}, {Pancino}, {Pauwels}, {Petit}, {Recio-Blanco}, {Richards}, {Rimoldini}, {Robin}, {Sarro}, {Siopis}, {Smith}, {Sozzetti}, {S{\"u}veges}, {Torra}, {van Reeven}, {Abbas}, {Abreu Aramburu}, {Accart}, {Aerts}, {Altavilla}, {{\'A}lvarez}, {Alvarez}, {Alves}, {Anderson}, {Andrei}, {Anglada Varela}, {Antiche}, {Antoja}, {Arcay}, {Astraatmadja}, {Bach}, {Baker}, {Balaguer-N{\'u}{\~n}ez}, {Balm}, {Barache}, {Barata}, {Barbato}, {Barblan}, {Barklem}, {Barrado}, {Barros}, {Barstow}, {Bartholom{\'e} Mu{\~n}oz}, {Bassilana}, {Becciani}, {Bellazzini}, {Berihuete}, {Bertone}, {Bianchi}, {Bienaym{\'e}}, {Blanco-Cuaresma}, {Boch}, {Boeche}, {Bombrun}, {Borrachero},
  {Bossini}, {Bouquillon}, {Bourda}, {Bragaglia}, {Bramante}, {Breddels}, {Bressan}, {Brouillet}, {Br{\"u}semeister}, {Brugaletta}, {Bucciarelli}, {Burlacu}, {Busonero}, {Butkevich}, {Buzzi}, {Caffau}, {Cancelliere}, {Cannizzaro}, {Cantat-Gaudin}, {Carballo}, {Carlucci}, {Carrasco}, {Casamiquela}, {Castellani}, {Castro-Ginard}, {Charlot}, {Chemin}, {Chiavassa}, {Cocozza}, {Costigan}, {Cowell}, {Crifo}, {Crosta}, {Crowley}, {Cuypers}, {Dafonte}, {Damerdji}, {Dapergolas}, {David}, {David}, {de Laverny}, {De Luise}, {De March}, {de Martino}, {de Souza}, {de Torres}, {Debosscher}, {del Pozo}, {Delbo}, {Delgado}, {Delgado}, {Di Matteo}, {Diakite}, {Diener}, {Distefano}, {Dolding}, {Drazinos}, {Dur{\'a}n}, {Edvardsson}, {Enke}, {Eriksson}, {Esquej}, {Eynard Bontemps}, {Fabre}, {Fabrizio}, {Faigler}, {Falc{\~a}o}, {Farr{\`a}s Casas}, {Federici}, {Fedorets}, {Fernique}, {Figueras}, {Filippi}, {Findeisen}, {Fonti}, {Fraile}, {Fraser}, {Fr{\'e}zouls}, {Gai}, {Galleti}, {Garabato}, {Garc{\'\i}a-Sedano}, {Garofalo},
  {Garralda}, {Gavel}, {Gavras}, {Gerssen}, {Geyer}, {Giacobbe}, {Gilmore}, {Girona}, {Giuffrida}, {Glass}, {Gomes}, {Granvik}, {Gueguen}, {Guerrier}, {Guiraud}, {Guti{\'e}rrez-S{\'a}nchez}, {Haigron}, {Hatzidimitriou}, {Hauser}, {Haywood}, {Heiter}, {Helmi}, {Heu}, {Hilger}, {Hobbs}, {Hofmann}, {Holland}, {Huckle}, {Hypki}, {Icardi}, {Jan{\ss}en}, {Jevardat de Fombelle}, {Jonker}, {Juh{\'a}sz}, {Julbe}, {Karampelas}, {Kewley}, {Klar}, {Kochoska}, {Kohley}, {Kolenberg}, {Kontizas}, {Kontizas}, {Koposov}, {Kordopatis}, {Kostrzewa-Rutkowska}, {Koubsky}, {Lambert}, {Lanza}, {Lasne}, {Lavigne}, {Le Fustec}, {Le Poncin-Lafitte}, {Lebreton}, {Leccia}, {Leclerc}, {Lecoeur-Taibi}, {Lenhardt}, {Leroux}, {Liao}, {Licata}, {Lindstr{\o}m}, {Lister}, {Livanou}, {Lobel}, {L{\'o}pez}, {Managau}, {Mann}, {Mantelet}, {Marchal}, {Marchant}, {Marconi}, {Marinoni}, {Marschalk{\'o}}, {Marshall}, {Martino}, {Marton}, {Mary}, {Massari}, {Matijevi{\v{c}}}, {Mazeh}, {McMillan}, {Messina}, {Michalik}, {Millar}, {Molina}, {Molinaro},
  {Moln{\'a}r}, {Montegriffo}, {Mor}, {Morbidelli}, {Morel}, {Morris}, {Mulone}, {Muraveva}, {Musella}, {Nelemans}, {Nicastro}, {Noval}, {O'Mullane}, {Ord{\'e}novic}, {Ord{\'o}{\~n}ez-Blanco}, {Osborne}, {Pagani}, {Pagano}, {Pailler}, {Palacin}, {Palaversa}, {Panahi}, {Pawlak}, {Piersimoni}, {Pineau}, {Plachy}, {Plum}, {Poggio}, {Poujoulet}, {Pr{\v{s}}a}, {Pulone}, {Racero}, {Ragaini}, {Rambaux}, {Ramos-Lerate}, {Regibo}, {Reyl{\'e}}, {Riclet}, {Ripepi}, {Riva}, {Rivard}, {Rixon}, {Roegiers}, {Roelens}, {Romero-G{\'o}mez}, {Rowell}, {Royer}, {Ruiz-Dern}, {Sadowski}, {Sagrist{\`a} Sell{\'e}s}, {Sahlmann}, {Salgado}, {Salguero}, {Sanna}, {Santana-Ros}, {Sarasso}, {Savietto}, {Schultheis}, {Sciacca}, {Segol}, {Segovia}, {S{\'e}gransan}, {Shih}, {Siltala}, {Silva}, {Smart}, {Smith}, {Solano}, {Solitro}, {Sordo}, {Soria Nieto}, {Souchay}, {Spagna}, {Spoto}, {Stampa}, {Steele}, {Steidelm{\"u}ller}, {Stephenson}, {Stoev}, {Suess}, {Surdej}, {Szabados}, {Szegedi-Elek}, {Tapiador}, {Taris}, {Tauran}, {Taylor},
  {Teixeira}, {Terrett}, {Teyssandier}, {Thuillot}, {Titarenko}, {Torra Clotet}, {Turon}, {Ulla}, {Utrilla}, {Uzzi}, {Vaillant}, {Valentini}, {Valette}, {van Elteren}, {Van Hemelryck}, {van Leeuwen}, {Vaschetto}, {Vecchiato}, {Veljanoski}, {Viala}, {Vicente}, {Vogt}, {von Essen}, {Voss}, {Votruba}, {Voutsinas}, {Walmsley}, {Weiler}, {Wertz}, {Wevers}, {Wyrzykowski}, {Yoldas}, {{\v{Z}}erjal}, {Ziaeepour}, {Zorec}, {Zschocke}, {Zucker}, {Zurbach}, \& {Zwitter}}]{GAIA2018}
{Gaia Collaboration}, {Brown}, A.~G.~A., {Vallenari}, A., {et~al.} 2018, \aap, 616, A1, \dodoi{10.1051/0004-6361/201833051}

\bibitem[{{Gaia Collaboration} {et~al.}(2021){Gaia Collaboration}, {Brown}, {Vallenari}, {Prusti}, {de Bruijne}, {Babusiaux}, {Biermann}, {Creevey}, {Evans}, {Eyer}, {Hutton}, {Jansen}, {Jordi}, {Klioner}, {Lammers}, {Lindegren}, {Luri}, {Mignard}, {Panem}, {Pourbaix}, {Randich}, {Sartoretti}, {Soubiran}, {Walton}, {Arenou}, {Bailer-Jones}, {Bastian}, {Cropper}, {Drimmel}, {Katz}, {Lattanzi}, {van Leeuwen}, {Bakker}, {Cacciari}, {Casta{\~n}eda}, {De Angeli}, {Ducourant}, {Fabricius}, {Fouesneau}, {Fr{\'e}mat}, {Guerra}, {Guerrier}, {Guiraud}, {Jean-Antoine Piccolo}, {Masana}, {Messineo}, {Mowlavi}, {Nicolas}, {Nienartowicz}, {Pailler}, {Panuzzo}, {Riclet}, {Roux}, {Seabroke}, {Sordo}, {Tanga}, {Th{\'e}venin}, {Gracia-Abril}, {Portell}, {Teyssier}, {Altmann}, {Andrae}, {Bellas-Velidis}, {Benson}, {Berthier}, {Blomme}, {Brugaletta}, {Burgess}, {Busso}, {Carry}, {Cellino}, {Cheek}, {Clementini}, {Damerdji}, {Davidson}, {Delchambre}, {Dell'Oro}, {Fern{\'a}ndez-Hern{\'a}ndez}, {Galluccio}, {Garc{\'\i}a-Lario},
  {Garcia-Reinaldos}, {Gonz{\'a}lez-N{\'u}{\~n}ez}, {Gosset}, {Haigron}, {Halbwachs}, {Hambly}, {Harrison}, {Hatzidimitriou}, {Heiter}, {Hern{\'a}ndez}, {Hestroffer}, {Hodgkin}, {Holl}, {Jan{\ss}en}, {Jevardat de Fombelle}, {Jordan}, {Krone-Martins}, {Lanzafame}, {L{\"o}ffler}, {Lorca}, {Manteiga}, {Marchal}, {Marrese}, {Moitinho}, {Mora}, {Muinonen}, {Osborne}, {Pancino}, {Pauwels}, {Petit}, {Recio-Blanco}, {Richards}, {Riello}, {Rimoldini}, {Robin}, {Roegiers}, {Rybizki}, {Sarro}, {Siopis}, {Smith}, {Sozzetti}, {Ulla}, {Utrilla}, {van Leeuwen}, {van Reeven}, {Abbas}, {Abreu Aramburu}, {Accart}, {Aerts}, {Aguado}, {Ajaj}, {Altavilla}, {{\'A}lvarez}, {{\'A}lvarez Cid-Fuentes}, {Alves}, {Anderson}, {Anglada Varela}, {Antoja}, {Audard}, {Baines}, {Baker}, {Balaguer-N{\'u}{\~n}ez}, {Balbinot}, {Balog}, {Barache}, {Barbato}, {Barros}, {Barstow}, {Bartolom{\'e}}, {Bassilana}, {Bauchet}, {Baudesson-Stella}, {Becciani}, {Bellazzini}, {Bernet}, {Bertone}, {Bianchi}, {Blanco-Cuaresma}, {Boch}, {Bombrun}, {Bossini},
  {Bouquillon}, {Bragaglia}, {Bramante}, {Breedt}, {Bressan}, {Brouillet}, {Bucciarelli}, {Burlacu}, {Busonero}, {Butkevich}, {Buzzi}, {Caffau}, {Cancelliere}, {C{\'a}novas}, {Cantat-Gaudin}, {Carballo}, {Carlucci}, {Carnerero}, {Carrasco}, {Casamiquela}, {Castellani}, {Castro-Ginard}, {Castro Sampol}, {Chaoul}, {Charlot}, {Chemin}, {Chiavassa}, {Cioni}, {Comoretto}, {Cooper}, {Cornez}, {Cowell}, {Crifo}, {Crosta}, {Crowley}, {Dafonte}, {Dapergolas}, {David}, \& {David}}]{Gaia-dr3}
---. 2021, \aap, 649, A1, \dodoi{10.1051/0004-6361/202039657}

\bibitem[{{Garling} {et~al.}(2020){Garling}, {Peter}, {Kochanek}, {Sand}, \& {Crnojevi{\'c}}}]{Garling2020}
{Garling}, C.~T., {Peter}, A. H.~G., {Kochanek}, C.~S., {Sand}, D.~J., \& {Crnojevi{\'c}}, D. 2020, \mnras, 492, 1713, \dodoi{10.1093/mnras/stz3526}

\bibitem[{{Garling} {et~al.}(2021){Garling}, {Peter}, {Kochanek}, {Sand}, \& {Crnojevi{\'c}}}]{Garling2021}
---. 2021, \mnras, 507, 4764, \dodoi{10.1093/mnras/stab2447}

\bibitem[{{Geha}(2026)}]{Geha2026}
{Geha}, M. 2026, arXiv e-prints, arXiv:2602.10202, \dodoi{10.48550/arXiv.2602.10202}

\bibitem[{{Geha} {et~al.}(2012){Geha}, {Blanton}, {Yan}, \& {Tinker}}]{Geha2012}
{Geha}, M., {Blanton}, M.~R., {Yan}, R., \& {Tinker}, J.~L. 2012, \apj, 757, 85, \dodoi{10.1088/0004-637X/757/1/85}

\bibitem[{{Geha} {et~al.}(2017){Geha}, {Wechsler}, {Mao}, {Tollerud}, {Weiner}, {Bernstein}, {Hoyle}, {Marchi}, {Marshall}, {Mu{\~n}oz}, \& {Lu}}]{SAGA-I}
{Geha}, M., {Wechsler}, R.~H., {Mao}, Y.-Y., {et~al.} 2017, \apj, 847, 4, \dodoi{10.3847/1538-4357/aa8626}

\bibitem[{{Geha} {et~al.}(2024){Geha}, {Mao}, {Wechsler}, {Asali}, {Kado-Fong}, {Kallivayalil}, {Nadler}, {Tollerud}, {Weiner}, {de los Reyes}, {Wang}, \& {Wu}}]{SAGA-IV}
{Geha}, M., {Mao}, Y.-Y., {Wechsler}, R.~H., {et~al.} 2024, \apj, 976, 118, \dodoi{10.3847/1538-4357/ad61e710.1134/S1063773708080082}

\bibitem[{{Greco} \& {Danieli}(2022)}]{artpop}
{Greco}, J.~P., \& {Danieli}, S. 2022, \apj, 941, 26, \dodoi{10.3847/1538-4357/ac75b7}

\bibitem[{{Greco} {et~al.}(2021){Greco}, {van Dokkum}, {Danieli}, {Carlsten}, \& {Conroy}}]{Greco2021}
{Greco}, J.~P., {van Dokkum}, P., {Danieli}, S., {Carlsten}, S.~G., \& {Conroy}, C. 2021, \apj, 908, 24, \dodoi{10.3847/1538-4357/abd030}

\bibitem[{{Greco} {et~al.}(2018){Greco}, {Greene}, {Strauss}, {Macarthur}, {Flowers}, {Goulding}, {Huang}, {Kim}, {Komiyama}, {Leauthaud}, {Leisman}, {Lupton}, {Sif{\'o}n}, \& {Wang}}]{Greco2018}
{Greco}, J.~P., {Greene}, J.~E., {Strauss}, M.~A., {et~al.} 2018, \apj, 857, 104, \dodoi{10.3847/1538-4357/aab842}

\bibitem[{{Green} {et~al.}(2022){Green}, {van den Bosch}, \& {Jiang}}]{Green2022}
{Green}, S.~B., {van den Bosch}, F.~C., \& {Jiang}, F. 2022, \mnras, 509, 2624, \dodoi{10.1093/mnras/stab3130}

\bibitem[{{Greene} {et~al.}(2023){Greene}, {Danieli}, {Carlsten}, {Beaton}, {Jiang}, \& {Li}}]{Greene2023}
{Greene}, J.~E., {Danieli}, S., {Carlsten}, S., {et~al.} 2023, \apj, 949, 94, \dodoi{10.3847/1538-4357/acc58c}

\bibitem[{{Guo} {et~al.}(2011){Guo}, {White}, {Boylan-Kolchin}, {De Lucia}, {Kauffmann}, {Lemson}, {Li}, {Springel}, \& {Weinmann}}]{Guo2011}
{Guo}, Q., {White}, S., {Boylan-Kolchin}, M., {et~al.} 2011, \mnras, 413, 101, \dodoi{10.1111/j.1365-2966.2010.18114.x}

\bibitem[{{Hargis} {et~al.}(2020){Hargis}, {Albers}, {Crnojevi{\'c}}, {Sand}, {Weisz}, {Carlin}, {Spekkens}, {Willman}, {Peter}, {Grillmair}, \& {Dolphin}}]{Hargis2020}
{Hargis}, J.~R., {Albers}, S., {Crnojevi{\'c}}, D., {et~al.} 2020, \apj, 888, 31, \dodoi{10.3847/1538-4357/ab58d2}

\bibitem[{Harris {et~al.}(2020)Harris, Millman, van~der Walt, Gommers, Virtanen, Cournapeau, Wieser, Taylor, Berg, Smith, Kern, Picus, Hoyer, van Kerkwijk, Brett, Haldane, del R{'{\i}}o, Wiebe, Peterson, G{'{e}}rard-Marchant, Sheppard, Reddy, Weckesser, Abbasi, Gohlke, \& Oliphant}]{Numpy}
Harris, C.~R., Millman, K.~J., van~der Walt, S.~J., {et~al.} 2020, Nature, 585, 357, \dodoi{10.1038/s41586-020-2649-2}

\bibitem[{{Helou} {et~al.}(1991){Helou}, {Madore}, {Schmitz}, {Bicay}, {Wu}, \& {Bennett}}]{NED}
{Helou}, G., {Madore}, B.~F., {Schmitz}, M., {et~al.} 1991, in Astrophysics and Space Science Library, Vol. 171, Databases and On-line Data in Astronomy, ed. M.~A. {Albrecht} \& D.~{Egret}, 89--106, \dodoi{10.1007/978-94-011-3250-3_10}

\bibitem[{{Hopkins} {et~al.}(2011){Hopkins}, {Quataert}, \& {Murray}}]{Hopkins2011}
{Hopkins}, P.~F., {Quataert}, E., \& {Murray}, N. 2011, \mnras, 417, 950, \dodoi{10.1111/j.1365-2966.2011.19306.x}

\bibitem[{{Hopkins} {et~al.}(2018){Hopkins}, {Wetzel}, {Kere{\v{s}}}, {Faucher-Gigu{\`e}re}, {Quataert}, {Boylan-Kolchin}, {Murray}, {Hayward}, {Garrison-Kimmel}, {Hummels}, {Feldmann}, {Torrey}, {Ma}, {Angl{\'e}s-Alc{\'a}zar}, {Su}, {Orr}, {Schmitz}, {Escala}, {Sanderson}, {Grudi{\'c}}, {Hafen}, {Kim}, {Fitts}, {Bullock}, {Wheeler}, {Chan}, {Elbert}, \& {Narayanan}}]{Hopkins2018}
{Hopkins}, P.~F., {Wetzel}, A., {Kere{\v{s}}}, D., {et~al.} 2018, \mnras, 480, 800, \dodoi{10.1093/mnras/sty1690}

\bibitem[{{Hunter}(2007)}]{matplotlib}
{Hunter}, J.~D. 2007, Computing in Science Engineering, 9, 90, \dodoi{10.1109/MCSE.2007.55}

\bibitem[{{Hunter} {et~al.}(2025){Hunter}, {Mutlu-Pakd{\.I}l}, {Sand}, {Bennet}, {Khim}, {Crnojevi{\'c}}, {Doliva-Dolinsky}, {Durodola}, {Fielder}, {Goebel-Bain}, {Jones}, {Karunakaran}, {Spekkens}, \& {Zaritsky}}]{Hunter2025}
{Hunter}, L.~C., {Mutlu-Pakd{\.I}l}, B., {Sand}, D.~J., {et~al.} 2025, \apj, 989, 58, \dodoi{10.3847/1538-4357/ade9a4}

\bibitem[{{Hunter} {et~al.}(2026){Hunter}, {Mutlu-Pakdil}, {Farnell}, {Sand}, {Bennet}, {Campana}, {Carlin}, {Crnojevi{\'c}}, {Doliva-Dolinsky}, {Durodola}, {Jones}, {Khim}, {Marin}, {Mendez}, {Prabhu}, {Spekkens}, \& {Zaritsky}}]{Hunter2026}
{Hunter}, L.~C., {Mutlu-Pakdil}, B., {Farnell}, M.~B., {et~al.} 2026, arXiv e-prints, arXiv:2603.18973, \dodoi{10.48550/arXiv.2603.18973}

\bibitem[{{Into} \& {Portinari}(2013)}]{Into2013}
{Into}, T., \& {Portinari}, L. 2013, \mnras, 430, 2715, \dodoi{10.1093/mnras/stt071}

\bibitem[{{Ivezi{\'c}} {et~al.}(2019){Ivezi{\'c}}, {Kahn}, {Tyson}, {Abel}, {Acosta}, {Allsman}, {Alonso}, {AlSayyad}, {Anderson}, {Andrew}, {Angel}, {Angeli}, {Ansari}, {Antilogus}, {Araujo}, {Armstrong}, {Arndt}, {Astier}, {Aubourg}, {Auza}, {Axelrod}, {Bard}, {Barr}, {Barrau}, {Bartlett}, {Bauer}, {Bauman}, {Baumont}, {Bechtol}, {Bechtol}, {Becker}, {Becla}, {Beldica}, {Bellavia}, {Bianco}, {Biswas}, {Blanc}, {Blazek}, {Blandford}, {Bloom}, {Bogart}, {Bond}, {Booth}, {Borgland}, {Borne}, {Bosch}, {Boutigny}, {Brackett}, {Bradshaw}, {Brandt}, {Brown}, {Bullock}, {Burchat}, {Burke}, {Cagnoli}, {Calabrese}, {Callahan}, {Callen}, {Carlin}, {Carlson}, {Chandrasekharan}, {Charles-Emerson}, {Chesley}, {Cheu}, {Chiang}, {Chiang}, {Chirino}, {Chow}, {Ciardi}, {Claver}, {Cohen-Tanugi}, {Cockrum}, {Coles}, {Connolly}, {Cook}, {Cooray}, {Covey}, {Cribbs}, {Cui}, {Cutri}, {Daly}, {Daniel}, {Daruich}, {Daubard}, {Daues}, {Dawson}, {Delgado}, {Dellapenna}, {de Peyster}, {de Val-Borro}, {Digel}, {Doherty}, {Dubois},
  {Dubois-Felsmann}, {Durech}, {Economou}, {Eifler}, {Eracleous}, {Emmons}, {Fausti Neto}, {Ferguson}, {Figueroa}, {Fisher-Levine}, {Focke}, {Foss}, {Frank}, {Freemon}, {Gangler}, {Gawiser}, {Geary}, {Gee}, {Geha}, {Gessner}, {Gibson}, {Gilmore}, {Glanzman}, {Glick}, {Goldina}, {Goldstein}, {Goodenow}, {Graham}, {Gressler}, {Gris}, {Guy}, {Guyonnet}, {Haller}, {Harris}, {Hascall}, {Haupt}, {Hernandez}, {Herrmann}, {Hileman}, {Hoblitt}, {Hodgson}, {Hogan}, {Howard}, {Huang}, {Huffer}, {Ingraham}, {Innes}, {Jacoby}, {Jain}, {Jammes}, {Jee}, {Jenness}, {Jernigan}, {Jevremovi{\'c}}, {Johns}, {Johnson}, {Johnson}, {Jones}, {Juramy-Gilles}, {Juri{\'c}}, {Kalirai}, {Kallivayalil}, {Kalmbach}, {Kantor}, {Karst}, {Kasliwal}, {Kelly}, {Kessler}, {Kinnison}, {Kirkby}, {Knox}, {Kotov}, {Krabbendam}, {Krughoff}, {Kub{\'a}nek}, {Kuczewski}, {Kulkarni}, {Ku}, {Kurita}, {Lage}, {Lambert}, {Lange}, {Langton}, {Le Guillou}, {Levine}, {Liang}, {Lim}, {Lintott}, {Long}, {Lopez}, {Lotz}, {Lupton}, {Lust}, {MacArthur}, {Mahabal},
  {Mandelbaum}, {Markiewicz}, {Marsh}, {Marshall}, {Marshall}, {May}, {McKercher}, {McQueen}, {Meyers}, {Migliore}, {Miller}, {Mills}, {Miraval}, {Moeyens}, {Moolekamp}, {Monet}, {Moniez}, {Monkewitz}, {Montgomery}, {Morrison}, {Mueller}, {Muller}, {Mu{\~n}oz Arancibia}, {Neill}, {Newbry}, {Nief}, {Nomerotski}, {Nordby}, {O'Connor}, {Oliver}, {Olivier}, {Olsen}, {O'Mullane}, {Ortiz}, {Osier}, {Owen}, {Pain}, {Palecek}, {Parejko}, {Parsons}, {Pease}, {Peterson}, {Peterson}, {Petravick}, {Libby Petrick}, {Petry}, {Pierfederici}, {Pietrowicz}, {Pike}, {Pinto}, {Plante}, {Plate}, {Plutchak}, {Price}, {Prouza}, {Radeka}, {Rajagopal}, {Rasmussen}, {Regnault}, {Reil}, {Reiss}, {Reuter}, {Ridgway}, {Riot}, {Ritz}, {Robinson}, {Roby}, {Roodman}, {Rosing}, {Roucelle}, {Rumore}, {Russo}, {Saha}, {Sassolas}, {Schalk}, {Schellart}, {Schindler}, {Schmidt}, {Schneider}, {Schneider}, {Schoening}, {Schumacher}, {Schwamb}, {Sebag}, {Selvy}, {Sembroski}, {Seppala}, {Serio}, {Serrano}, {Shaw}, {Shipsey}, {Sick}, {Silvestri},
  {Slater}, {Smith}, {Smith}, {Sobhani}, {Soldahl}, {Storrie-Lombardi}, {Stover}, {Strauss}, {Street}, {Stubbs}, {Sullivan}, {Sweeney}, {Swinbank}, {Szalay}, {Takacs}, {Tether}, {Thaler}, {Thayer}, {Thomas}, {Thornton}, {Thukral}, {Tice}, {Trilling}, {Turri}, {Van Berg}, {Vanden Berk}, {Vetter}, {Virieux}, {Vucina}, {Wahl}, {Walkowicz}, {Walsh}, {Walter}, {Wang}, {Wang}, {Warner}, {Wiecha}, {Willman}, {Winters}, {Wittman}, {Wolff}, {Wood-Vasey}, {Wu}, {Xin}, {Yoachim}, \& {Zhan}}]{LSST2019}
{Ivezi{\'c}}, {\v{Z}}., {Kahn}, S.~M., {Tyson}, J.~A., {et~al.} 2019, \apj, 873, 111, \dodoi{10.3847/1538-4357/ab042c}

\bibitem[{{Jahn} {et~al.}(2019){Jahn}, {Sales}, {Wetzel}, {Boylan-Kolchin}, {Chan}, {El-Badry}, {Lazar}, \& {Bullock}}]{Jahn2019}
{Jahn}, E.~D., {Sales}, L.~V., {Wetzel}, A., {et~al.} 2019, \mnras, 489, 5348, \dodoi{10.1093/mnras/stz2457}

\bibitem[{{Jahn} {et~al.}(2022){Jahn}, {Sales}, {Wetzel}, {Samuel}, {El-Badry}, {Boylan-Kolchin}, \& {Bullock}}]{Jahn2022}
---. 2022, \mnras, 513, 2673, \dodoi{10.1093/mnras/stac811}

\bibitem[{{Jang} \& {Lee}(2017)}]{Jang2017}
{Jang}, I.~S., \& {Lee}, M.~G. 2017, \apj, 835, 28, \dodoi{10.3847/1538-4357/835/1/28}

\bibitem[{{Jarrett} {et~al.}(2000){Jarrett}, {Chester}, {Cutri}, {Schneider}, {Skrutskie}, \& {Huchra}}]{Jarrett2000}
{Jarrett}, T.~H., {Chester}, T., {Cutri}, R., {et~al.} 2000, \aj, 119, 2498, \dodoi{10.1086/301330}

\bibitem[{{Jiang} {et~al.}(2021){Jiang}, {Dekel}, {Freundlich}, {van den Bosch}, {Green}, {Hopkins}, {Benson}, \& {Du}}]{Jiang2021}
{Jiang}, F., {Dekel}, A., {Freundlich}, J., {et~al.} 2021, \mnras, 502, 621, \dodoi{10.1093/mnras/staa4034}

\bibitem[{{Kado-Fong} {et~al.}(2025){Kado-Fong}, {Mao}, {Asali}, {Geha}, {Wechsler}, {de los Reyes}, {Wang}, {Nadler}, {Kallivayalil}, {Tollerud}, \& {Weiner}}]{Kado-Fong2025}
{Kado-Fong}, E., {Mao}, Y.-Y., {Asali}, Y., {et~al.} 2025, \apj, 994, 231, \dodoi{10.3847/1538-4357/ae102d}

\bibitem[{{Kallivayalil} {et~al.}(2018){Kallivayalil}, {Sales}, {Zivick}, {Fritz}, {Del Pino}, {Sohn}, {Besla}, {van der Marel}, {Navarro}, \& {Sacchi}}]{Kallivayalil2018}
{Kallivayalil}, N., {Sales}, L.~V., {Zivick}, P., {et~al.} 2018, \apj, 867, 19, \dodoi{10.3847/1538-4357/aadfee}

\bibitem[{{Karachentsev} {et~al.}(2025){Karachentsev}, {Karachentseva}, {Vladimirova}, \& {Kozyrev}}]{Karachentsev2025b}
{Karachentsev}, I., {Karachentseva}, V., {Vladimirova}, K., \& {Kozyrev}, C. 2025, \pasa, 42, e026, \dodoi{10.1017/pasa.2025.7}

\bibitem[{{Karachentsev} \& {Kaisina}(2022)}]{Karachentsev2022}
{Karachentsev}, I.~D., \& {Kaisina}, E.~I. 2022, Astrophysical Bulletin, 77, 372, \dodoi{10.1134/S1990341322040058}

\bibitem[{{Karachentsev} {et~al.}(2013){Karachentsev}, {Makarov}, \& {Kaisina}}]{Karachentsev2013}
{Karachentsev}, I.~D., {Makarov}, D.~I., \& {Kaisina}, E.~I. 2013, \aj, 145, 101, \dodoi{10.1088/0004-6256/145/4/101}

\bibitem[{{Karachentsev} {et~al.}(2020){Karachentsev}, {Makarova}, {Brent Tully}, {Anand}, {Rizzi}, \& {Shaya}}]{Karachentsev2020}
{Karachentsev}, I.~D., {Makarova}, L.~N., {Brent Tully}, R., {et~al.} 2020, \aap, 643, A124, \dodoi{10.1051/0004-6361/202038928}

\bibitem[{{Karachentsev} {et~al.}(2017){Karachentsev}, {Makarova}, {Tully}, {Rizzi}, {Karachentseva}, \& {Shaya}}]{Karachentsev2017_DDO161}
{Karachentsev}, I.~D., {Makarova}, L.~N., {Tully}, R.~B., {et~al.} 2017, \mnras, 469, L113, \dodoi{10.1093/mnrasl/slx061}

\bibitem[{{Karachentsev} {et~al.}(2015){Karachentsev}, {Tully}, {Makarova}, {Makarov}, \& {Rizzi}}]{Karachentsev2015LeoSpur}
{Karachentsev}, I.~D., {Tully}, R.~B., {Makarova}, L.~N., {Makarov}, D.~I., \& {Rizzi}, L. 2015, \apj, 805, 144, \dodoi{10.1088/0004-637X/805/2/144}

\bibitem[{{Karachentsev} {et~al.}(1999){Karachentsev}, {Sharina}, {Grebel}, {Dolphin}, {Geisler}, {Guhathakurta}, {Hodge}, {Karachentseva}, {Sarajedini}, \& {Seitzer}}]{Karachentsev1999DDO044}
{Karachentsev}, I.~D., {Sharina}, M.~E., {Grebel}, E.~K., {et~al.} 1999, \aap, 352, 399, \dodoi{10.48550/arXiv.astro-ph/9910402}

\bibitem[{{Kim} {et~al.}(2022){Kim}, {Kang}, {Lee}, \& {Jang}}]{Kim2022}
{Kim}, Y.~J., {Kang}, J., {Lee}, M.~G., \& {Jang}, I.~S. 2022, \apj, 929, 36, \dodoi{10.3847/1538-4357/ac58f3}

\bibitem[{{Kim} \& {Lee}(2021)}]{Kim2021}
{Kim}, Y.~J., \& {Lee}, M.~G. 2021, \apj, 923, 152, \dodoi{10.3847/1538-4357/ac2d94}

\bibitem[{{Klypin} {et~al.}(1999){Klypin}, {Kravtsov}, {Valenzuela}, \& {Prada}}]{Klypin1999}
{Klypin}, A., {Kravtsov}, A.~V., {Valenzuela}, O., \& {Prada}, F. 1999, \apj, 522, 82, \dodoi{10.1086/307643}

\bibitem[{{Koribalski} {et~al.}(2018){Koribalski}, {Wang}, {Kamphuis}, {Westmeier}, {Staveley-Smith}, {Oh}, {L{\'o}pez-S{\'a}nchez}, {Wong}, {Ott}, {de Blok}, \& {Shao}}]{Koribalski2018}
{Koribalski}, B.~S., {Wang}, J., {Kamphuis}, P., {et~al.} 2018, \mnras, 478, 1611, \dodoi{10.1093/mnras/sty479}

\bibitem[{{Kroupa}(2001)}]{Kroupa2001}
{Kroupa}, P. 2001, \mnras, 322, 231, \dodoi{10.1046/j.1365-8711.2001.04022.x}

\bibitem[{{Labrie} {et~al.}(2023){Labrie}, {Simpson}, {Cardenes}, {Turner}, {Soraisam}, {Quint}, {Oberdorf}, {Placco}, {Berke}, {Smirnova}, {Conseil}, {Vacca}, \& {Thomas-Osip}}]{DRAGONS_paper}
{Labrie}, K., {Simpson}, C., {Cardenes}, R., {et~al.} 2023, Research Notes of the American Astronomical Society, 7, 214, \dodoi{10.3847/2515-5172/ad0044}

\bibitem[{{Lang} {et~al.}(2010){Lang}, {Hogg}, {Mierle}, {Blanton}, \& {Roweis}}]{Lang2010}
{Lang}, D., {Hogg}, D.~W., {Mierle}, K., {Blanton}, M., \& {Roweis}, S. 2010, \aj, 139, 1782, \dodoi{10.1088/0004-6256/139/5/1782}

\bibitem[{{Lee} {et~al.}(1993){Lee}, {Freedman}, \& {Madore}}]{Lee1993}
{Lee}, M.~G., {Freedman}, W.~L., \& {Madore}, B.~F. 1993, \apj, 417, 553, \dodoi{10.1086/173334}

\bibitem[{{Leroy} {et~al.}(2019){Leroy}, {Sandstrom}, {Lang}, {Lewis}, {Salim}, {Behrens}, {Chastenet}, {Chiang}, {Gallagher}, {Kessler}, \& {Utomo}}]{Leroy2019}
{Leroy}, A.~K., {Sandstrom}, K.~M., {Lang}, D., {et~al.} 2019, \apjs, 244, 24, \dodoi{10.3847/1538-4365/ab3925}

\bibitem[{{Li} {et~al.}(2024){Li}, {Greene}, {Carlsten}, \& {Danieli}}]{Li2024}
{Li}, J., {Greene}, J.~E., {Carlsten}, S.~G., \& {Danieli}, S. 2024, \apjl, 975, L23, \dodoi{10.3847/2041-8213/ad5b59}

\bibitem[{{Li} {et~al.}(2026{\natexlab{a}}){Li}, {Greene}, {Danieli}, {Carlsten}, \& {Geha}}]{Li2026_DDO161}
{Li}, J., {Greene}, J.~E., {Danieli}, S., {Carlsten}, S.~G., \& {Geha}, M. 2026{\natexlab{a}}, \apjl, 998, L24, \dodoi{10.3847/2041-8213/ae3ddd}

\bibitem[{{Li} {et~al.}(2026{\natexlab{b}}){Li}, {Greene}, {Danieli}, {Carlsten}, {Geha}, {Jiang}, \& {Tanaka}}]{Li2025}
{Li}, J., {Greene}, J.~E., {Danieli}, S., {et~al.} 2026{\natexlab{b}}, \apj, 1002, 75, \dodoi{10.3847/1538-4357/ae4495}

\bibitem[{{Li} {et~al.}(2023{\natexlab{a}}){Li}, {Greene}, {Greco}, {Beaton}, {Danieli}, {Goulding}, {Huang}, \& {Kado-Fong}}]{Li2023}
{Li}, J., {Greene}, J.~E., {Greco}, J., {et~al.} 2023{\natexlab{a}}, \apj, 955, 2, \dodoi{10.3847/1538-4357/ace4c5}

\bibitem[{{Li} {et~al.}(2023{\natexlab{b}}){Li}, {Greene}, {Greco}, {Huang}, {Melchior}, {Beaton}, {Casey}, {Danieli}, {Goulding}, {Joseph}, {Kado-Fong}, {Kim}, \& {MacArthur}}]{Li2022}
{Li}, J., {Greene}, J.~E., {Greco}, J.~P., {et~al.} 2023{\natexlab{b}}, \apj, 955, 1, \dodoi{10.3847/1538-4357/ace829}

\bibitem[{{Liu}(2000)}]{Liu2000}
{Liu}, M.~C. 2000, PhD thesis, University of California, Berkeley

\bibitem[{{Lovell} {et~al.}(2021){Lovell}, {Cautun}, {Frenk}, {Hellwing}, \& {Newton}}]{Lovell2021}
{Lovell}, M.~R., {Cautun}, M., {Frenk}, C.~S., {Hellwing}, W.~A., \& {Newton}, O. 2021, \mnras, 507, 4826, \dodoi{10.1093/mnras/stab2452}

\bibitem[{{Mao} {et~al.}(2024){Mao}, {Geha}, {Wechsler}, {Asali}, {Wang}, {Kado-Fong}, {Kallivayalil}, {Nadler}, {Tollerud}, {Weiner}, {de los Reyes}, \& {Wu}}]{SAGA-III}
{Mao}, Y.-Y., {Geha}, M., {Wechsler}, R.~H., {et~al.} 2024, \apj, 976, 117, \dodoi{10.3847/1538-4357/ad64c410.1134/S1063773708080057}

\bibitem[{{Mart{\'\i}nez-Delgado} {et~al.}(2012){Mart{\'\i}nez-Delgado}, {Romanowsky}, {Gabany}, {Annibali}, {Arnold}, {Fliri}, {Zibetti}, {van der Marel}, {Rix}, {Chonis}, {Carballo-Bello}, {Aloisi}, {Macci{\`o}}, {Gallego-Laborda}, {Brodie}, \& {Merrifield}}]{Martinez-Delgado2012}
{Mart{\'\i}nez-Delgado}, D., {Romanowsky}, A.~J., {Gabany}, R.~J., {et~al.} 2012, \apjl, 748, L24, \dodoi{10.1088/2041-8205/748/2/L24}

\bibitem[{{Mateo}(1998)}]{Mateo1998}
{Mateo}, M.~L. 1998, \araa, 36, 435, \dodoi{10.1146/annurev.astro.36.1.435}

\bibitem[{{McConnachie}(2012)}]{McConnachie2012}
{McConnachie}, A.~W. 2012, \aj, 144, 4, \dodoi{10.1088/0004-6256/144/1/4}

\bibitem[{McCully {et~al.}(2018)McCully, Crawford, Kovacs, Tollerud, Betts, Bradley, Craig, Turner, Streicher, Sipocz, Robitaille, \& Deil}]{astroscrappy}
McCully, C., Crawford, S., Kovacs, G., {et~al.} 2018, astropy/astroscrappy: v1.0.5 Zenodo Release, v1.0.5,  Zenodo, \dodoi{10.5281/zenodo.1482019}

\bibitem[{{McDonough} \& {Brainerd}(2022)}]{McDonough2022}
{McDonough}, B., \& {Brainerd}, T.~G. 2022, \apj, 933, 161, \dodoi{10.3847/1538-4357/ac752d}

\bibitem[{{McNanna} {et~al.}(2024){McNanna}, {Bechtol}, {Mau}, {Nadler}, {Medoff}, {Drlica-Wagner}, {Cerny}, {Crnojevi{\'c}}, {Mutlu-Pakd{\i}l}, {Vivas}, {Pace}, {Carlin}, {Collins}, {Ferguson}, {Mart{\'\i}nez-Delgado}, {Mart{\'\i}nez-V{\'a}zquez}, {Noel}, {Riley}, {Sand}, {Smercina}, {Tollerud}, {Wechsler}, {Abbott}, {Aguena}, {Alves}, {Bacon}, {Bom}, {Brooks}, {Burke}, {Carballo-Bello}, {Carnero Rosell}, {Carretero}, {da Costa}, {Davis}, {de Vicente}, {Diehl}, {Doel}, {Ferrero}, {Frieman}, {Giannini}, {Gruen}, {Gutierrez}, {Gruendl}, {Hinton}, {Hollowood}, {Honscheid}, {James}, {Kuehn}, {Marshall}, {Mena-Fern{\'a}ndez}, {Miquel}, {Pereira}, {Pieres}, {Malag{\'o}n}, {Sakowska}, {Sanchez}, {Sanchez Cid}, {Santiago}, {Sevilla-Noarbe}, {Smith}, {Stringfellow}, {Suchyta}, {Swanson}, {Tarle}, {Weaverdyck}, {Wiseman}, {DES Collaboration}, \& {DELVE Collaboration}}]{McNanna2024}
{McNanna}, M., {Bechtol}, K., {Mau}, S., {et~al.} 2024, \apj, 961, 126, \dodoi{10.3847/1538-4357/ad07d0}

\bibitem[{{McQuinn} {et~al.}(2015{\natexlab{a}}){McQuinn}, {Lelli}, {Skillman}, {Dolphin}, {McGaugh}, \& {Williams}}]{McQuinn2015starburst}
{McQuinn}, K. B.~W., {Lelli}, F., {Skillman}, E.~D., {et~al.} 2015{\natexlab{a}}, \mnras, 450, 3886, \dodoi{10.1093/mnras/stv841}

\bibitem[{{McQuinn} {et~al.}(2017){McQuinn}, {Skillman}, {Dolphin}, {Berg}, \& {Kennicutt}}]{McQuinn2017}
{McQuinn}, K. B.~W., {Skillman}, E.~D., {Dolphin}, A.~E., {Berg}, D., \& {Kennicutt}, R. 2017, \aj, 154, 51, \dodoi{10.3847/1538-3881/aa7aad}

\bibitem[{{McQuinn} {et~al.}(2010){McQuinn}, {Skillman}, {Cannon}, {Dalcanton}, {Dolphin}, {Hidalgo-Rodr{\'\i}guez}, {Holtzman}, {Stark}, {Weisz}, \& {Williams}}]{McQuinn2010}
{McQuinn}, K. B.~W., {Skillman}, E.~D., {Cannon}, J.~M., {et~al.} 2010, \apj, 721, 297, \dodoi{10.1088/0004-637X/721/1/297}

\bibitem[{{McQuinn} {et~al.}(2014){McQuinn}, {Cannon}, {Dolphin}, {Skillman}, {Salzer}, {Haynes}, {Adams}, {Cave}, {Elson}, {Giovanelli}, {Ott}, \& {Saintonge}}]{McQuinn2014}
{McQuinn}, K. B.~W., {Cannon}, J.~M., {Dolphin}, A.~E., {et~al.} 2014, \apj, 785, 3, \dodoi{10.1088/0004-637X/785/1/3}

\bibitem[{{McQuinn} {et~al.}(2015{\natexlab{b}}){McQuinn}, {Skillman}, {Dolphin}, {Cannon}, {Salzer}, {Rhode}, {Adams}, {Berg}, {Giovanelli}, {Girardi}, \& {Haynes}}]{McQuinn2015}
{McQuinn}, K. B.~W., {Skillman}, E.~D., {Dolphin}, A., {et~al.} 2015{\natexlab{b}}, \apj, 812, 158, \dodoi{10.1088/0004-637X/812/2/158}

\bibitem[{{Medoff} {et~al.}(2025){Medoff}, {Mutlu-Pakdil}, {Carlin}, {Drlica-Wagner}, {Tollerud}, {Doliva-Dolinsky}, {Sand}, {Mart{\'\i}nez-V{\'a}zquez}, {Stringfellow}, {Cerny}, {Crnojevi{\'c}}, {Ferguson}, {Fielder}, {Chaturvedi}, {Kallivayalil}, {No{\"e}l}, {Vivas}, {Walker}, {Adam{\'o}w}, {Bom}, {Carballo-Bello}, {Choi}, {Medina}, {Navabi}, {Pace}, {Riley}, \& {Sakowska}}]{Medoff2025}
{Medoff}, J., {Mutlu-Pakdil}, B., {Carlin}, J.~L., {et~al.} 2025, \apj, 990, 108, \dodoi{10.3847/1538-4357/adf211}

\bibitem[{{Mei} {et~al.}(2005){Mei}, {Blakeslee}, {Tonry}, {Jord{\'a}n}, {Peng}, {C{\^o}t{\'e}}, {Ferrarese}, {Merritt}, {Milosavljevi{\'c}}, \& {West}}]{Mei2005}
{Mei}, S., {Blakeslee}, J.~P., {Tonry}, J.~L., {et~al.} 2005, \apjs, 156, 113, \dodoi{10.1086/426544}

\bibitem[{{Mercado} {et~al.}(2026){Mercado}, {Baxter}, {Rodriguez Wimberly}, {Moreno}, {Wheeler}, {Gandhi}, {Wetzel}, {Feldmann}, {Tortora}, \& {Samuel}}]{Mercado2026}
{Mercado}, F.~J., {Baxter}, D.~C., {Rodriguez Wimberly}, M.~K., {et~al.} 2026, \apj, 1004, 217, \dodoi{10.3847/1538-4357/ae7360}

\bibitem[{{Miyazaki} {et~al.}(2018){Miyazaki}, {Komiyama}, {Kawanomoto}, {Doi}, {Furusawa}, {Hamana}, {Hayashi}, {Ikeda}, {Kamata}, {Karoji}, {Koike}, {Kurakami}, {Miyama}, {Morokuma}, {Nakata}, {Namikawa}, {Nakaya}, {Nariai}, {Obuchi}, {Oishi}, {Okada}, {Okura}, {Tait}, {Takata}, {Tanaka}, {Tanaka}, {Terai}, {Tomono}, {Uraguchi}, {Usuda}, {Utsumi}, {Yamada}, {Yamanoi}, {Aihara}, {Fujimori}, {Mineo}, {Miyatake}, {Oguri}, {Uchida}, {Tanaka}, {Yasuda}, {Takada}, {Murayama}, {Nishizawa}, {Sugiyama}, {Chiba}, {Futamase}, {Wang}, {Chen}, {Ho}, {Liaw}, {Chiu}, {Ho}, {Lai}, {Lee}, {Jeng}, {Iwamura}, {Armstrong}, {Bickerton}, {Bosch}, {Gunn}, {Lupton}, {Loomis}, {Price}, {Smith}, {Strauss}, {Turner}, {Suzuki}, {Miyazaki}, {Muramatsu}, {Yamamoto}, {Endo}, {Ezaki}, {Ito}, {Kawaguchi}, {Sofuku}, {Taniike}, {Akutsu}, {Dojo}, {Kasumi}, {Matsuda}, {Imoto}, {Miwa}, {Suzuki}, {Takeshi}, \& {Yokota}}]{Miyazaki2018}
{Miyazaki}, S., {Komiyama}, Y., {Kawanomoto}, S., {et~al.} 2018, \pasj, 70, S1, \dodoi{10.1093/pasj/psx063}

\bibitem[{{Moore} {et~al.}(1999){Moore}, {Ghigna}, {Governato}, {Lake}, {Quinn}, {Stadel}, \& {Tozzi}}]{Moore1999}
{Moore}, B., {Ghigna}, S., {Governato}, F., {et~al.} 1999, \apjl, 524, L19, \dodoi{10.1086/312287}

\bibitem[{{Moustakas} \& {Lang}(in prep)}]{Moustakas2025_inprep}
{Moustakas}, J., \& {Lang}, D. in prep

\bibitem[{{Moustakas} {et~al.}(2023){Moustakas}, {Lang}, {Dey}, {Juneau}, {Meisner}, {Myers}, {Schlafly}, {Schlegel}, {Valdes}, {Weaver}, \& {Zhou}}]{Moustakas2023}
{Moustakas}, J., {Lang}, D., {Dey}, A., {et~al.} 2023, \apjs, 269, 3, \dodoi{10.3847/1538-4365/acfaa2}

\bibitem[{{M{\"u}ller} {et~al.}(2023){M{\"u}ller}, {Heesters}, {Jerjen}, {Anand}, \& {Revaz}}]{Muller2023}
{M{\"u}ller}, O., {Heesters}, N., {Jerjen}, H., {Anand}, G., \& {Revaz}, Y. 2023, \aap, 673, A160, \dodoi{10.1051/0004-6361/202345953}

\bibitem[{{M{\"u}ller} \& {Jerjen}(2020)}]{Muller2020}
{M{\"u}ller}, O., \& {Jerjen}, H. 2020, \aap, 644, A91, \dodoi{10.1051/0004-6361/202038862}

\bibitem[{{M{\"u}ller} {et~al.}(2018){M{\"u}ller}, {Pawlowski}, {Jerjen}, \& {Lelli}}]{Muller2018}
{M{\"u}ller}, O., {Pawlowski}, M.~S., {Jerjen}, H., \& {Lelli}, F. 2018, Science, 359, 534, \dodoi{10.1126/science.aao1858}

\bibitem[{{Munshi} {et~al.}(2019){Munshi}, {Brooks}, {Christensen}, {Applebaum}, {Holley-Bockelmann}, {Quinn}, \& {Wadsley}}]{Munshi2019}
{Munshi}, F., {Brooks}, A.~M., {Christensen}, C., {et~al.} 2019, \apj, 874, 40, \dodoi{10.3847/1538-4357/ab0085}

\bibitem[{{Mutlu-Pakdil} {et~al.}(2024){Mutlu-Pakdil}, {Sand}, {Crnojevi{\'c}}, {Bennet}, {Jones}, {Spekkens}, {Karunakaran}, {Zaritsky}, {Caldwell}, {Fielder}, {Guhathakurta}, {Seth}, {Simon}, {Strader}, \& {Toloba}}]{Mutlu-Pakdil2024}
{Mutlu-Pakdil}, B., {Sand}, D.~J., {Crnojevi{\'c}}, D., {et~al.} 2024, \apj, 966, 188, \dodoi{10.3847/1538-4357/ad36c4}

\bibitem[{{Nadler} {et~al.}(2020){Nadler}, {Wechsler}, {Bechtol}, {Mao}, {Green}, {Drlica-Wagner}, {McNanna}, {Mau}, {Pace}, {Simon}, {Kravtsov}, {Dodelson}, {Li}, {Riley}, {Wang}, {Abbott}, {Aguena}, {Allam}, {Annis}, {Avila}, {Bernstein}, {Bertin}, {Brooks}, {Burke}, {Rosell}, {Kind}, {Carretero}, {Costanzi}, {da Costa}, {De Vicente}, {Desai}, {Evrard}, {Flaugher}, {Fosalba}, {Frieman}, {Garc{\'\i}a-Bellido}, {Gaztanaga}, {Gerdes}, {Gruen}, {Gschwend}, {Gutierrez}, {Hartley}, {Hinton}, {Honscheid}, {Krause}, {Kuehn}, {Kuropatkin}, {Lahav}, {Maia}, {Marshall}, {Menanteau}, {Miquel}, {Palmese}, {Paz-Chinch{\'o}n}, {Plazas}, {Romer}, {Sanchez}, {Santiago}, {Scarpine}, {Serrano}, {Smith}, {Soares-Santos}, {Suchyta}, {Tarle}, {Thomas}, {Varga}, {Walker}, \& {DES Collaboration}}]{Nadler2020}
{Nadler}, E.~O., {Wechsler}, R.~H., {Bechtol}, K., {et~al.} 2020, \apj, 893, 48, \dodoi{10.3847/1538-4357/ab846a}

\bibitem[{{Nadler} {et~al.}(2021){Nadler}, {Drlica-Wagner}, {Bechtol}, {Mau}, {Wechsler}, {Gluscevic}, {Boddy}, {Pace}, {Li}, {McNanna}, {Riley}, {Garc{\'\i}a-Bellido}, {Mao}, {Green}, {Burke}, {Peter}, {Jain}, {Abbott}, {Aguena}, {Allam}, {Annis}, {Avila}, {Brooks}, {Carrasco Kind}, {Carretero}, {Costanzi}, {da Costa}, {De Vicente}, {Desai}, {Diehl}, {Doel}, {Everett}, {Evrard}, {Flaugher}, {Frieman}, {Gerdes}, {Gruen}, {Gruendl}, {Gschwend}, {Gutierrez}, {Hinton}, {Honscheid}, {Huterer}, {James}, {Krause}, {Kuehn}, {Kuropatkin}, {Lahav}, {Maia}, {Marshall}, {Menanteau}, {Miquel}, {Palmese}, {Paz-Chinch{\'o}n}, {Plazas}, {Romer}, {Sanchez}, {Scarpine}, {Serrano}, {Sevilla-Noarbe}, {Smith}, {Soares-Santos}, {Suchyta}, {Swanson}, {Tarle}, {Tucker}, {Walker}, {Wester}, \& {DES Collaboration}}]{Nadler2021}
{Nadler}, E.~O., {Drlica-Wagner}, A., {Bechtol}, K., {et~al.} 2021, \prl, 126, 091101, \dodoi{10.1103/PhysRevLett.126.091101}

\bibitem[{{Nadler} {et~al.}(2023){Nadler}, {Mansfield}, {Wang}, {Du}, {Adhikari}, {Banerjee}, {Benson}, {Darragh-Ford}, {Mao}, {Wagner-Carena}, {Wechsler}, \& {Wu}}]{Nadler2023}
{Nadler}, E.~O., {Mansfield}, P., {Wang}, Y., {et~al.} 2023, \apj, 945, 159, \dodoi{10.3847/1538-4357/acb68c}

\bibitem[{{Nagai} \& {Kravtsov}(2005)}]{Nagai2005}
{Nagai}, D., \& {Kravtsov}, A.~V. 2005, \apj, 618, 557, \dodoi{10.1086/426016}

\bibitem[{{Nelson} {et~al.}(2019){Nelson}, {Pillepich}, {Springel}, {Pakmor}, {Weinberger}, {Genel}, {Torrey}, {Vogelsberger}, {Marinacci}, \& {Hernquist}}]{Nelson2019}
{Nelson}, D., {Pillepich}, A., {Springel}, V., {et~al.} 2019, \mnras, 490, 3234, \dodoi{10.1093/mnras/stz2306}

\bibitem[{{Nidever} {et~al.}(2021){Nidever}, {Olsen}, {Choi}, {Ruiz-Lara}, {Miller}, {Johnson}, {Bell}, {Blum}, {Cioni}, {Gallart}, {Majewski}, {Martin}, {Massana}, {Monachesi}, {No{\"e}l}, {Sakowska}, {van der Marel}, {Walker}, {Zaritsky}, {Bell}, {Conn}, {de Boer}, {Gruendl}, {Monelli}, {Mu{\~n}oz}, {Saha}, {Vivas}, {Bernard}, {Besla}, {Carballo-Bello}, {Dorta}, {Martinez-Delgado}, {Goater}, {Rusakov}, \& {Stringfellow}}]{Nidever2021}
{Nidever}, D.~L., {Olsen}, K., {Choi}, Y., {et~al.} 2021, \aj, 161, 74, \dodoi{10.3847/1538-3881/abceb7}

\bibitem[{{Oke} \& {Gunn}(1983)}]{Oke1983}
{Oke}, J.~B., \& {Gunn}, J.~E. 1983, \apj, 266, 713, \dodoi{10.1086/160817}

\bibitem[{{Pace}(2025)}]{Pace2024}
{Pace}, A.~B. 2025, The Open Journal of Astrophysics, 8, 142, \dodoi{10.33232/001c.144859}

\bibitem[{{Pan} {et~al.}(2023){Pan}, {Simpson}, {Kravtsov}, {G{\'o}mez}, {Grand}, {Marinacci}, {Pakmor}, {Manwadkar}, \& {Esmerian}}]{Pan2022}
{Pan}, Y., {Simpson}, C.~M., {Kravtsov}, A., {et~al.} 2023, \mnras, 519, 4499, \dodoi{10.1093/mnras/stac3663}

\bibitem[{{Pan} {et~al.}(2026){Pan}, {Danieli}, {Greene}, {Li}, {Leauthaud}, {Kado-Fong}, {Luo}, {Mintz}, {Brooks}, {Huang}, {Peter}, {Bhattacharyya}, \& {Kelvin}}]{Pan2026}
{Pan}, Y., {Danieli}, S., {Greene}, J.~E., {et~al.} 2026, \apj, 1000, 119, \dodoi{10.3847/1538-4357/ae42cb}

\bibitem[{{Pardy} {et~al.}(2020){Pardy}, {D'Onghia}, {Navarro}, {Grand}, {G{\'o}mez}, {Marinacci}, {Pakmor}, {Simpson}, \& {Springel}}]{Pardy2020}
{Pardy}, S.~A., {D'Onghia}, E., {Navarro}, J.~F., {et~al.} 2020, \mnras, 492, 1543, \dodoi{10.1093/mnras/stz3192}

\bibitem[{{Patel} {et~al.}(2020){Patel}, {Kallivayalil}, {Garavito-Camargo}, {Besla}, {Weisz}, {van der Marel}, {Boylan-Kolchin}, {Pawlowski}, \& {G{\'o}mez}}]{Patel2020}
{Patel}, E., {Kallivayalil}, N., {Garavito-Camargo}, N., {et~al.} 2020, \apj, 893, 121, \dodoi{10.3847/1538-4357/ab7b75}

\bibitem[{{Paudel} {et~al.}(2026){Paudel}, {Sabiu}, {Yoon}, {Chhatkuli}, {Yoo}, \& {Pokhrel}}]{Paudel2026}
{Paudel}, S., {Sabiu}, C.~G., {Yoon}, S.-J., {et~al.} 2026, arXiv e-prints, arXiv:2606.16299, \dodoi{10.48550/arXiv.2606.16299}

\bibitem[{{Pillepich} {et~al.}(2018){Pillepich}, {Springel}, {Nelson}, {Genel}, {Naiman}, {Pakmor}, {Hernquist}, {Torrey}, {Vogelsberger}, {Weinberger}, \& {Marinacci}}]{Pillepich2018}
{Pillepich}, A., {Springel}, V., {Nelson}, D., {et~al.} 2018, \mnras, 473, 4077, \dodoi{10.1093/mnras/stx2656}

\bibitem[{{Pillepich} {et~al.}(2019){Pillepich}, {Nelson}, {Springel}, {Pakmor}, {Torrey}, {Weinberger}, {Vogelsberger}, {Marinacci}, {Genel}, {van der Wel}, \& {Hernquist}}]{Pillepich2019}
{Pillepich}, A., {Nelson}, D., {Springel}, V., {et~al.} 2019, \mnras, 490, 3196, \dodoi{10.1093/mnras/stz2338}

\bibitem[{{Putman} {et~al.}(2021){Putman}, {Zheng}, {Price-Whelan}, {Grcevich}, {Johnson}, {Tollerud}, \& {Peek}}]{Putman2021}
{Putman}, M.~E., {Zheng}, Y., {Price-Whelan}, A.~M., {et~al.} 2021, \apj, 913, 53, \dodoi{10.3847/1538-4357/abe391}

\bibitem[{{Rodr{\'\i}guez-Puebla} {et~al.}(2017){Rodr{\'\i}guez-Puebla}, {Primack}, {Avila-Reese}, \& {Faber}}]{RP2017}
{Rodr{\'\i}guez-Puebla}, A., {Primack}, J.~R., {Avila-Reese}, V., \& {Faber}, S.~M. 2017, \mnras, 470, 651, \dodoi{10.1093/mnras/stx1172}

\bibitem[{{Sabbi} {et~al.}(2018){Sabbi}, {Calzetti}, {Ubeda}, {Adamo}, {Cignoni}, {Thilker}, {Aloisi}, {Elmegreen}, {Elmegreen}, {Gouliermis}, {Grebel}, {Messa}, {Smith}, {Tosi}, {Dolphin}, {Andrews}, {Ashworth}, {Bright}, {Brown}, {Chandar}, {Christian}, {Clayton}, {Cook}, {Dale}, {de Mink}, {Dobbs}, {Evans}, {Fumagalli}, {Gallagher}, {Grasha}, {Herrero}, {Hunter}, {Johnson}, {Kahre}, {Kennicutt}, {Kim}, {Krumholz}, {Lee}, {Lennon}, {Martin}, {Nair}, {Nota}, {{\"O}stlin}, {Pellerin}, {Prieto}, {Regan}, {Ryon}, {Sacchi}, {Schaerer}, {Schiminovich}, {Shabani}, {Van Dyk}, {Walterbos}, {Whitmore}, \& {Wofford}}]{Sabbi2018}
{Sabbi}, E., {Calzetti}, D., {Ubeda}, L., {et~al.} 2018, \apjs, 235, 23, \dodoi{10.3847/1538-4365/aaa8e5}

\bibitem[{{Sacchi} {et~al.}(2021){Sacchi}, {Richstein}, {Kallivayalil}, {van der Marel}, {Libralato}, {Zivick}, {Besla}, {Brown}, {Choi}, {Deason}, {Fritz}, {Geha}, {Guhathakurta}, {Jeon}, {Kirby}, {Majewski}, {Patel}, {Simon}, {Tony Sohn}, {Tollerud}, \& {Wetzel}}]{Sacchi2021}
{Sacchi}, E., {Richstein}, H., {Kallivayalil}, N., {et~al.} 2021, \apjl, 920, L19, \dodoi{10.3847/2041-8213/ac2aa3}

\bibitem[{{Sales} {et~al.}(2017){Sales}, {Navarro}, {Kallivayalil}, \& {Frenk}}]{Sales2017}
{Sales}, L.~V., {Navarro}, J.~F., {Kallivayalil}, N., \& {Frenk}, C.~S. 2017, \mnras, 465, 1879, \dodoi{10.1093/mnras/stw2816}

\bibitem[{{Sales} {et~al.}(2013){Sales}, {Wang}, {White}, \& {Navarro}}]{Sales2013}
{Sales}, L.~V., {Wang}, W., {White}, S. D.~M., \& {Navarro}, J.~F. 2013, \mnras, 428, 573, \dodoi{10.1093/mnras/sts054}

\bibitem[{{Sales} {et~al.}(2022){Sales}, {Wetzel}, \& {Fattahi}}]{Sales2022}
{Sales}, L.~V., {Wetzel}, A., \& {Fattahi}, A. 2022, Nature Astronomy, 6, 897, \dodoi{10.1038/s41550-022-01689-w}

\bibitem[{{Samuel} {et~al.}(2020){Samuel}, {Wetzel}, {Tollerud}, {Garrison-Kimmel}, {Loebman}, {El-Badry}, {Hopkins}, {Boylan-Kolchin}, {Faucher-Gigu{\`e}re}, {Bullock}, {Benincasa}, \& {Bailin}}]{Samuel2020}
{Samuel}, J., {Wetzel}, A., {Tollerud}, E., {et~al.} 2020, \mnras, 491, 1471, \dodoi{10.1093/mnras/stz3054}

\bibitem[{{Sand} {et~al.}(2015){Sand}, {Spekkens}, {Crnojevi{\'c}}, {Hargis}, {Willman}, {Strader}, \& {Grillmair}}]{Sand2015}
{Sand}, D.~J., {Spekkens}, K., {Crnojevi{\'c}}, D., {et~al.} 2015, \apjl, 812, L13, \dodoi{10.1088/2041-8205/812/1/L13}

\bibitem[{{Sand} {et~al.}(2024){Sand}, {Mutlu-Pakdil}, {Jones}, {Karunakaran}, {Andrews}, {Bennet}, {Crnojevi{\'c}}, {Donatiello}, {Drlica-Wagner}, {Fielder}, {Mart{\'\i}nez-Delgado}, {Mart{\'\i}nez-V{\'a}zquez}, {Spekkens}, {Doliva-Dolinsky}, {Hunter}, {Carlin}, {Cerny}, {Hai}, {McQuinn}, {Pace}, \& {Smercina}}]{Sand2024}
{Sand}, D.~J., {Mutlu-Pakdil}, B., {Jones}, M.~G., {et~al.} 2024, \apjl, 977, L5, \dodoi{10.3847/2041-8213/ad927c}

\bibitem[{{Santos-Santos} {et~al.}(2021){Santos-Santos}, {Fattahi}, {Sales}, \& {Navarro}}]{Santos-Santos2021}
{Santos-Santos}, I. M.~E., {Fattahi}, A., {Sales}, L.~V., \& {Navarro}, J.~F. 2021, \mnras, 504, 4551, \dodoi{10.1093/mnras/stab1020}

\bibitem[{{Santos-Santos} {et~al.}(2022){Santos-Santos}, {Sales}, {Fattahi}, \& {Navarro}}]{Santos-Santos2022}
{Santos-Santos}, I. M.~E., {Sales}, L.~V., {Fattahi}, A., \& {Navarro}, J.~F. 2022, \mnras, 515, 3685, \dodoi{10.1093/mnras/stac2057}

\bibitem[{{Savino} {et~al.}(2025){Savino}, {Weisz}, {Dolphin}, {Durbin}, {Kallivayalil}, {Wetzel}, {Anderson}, {Besla}, {Boylan-Kolchin}, {Brown}, {Bullock}, {Cole}, {Collins}, {Cooper}, {Deason}, {Dotter}, {Fardal}, {Ferguson}, {Fritz}, {Geha}, {Gilbert}, {Guhathakurta}, {Ibata}, {Irwin}, {Jeon}, {Kirby}, {Lewis}, {Mackey}, {Majewski}, {Martin}, {McConnachie}, {Patel}, {Rich}, {Skillman}, {Simon}, {Sohn}, {Tollerud}, \& {van der Marel}}]{Savino2025}
{Savino}, A., {Weisz}, D.~R., {Dolphin}, A.~E., {et~al.} 2025, \apj, 979, 205, \dodoi{10.3847/1538-4357/ada24f}

\bibitem[{{Schlafly} \& {Finkbeiner}(2011)}]{Schlafly2011}
{Schlafly}, E.~F., \& {Finkbeiner}, D.~P. 2011, \apj, 737, 103, \dodoi{10.1088/0004-637X/737/2/103}

\bibitem[{{Schlegel} {et~al.}(1998){Schlegel}, {Finkbeiner}, \& {Davis}}]{SFD1998}
{Schlegel}, D.~J., {Finkbeiner}, D.~P., \& {Davis}, M. 1998, \apj, 500, 525, \dodoi{10.1086/305772}

\bibitem[{{Shi} {et~al.}(2020){Shi}, {Wang}, {Mo}, {Vogelsberger}, {Ho}, {Du}, {Nelson}, {Pillepich}, \& {Hernquist}}]{Shi2020}
{Shi}, J., {Wang}, H., {Mo}, H., {et~al.} 2020, \apj, 893, 139, \dodoi{10.3847/1538-4357/ab8464}

\bibitem[{{Simon}(2019)}]{Simon2019}
{Simon}, J.~D. 2019, \araa, 57, 375, \dodoi{10.1146/annurev-astro-091918-104453}

\bibitem[{Simpson {et~al.}(2024)Simpson, Labrie, Teal, Berke, Turner, Smirnova, \& Vacca}]{DRAGONS_software}
Simpson, C., Labrie, K., Teal, D., {et~al.} 2024, DRAGONS, 3.2.2,  Zenodo, \dodoi{10.5281/zenodo.13821517}

\bibitem[{{Skibba} {et~al.}(2012){Skibba}, {Engelbracht}, {Aniano}, {Babler}, {Bernard}, {Bot}, {Carlson}, {Galametz}, {Galliano}, {Gordon}, {Hony}, {Israel}, {Lebouteiller}, {Li}, {Madden}, {Meixner}, {Misselt}, {Montiel}, {Okumura}, {Panuzzo}, {Paradis}, {Roman-Duval}, {Rubio}, {Sauvage}, {Seale}, {Srinivasan}, \& {van Loon}}]{Skibba2012LMC}
{Skibba}, R.~A., {Engelbracht}, C.~W., {Aniano}, G., {et~al.} 2012, \apj, 761, 42, \dodoi{10.1088/0004-637X/761/1/42}

\bibitem[{{Smercina} {et~al.}(2018){Smercina}, {Bell}, {Price}, {D'Souza}, {Slater}, {Bailin}, {Monachesi}, \& {Nidever}}]{Smercina2018}
{Smercina}, A., {Bell}, E.~F., {Price}, P.~A., {et~al.} 2018, \apj, 863, 152, \dodoi{10.3847/1538-4357/aad2d6}

\bibitem[{{Spergel} {et~al.}(2015){Spergel}, {Gehrels}, {Baltay}, {Bennett}, {Breckinridge}, {Donahue}, {Dressler}, {Gaudi}, {Greene}, {Guyon}, {Hirata}, {Kalirai}, {Kasdin}, {Macintosh}, {Moos}, {Perlmutter}, {Postman}, {Rauscher}, {Rhodes}, {Wang}, {Weinberg}, {Benford}, {Hudson}, {Jeong}, {Mellier}, {Traub}, {Yamada}, {Capak}, {Colbert}, {Masters}, {Penny}, {Savransky}, {Stern}, {Zimmerman}, {Barry}, {Bartusek}, {Carpenter}, {Cheng}, {Content}, {Dekens}, {Demers}, {Grady}, {Jackson}, {Kuan}, {Kruk}, {Melton}, {Nemati}, {Parvin}, {Poberezhskiy}, {Peddie}, {Ruffa}, {Wallace}, {Whipple}, {Wollack}, \& {Zhao}}]{Spergel2015}
{Spergel}, D., {Gehrels}, N., {Baltay}, C., {et~al.} 2015, arXiv e-prints, arXiv:1503.03757, \dodoi{10.48550/arXiv.1503.03757}

\bibitem[{{Springel} {et~al.}(2001){Springel}, {White}, {Tormen}, \& {Kauffmann}}]{Springel2001}
{Springel}, V., {White}, S. D.~M., {Tormen}, G., \& {Kauffmann}, G. 2001, \mnras, 328, 726, \dodoi{10.1046/j.1365-8711.2001.04912.x}

\bibitem[{{Stierwalt} {et~al.}(2026){Stierwalt}, {Luber}, {Goldberg Little}, {Privon}, {Besla}, {Johnson}, {Kallivayalil}, {Patton}, {Putman}, \& {Simpson-Heil}}]{Stierwalt2026}
{Stierwalt}, S., {Luber}, N., {Goldberg Little}, Z., {et~al.} 2026, \apj, 997, 132, \dodoi{10.3847/1538-4357/ae23cc}

\bibitem[{{Tan} {et~al.}(2026){Tan}, {Drlica-Wagner}, {Pace}, {Cerny}, {Nadler}, {Doliva-Dolinsky}, {Anbajagane}, {Li}, {Simon}, {Vivas}, {Walker}, {Adam{\'o}w}, {Bechtol}, {Carlin}, {Casey}, {Chang}, {Chaturvedi}, {Cheng}, {Chiti}, {Choi}, {Crnojevi{\'c}}, {Ferguson}, {Gruendl}, {Ji}, {Limberg}, {Medina}, {Mutlu-Pakdil}, {No{\"e}l}, {Overdeck}, {Placco}, {Riley}, {Sand}, {Sharp}, {Sherman}, {Stringfellow}, {Wechsler}, {Aguena}, {Allam}, {Alves}, {Bacon}, {Brooks}, {Burke}, {Camilleri}, {Carballo-Bello}, {Carnero Rosell}, {Carretero}, {da Costa}, {da Silva Pereira}, {Davis}, {de Vicente}, {Desai}, {Everett}, {Flaugher}, {Frieman}, {Garc{\'\i}a-Bellido}, {Gruen}, {Gutierrez}, {Herner}, {Hinton}, {Hollowood}, {James}, {Kuehn}, {Lahav}, {Lee}, {Marshall}, {Mart{\'\i}nez-V{\'a}zquez}, {Massana}, {Mena-Fern{\'a}ndez}, {Miquel}, {Muir}, {Myles}, {Ogando}, {Plazas Malag{\'o}n}, {Porredon}, {Sanchez}, {Sanchez Cid}, {Sevilla-Noarbe}, {Smith}, {Suchyta}, {Swanson}, {To}, {Tollerud}, {Tucker}, {Vikram}, {Weaverdyck},
  {Yamamoto}, {Zenteno}, {Delve Collaboration}, \& {Des Collaboration}}]{Tan2025}
{Tan}, C.~Y., {Drlica-Wagner}, A., {Pace}, A.~B., {et~al.} 2026, \apj, 1000, 87, \dodoi{10.3847/1538-4357/ae4479}

\bibitem[{{Tanaka} {et~al.}(2018){Tanaka}, {Chiba}, {Hayashi}, {Komiyama}, {Okamoto}, {Cooper}, {Okamoto}, \& {Spitler}}]{Tanaka2018}
{Tanaka}, M., {Chiba}, M., {Hayashi}, K., {et~al.} 2018, \apj, 865, 125, \dodoi{10.3847/1538-4357/aad9fe}

\bibitem[{{Tanaka} {et~al.}(2021){Tanaka}, {Ikeda}, {Murata}, {Takita}, {Mineo}, {Koike}, {Okura}, \& {Harasawa}}]{Tanaka2021}
{Tanaka}, M., {Ikeda}, H., {Murata}, K., {et~al.} 2021, \pasj, 73, 735, \dodoi{10.1093/pasj/psab034}

\bibitem[{{Tonry} \& {Schneider}(1988)}]{Tonry1988}
{Tonry}, J., \& {Schneider}, D.~P. 1988, \aj, 96, 807, \dodoi{10.1086/114847}

\bibitem[{{Tully} {et~al.}(2016){Tully}, {Courtois}, \& {Sorce}}]{Tully2016}
{Tully}, R.~B., {Courtois}, H.~M., \& {Sorce}, J.~G. 2016, \aj, 152, 50, \dodoi{10.3847/0004-6256/152/2/50}

\bibitem[{{Tully} {et~al.}(2009){Tully}, {Rizzi}, {Shaya}, {Courtois}, {Makarov}, \& {Jacobs}}]{Tully2009}
{Tully}, R.~B., {Rizzi}, L., {Shaya}, E.~J., {et~al.} 2009, \aj, 138, 323, \dodoi{10.1088/0004-6256/138/2/323}

\bibitem[{{Tully} {et~al.}(2013){Tully}, {Courtois}, {Dolphin}, {Fisher}, {H{\'e}raudeau}, {Jacobs}, {Karachentsev}, {Makarov}, {Makarova}, {Mitronova}, {Rizzi}, {Shaya}, {Sorce}, \& {Wu}}]{Tully2013}
{Tully}, R.~B., {Courtois}, H.~M., {Dolphin}, A.~E., {et~al.} 2013, \aj, 146, 86, \dodoi{10.1088/0004-6256/146/4/86}

\bibitem[{{Tully} {et~al.}(2023){Tully}, {Kourkchi}, {Courtois}, {Anand}, {Blakeslee}, {Brout}, {Jaeger}, {Dupuy}, {Guinet}, {Howlett}, {Jensen}, {Pomar{\`e}de}, {Rizzi}, {Rubin}, {Said}, {Scolnic}, \& {Stahl}}]{Tully2023}
{Tully}, R.~B., {Kourkchi}, E., {Courtois}, H.~M., {et~al.} 2023, \apj, 944, 94, \dodoi{10.3847/1538-4357/ac94d8}

\bibitem[{{van Dokkum}(2001)}]{LACosmic}
{van Dokkum}, P.~G. 2001, \pasp, 113, 1420, \dodoi{10.1086/323894}

\bibitem[{Virtanen {et~al.}(2020)Virtanen, Gommers, Oliphant, Haberland, Reddy, Cournapeau, Burovski, Peterson, Weckesser, Bright, {van der Walt}, Brett, Wilson, Millman, Mayorov, Nelson, Jones, Kern, Larson, Carey, Polat, Feng, Moore, {VanderPlas}, Laxalde, Perktold, Cimrman, Henriksen, Quintero, Harris, Archibald, Ribeiro, Pedregosa, {van Mulbregt}, \& {SciPy 1.0 Contributors}}]{scipy}
Virtanen, P., Gommers, R., Oliphant, T.~E., {et~al.} 2020, Nature Methods, 17, 261, \dodoi{10.1038/s41592-019-0686-2}

\bibitem[{{Wang} {et~al.}(2012){Wang}, {Frenk}, {Navarro}, {Gao}, \& {Sawala}}]{Wang2012}
{Wang}, J., {Frenk}, C.~S., {Navarro}, J.~F., {Gao}, L., \& {Sawala}, T. 2012, \mnras, 424, 2715, \dodoi{10.1111/j.1365-2966.2012.21357.x}

\bibitem[{{Wang} {et~al.}(2021){Wang}, {Takada}, {Li}, {Carlsten}, {Lan}, {Shi}, {Miyatake}, {More}, {Beaton}, {Lupton}, {Lin}, {Qiu}, \& {Luo}}]{Wang2021}
{Wang}, W., {Takada}, M., {Li}, X., {et~al.} 2021, \mnras, 500, 3776, \dodoi{10.1093/mnras/staa3495}

\bibitem[{{Wang} {et~al.}(2025){Wang}, {Yang}, {Jing}, {Ross}, {Siudek}, {Moustakas}, {Moore}, {Cole}, {Frenk}, {Yu}, {Koposov}, {Han}, {Tan}, {Xu}, {Gu}, {Wang}, {Gnedin}, {Aguilar}, {Ahlen}, {Bianchi}, {Brooks}, {Claybaugh}, {de la Macorra}, {Dey}, {Doel}, {Forero-Romero}, {Gazta{\~n}aga}, {Gontcho A Gontcho}, {Gutierrez}, {Honscheid}, {Ishak}, {Kisner}, {Landriau}, {Le Guillou}, {Manera}, {Meisner}, {Miquel}, {Nadathur}, {Poppett}, {Prada}, {P{\'e}rez-R{\`a}fols}, {Rossi}, {Sanchez}, {Schlegel}, {Seo}, {Silber}, {Sprayberry}, {Tarl{\'e}}, {Weaver}, \& {Zou}}]{Wang2025}
{Wang}, W., {Yang}, X., {Jing}, Y., {et~al.} 2025, \apj, 986, 218, \dodoi{10.3847/1538-4357/add5df}

\bibitem[{{Wang} {et~al.}(2024){Wang}, {Nadler}, {Mao}, {Wechsler}, {Abel}, {Behroozi}, {Geha}, {Asali}, {de los Reyes}, {Kado-Fong}, {Kallivayalil}, {Tollerud}, {Weiner}, \& {Wu}}]{SAGA-V}
{Wang}, Y., {Nadler}, E.~O., {Mao}, Y.-Y., {et~al.} 2024, \apj, 976, 119, \dodoi{10.3847/1538-4357/ad7f4c}

\bibitem[{{Weisz} {et~al.}(2011){Weisz}, {Dalcanton}, {Williams}, {Gilbert}, {Skillman}, {Seth}, {Dolphin}, {McQuinn}, {Gogarten}, {Holtzman}, {Rosema}, {Cole}, {Karachentsev}, \& {Zaritsky}}]{Weisz2011}
{Weisz}, D.~R., {Dalcanton}, J.~J., {Williams}, B.~F., {et~al.} 2011, \apj, 739, 5, \dodoi{10.1088/0004-637X/739/1/5}

\bibitem[{{Wenger} {et~al.}(2000){Wenger}, {Ochsenbein}, {Egret}, {Dubois}, {Bonnarel}, {Borde}, {Genova}, {Jasniewicz}, {Lalo{\"e}}, {Lesteven}, \& {Monier}}]{SIMBAD}
{Wenger}, M., {Ochsenbein}, F., {Egret}, D., {et~al.} 2000, \aaps, 143, 9, \dodoi{10.1051/aas:2000332}

\bibitem[{{Wetzel} {et~al.}(2015){Wetzel}, {Tollerud}, \& {Weisz}}]{Wetzel2015}
{Wetzel}, A.~R., {Tollerud}, E.~J., \& {Weisz}, D.~R. 2015, \apjl, 808, L27, \dodoi{10.1088/2041-8205/808/1/L27}

\bibitem[{{Wheeler} {et~al.}(2015){Wheeler}, {O{\~n}orbe}, {Bullock}, {Boylan-Kolchin}, {Elbert}, {Garrison-Kimmel}, {Hopkins}, \& {Kere{\v{s}}}}]{Wheeler2015}
{Wheeler}, C., {O{\~n}orbe}, J., {Bullock}, J.~S., {et~al.} 2015, \mnras, 453, 1305, \dodoi{10.1093/mnras/stv1691}

\bibitem[{{Willmer}(2018)}]{Willmer2018}
{Willmer}, C. N.~A. 2018, \apjs, 236, 47, \dodoi{10.3847/1538-4365/aabfdf}

\bibitem[{{Zaritsky} {et~al.}(2023){Zaritsky}, {Donnerstein}, {Dey}, {Karunakaran}, {Kadowaki}, {Khim}, {Spekkens}, \& {Zhang}}]{Zaritsky2023}
{Zaritsky}, D., {Donnerstein}, R., {Dey}, A., {et~al.} 2023, \apjs, 267, 27, \dodoi{10.3847/1538-4365/acdd71}

\bibitem[{{Zhang} {et~al.}(2021){Zhang}, {Mackey}, \& {Da Costa}}]{Zhang2021_NGC6822}
{Zhang}, S., {Mackey}, D., \& {Da Costa}, G.~S. 2021, \mnras, 508, 2098, \dodoi{10.1093/mnras/stab2642}

\bibitem[{{Zheng} {et~al.}(2024){Zheng}, {Faerman}, {Oppenheimer}, {Putman}, {McQuinn}, {Kirby}, {Burchett}, {Telford}, {Werk}, \& {Kim}}]{Zheng2024}
{Zheng}, Y., {Faerman}, Y., {Oppenheimer}, B.~D., {et~al.} 2024, \apj, 960, 55, \dodoi{10.3847/1538-4357/acfe6b}

\bibitem[{{Zhu} {et~al.}(2025){Zhu}, {Asali}, {Putman}, {Westmeier}, {de Blok}, {Catinella}, {Deg}, {For}, {Kleiner}, {Lee-Waddell}, {Maccagni}, {Pisano}, {Shen}, {Spekkens}, \& {Staveley-Smith}}]{Zhu2025}
{Zhu}, J., {Asali}, Y., {Putman}, M., {et~al.} 2025, arXiv e-prints, arXiv:2510.27019, \dodoi{10.48550/arXiv.2510.27019}

\bibitem[{{Zou} {et~al.}(2017){Zou}, {Zhou}, {Fan}, {Zhang}, {Zhou}, {Nie}, {Peng}, {McGreer}, {Jiang}, {Dey}, {Fan}, {He}, {Jiang}, {Lang}, {Lesser}, {Ma}, {Mao}, {Schlegel}, \& {Wang}}]{BASS}
{Zou}, H., {Zhou}, X., {Fan}, X., {et~al.} 2017, \pasp, 129, 064101, \dodoi{10.1088/1538-3873/aa65ba}

\end{thebibliography}
\bibliographystyle{aasjournal}

\newpage
\appendix

\end{CJK*}
\end{document}